\documentclass[11pt,a4paper]{article}
\pdfoutput=1
\usepackage{jheppub}

\usepackage{amsmath,amssymb,calc}
\usepackage{color}
\usepackage{graphicx}
\makeatletter
\renewcommand\section{\@startsection {section}{1}{\z@}%
                                   {-3.5ex \@plus -1ex \@minus -.2ex}%nn
                                   {2.3ex \@plus.2ex}%
                                   {\normalfont\large\bfseries}}
\renewcommand\subsection{\@startsection{subsection}{2}{\z@}%
                                     {-3.25ex\@plus -1ex \@minus -.2ex}%
                                     {1.5ex \@plus .2ex}%
                                     {\normalfont\bfseries}}
\makeatother

\let\h=\eta

\newcommand{\bea}{\begin{eqnarray}}
\newcommand{\eea}{\end{eqnarray}}
\newcommand{\be}{\begin{equation}}
\newcommand{\ee}{\end{equation}}
\newcommand{\bma}{\begin{pmatrix}}
\newcommand{\ema}{\end{pmatrix}}

\newcommand{\cF}{{\cal F}}

\def\barray{\begin{array}}
\def\earray{\end{array}}
\def\be{\begin{equation}}
\def\ee{\end{equation}}
\def\ben{\begin{equation} \nonumber}
\def\een{\end{equation}}
\def\ban{\begin{eqnarray*}}
\def\ean{\end{eqnarray*}}
\def\ba{\begin{eqnarray}}
\def\ea{\end{eqnarray}}

\def\curv{\mathcal{R}}

\def\({\left(}
\def\){\right)}
\def\[{\left[}
\def\]{\right]}
\def\half{{1\over2}}
\def\tr{{\rm Tr}}
\def\nn{\nonumber}

\def\h{\mathcal{H}}

\def\hz{\hat{z}}
\def\hdp{\hat{\delta\phi}}

\def\v{\mathcal{V}}
\def\mpsi{m_{\psi}}

\def\tr{{\rm Tr}}
\def\hg{{\hat\gamma}}
\def\hy{{\hat t}}

\def\MM{{\cal M}}
\def\GG{{\cal G}}

\def\AA{{\cal A}}

\def\zhat{{\hat z}}
\def\phihat{{\hat \phi}}
\def\One{{\hbox{ 1\kern-.8mm l}}}

\def\dphi{{\delta\phi}}
\def\dphib{{\delta\bar\phi}}

\def\ada{{\hat X}}
\def\adab{\hat {\bar X}}
\def\zhat{{\hat z}}
\def\zhatb{{\hat {\bar z}}}
\def\phihat{{\hat{\delta\!\phi}}}
\def\phihatb{\hat{\bar{\delta\!\phi}}}
\def\varphihat{{\hat\varphi}}

\def\axsub{{\hat x}}
\def\slow{{\rm slow}}

\def\infsub{{\!\infty}}

\def\phitil{{\boldsymbol\varphi}}
\def\phitilb{{\bar{\boldsymbol\varphi}}}
\def\ztil{{\boldsymbol z}}
\def\ztilb{{\bar{\boldsymbol z}}}

\def\ie{{\it i.e.}}

\newcommand {\bflam}{{\boldsymbol\Lambda}}

\newcommand{\bfgam}{{\boldsymbol\Gamma}}
\newcommand {\bfnu}{{\boldsymbol\nu}}
\newcommand{\bfxi}{{\boldsymbol\xi}}
\newcommand {\bfkap}{{\boldsymbol\kappa}}
\newcommand {\bfbeta}{{\boldsymbol\beta}}

\def\da{\Psi}
\def\dN{\delta\!N}

\def\axion{\mathcal{X}}
\def\daxion{\delta\!\mathcal{X}}

\definecolor{darkgreen}{cmyk}{0.85,0.2,1.00,0.2}

\newcommand{\pref}[1]{$(\ref{#1})$}

\def\be{\begin{equation}}
\def\ee{\end{equation}}
\def\bea{\begin{eqnarray}}
\def\eea{\end{eqnarray}}

\def\ie{{\it i.e.}}

\def\IZ{\relax\ifmmode\mathchoice
{\hbox{\cmss Z\kern-.4em Z}}{\hbox{\cmss Z\kern-.4em Z}}
{\lower.9pt\hbox{\cmsss Z\kern-.4em Z}} {\lower1.2pt\hbox{\cmsss
Z\kern-.4em Z}}\else{\cmss Z\kern-.4em Z}\fi}
\def\IR{\relax{\rm I\kern-.18em R}}
\def\One{{\hbox{ 1\kern-.8mm l}}}

\def\tr{{\rm Tr\,}}

\newlength{\bredde}
\def\slash#1{\settowidth{\bredde}{$#1$}\ifmmode\,\raisebox{.15ex}{/}
\hspace*{-\bredde} #1\else$\,\raisebox{.15ex}{/}\hspace*{-\bredde}
#1$\fi}

\newsavebox{\zzzbar}
\sbox{\zzzbar}
  {\setlength{\unitlength}{0.9em}
  \begin{picture}(0.6,0.7)
  \thinlines
  \put(0,0){\line(1,0){0.6}}
  \put(0,0.75){\line(1,0){0.575}}
  \multiput(0,0)(0.0125,0.025){30}{\rule{0.3pt}{0.3pt}}
  \multiput(0.2,0)(0.0125,0.025){30}{\rule{0.3pt}{0.3pt}}
  \put(0,0.75){\line(0,-1){0.15}}
  \put(0.015,0.75){\line(0,-1){0.1}}
  \put(0.03,0.75){\line(0,-1){0.075}}
  \put(0.045,0.75){\line(0,-1){0.05}}
  \put(0.05,0.75){\line(0,-1){0.025}}
  \put(0.6,0){\line(0,1){0.15}}
  \put(0.585,0){\line(0,1){0.1}}
  \put(0.57,0){\line(0,1){0.075}}
  \put(0.555,0){\line(0,1){0.05}}
  \put(0.55,0){\line(0,1){0.025}}
  \end{picture}}

\def\Im{{\rm Im ~}}

\newcommand{\ena}{\end{eqnarray}}
\newcommand{\beqa}{\begin{eqnarray}}
\newcommand{\eeqa}{\end{eqnarray}}

\newfont{\goth}{ygoth.tfm scaled 1200}                   % gothic font (usual)
\numberwithin{equation}{section}

\renewcommand{\thesection}{\arabic{section}}

\title{Perturbations in Chromo-Natural Inflation}

\author[a,b]{Peter Adshead}
\author[a,c]{Emil Martinec}
\author[a,b,d]{Mark Wyman}

\affiliation[a]{Enrico Fermi Institute} 
\affiliation[b]{Kavli Institute for Cosmological Physics}
\affiliation[c]{Department of Physics}
\affiliation[d]{Department of Astronomy and Astrophysics\\ University of Chicago, Chicago, IL 60637, USA}

\emailAdd{adshead@kicp.uchicago.edu} 
\emailAdd{ejmartin@uchicago.edu} 
\emailAdd{markwy@oddjob.uchicago.edu}

\abstract{
Chromo-Natural Inflation is the first worked example of a model of inflation in which slow-roll inflation is achieved by ``magnetic drift" as opposed to Hubble friction. In this work, we give an account of the perturbations at linear order in this model. Our analysis uncovers two novel phenomena. First, the amplitude of scalar curvature perturbations is not directly tied to the shape of the inflationary potential. This allows the theory to violate na\"ive formulations of the Lyth bound. Second, the tensor sector of the theory is significantly altered from the usual case: the non-Abelian gauge field perturbations have a tensor degree of freedom. One chirality of the this tensor can be exponentially enhanced by a temporary instability near horizon crossing; this chiral instability exists because of the classical gauge field background, which violates parity.  These tensor fluctuations of the gauge field also couple to gravitational waves at linear order in perturbation theory and 
source a chiral spectrum of gravitational waves. This spectrum can be exponentially enhanced over the usual inflationary spectrum due to the instability in the gauge sector.  These new features cause the theory in its present form to be in significant tension with current observational data. 
 This is because the new scalar physics leads to a significant reddening of the spectral tilt in the same region of parameter space where the exponential enhancement of the gravitational wave amplitude is small enough to satisfy current constraints on the tensor-to-scalar index.  Hence,  the model either predicts a spectral tilt that is too red, or it overproduces gravitational waves, or both.
  }

\begin{document}

\maketitle
\flushbottom

\section{Introduction}

Inflation \cite{Guth:1980zm, Linde:1981mu, Albrecht:1982wi} is a remarkably successful paradigm, simultaneously solving fine tuning problems associated with the initial conditions of the standard hot big bang scenario while providing a mechanism for generating primordial fluctuations with the right amplitude and scale dependence to seed structure formation \cite{Mukhanov:1981xt,Chibisov:1982nx}, as well as possibly producing primordial gravitational waves \cite{Starobinsky:1979ty}.

At the classical level, generating a period of accelerated expansion is simple to achieve by arranging for the potential energy of a scalar field to dominate the Universe's energy density. 
Extending this period to a sufficient length to solve the flatness and horizon problems requires that the scalar must remain potential dominated for a long time. In the case of a single scalar field with a canonical kinetic term, this translates into a requirement that the potential be very flat, or slowly varying, relative to its height.
Maintaining this epoch of ``slow roll" also requires that the curvature of the potential also be small.  It is by now well established that there are many functional forms and parameters for potentials which have the desired properties. However, since these are scalar field theories, they generally lack symmetries that protect their potentials against radiative corrections. These radiative corrections are generically large, and spoil the flatness of the inflationary potential. Tuning these corrections away presents a hierarchy problem. 
In inflationary cosmology, this hierarchy problem is known as the `eta' problem, since the curvature of the potential is represented symbolically by the greek letter, $\eta$, in much of the inflationary literature. Many attempts to evade this problem exist. One that is particularly relevant for the present study is the model known as Natural Inflation, which solves the problem by using an axion as the inflationary scalar. \cite{Freese:1990rb, Adams:1992bn}. One can then argue that the form and shallowness of the potential are protected 
from quantum corrections by the axion's residual shift symmetry. Unfortunately for Natural Inflation, matching cosmic microwave background observations requires the model to have a Planck-scale axion decay constant, $f$ \cite{Freese:2004un}. It has thus far proved impossible to realize such a set-up in a UV-complete model such as string theory \cite{Banks:2003sx}. Recently, models have been found in string theory where an axion undergoes monodromy, which appears to be a more promising direction for realizing axion inflation 
in a way that is under theoretical control \cite{Silverstein:2008sg, McAllister:2008hb}.

There is an alternative means for achieving a long period of slow roll, where the interaction of a pseudo-scalar with gauge fields replaces a flat scalar potential. Since the resulting theories can have steep scalar potentials, and since the interactions of gauge theories are constrained by gauge symmetry, one might hope that realizing an inflationary epoch in this way might evade the `eta' problem. The (copious) emission of Abelian gauge quanta to permit axion inflation to proceed slowly despite the presence of a steep potential was proposed by \cite{Anber:2009ua}. In such models, producing observationally viable levels of gaussian, red-tilted fluctuations requires the introduction of $\mathcal{O}(10^5)$ gauge fields, a requirement which renders this particular approach somewhat less appealing. In our recently proposed theory of Chromo-Natural Inflation \cite{Adshead:2012kp, Adshead:2012qe, Martinec:2012bv}, we demonstrated that it may be possible to alleviate this problem by coupling the axion to non-Abelian gauge fields arranged in a classical, rotationally invariant configuration.  The interactions between the axion and the gauge fields generate a slowly rolling inflationary solution for a wide range of parameters, including
both large and small values for the axion decay constant. Importantly, this range includes values $f \ll M_{\rm pl}$. 
In a related approach, the authors of \cite{Maleknejad:2011jw, Maleknejad:2011sq} proposed a model of inflation which does not involve scalar fields at all. The relationship between these models was demonstrated in \cite{SheikhJabbari:2012qf, Adshead:2012qe}, and amounts to the observation that the axionic scalar can be integrated out of the dynamics when its mass is very large. For a recent review of these and other models, see \cite{Maleknejad:2012fw}.

The existence of rotationally invariant classical solutions to the Einstein-Yang Mills equations was originally noticed quite some time ago \cite{Cervero:1978db, Henneaux:1982vs, Hosotani:1984wj, Moniz:1990hf}. Their only previous appearance in inflation studies was in \cite{Moniz:1991kx}, who studied the situation in which the inflaton was charged under the gauge group but for whom the gauge fields otherwise played no role in generating slow-roll inflation. Outside of the inflationary context, the same gauge field configuration has been invoked in models for dark energy \cite{ Gal'tsov:2010dd, Galtsov:2011aa,Elizalde:2012yk}.

Although Chromo-Natural Inflation involves vector gauge fields and axions, it differs in important ways from the many other models that involve either axions, vector-like fields, or both. The seminal example of an axionic inflationary theory, 
Natural Inflation, requires $f \sim M_{\rm pl}$ as we noted before \cite{Freese:1990rb}. Within axionic theories, aside from the monodromy models noted above, there have been a variety of attempts to cure the need for a Planckian decay constant, e.g. \cite{Kim:2004rp, Dimopoulos:2005ac, Easther:2005zr, Germani:2010hd}; however,
each of these methods has relied on various additions and complications of the scalar sector, rather than on fundamentally new physics. Inflationary models involving vector or gauge fields, on the other hand, have explored a wide variety of different approaches. Closest to our study are models that couple gauge fields to scalars. For instance, a uniform gauge-kinetic coupling of the inflaton to multiple vector fields is studied in \cite{Gumrukcuoglu:2010yc,Kanno:2010nr,Watanabe:2009ct,Watanabe:2010fh, Yamamoto:2012sq, Maeda:2012eg}; in a different direction, \cite{Dimopoulos:2008yv} studied a case wherein  the vacuum fluctuations of a vector multiplet played a role
in generating the curvature perturbations. There are also many models in which the vector fields themselves generate
inflation (e.g. \cite{Ford:1989me,ArmendarizPicon:2004pm,  Koivisto:2008xf, Golovnev:2008cf,Golovnev:2009ks, Alexander:2011hz}). In such models, the inherent anisotropy of the vector fields is typically ameliorated by invoking many fields whose effects average out, leaving a statistically homogenous universe. However, many of these vector models \cite{Ford:1989me,ArmendarizPicon:2004pm,  Golovnev:2008cf,Golovnev:2009ks}, though not \cite{Alexander:2011hz}, are inherently unstable \cite{Himmetoglu:2008zp,Himmetoglu:2008hx,Golovnev:2009rm,Himmetoglu:2009qi}.

In this paper we provide a full account of the perturbations in Chromo-Natural Inflation \cite{Adshead:2012kp} that we first sketched in  \cite{Adshead:2013qp}. We show that models of inflation making use of classical non-Abelian gauge fields produce chiral gravitational waves due to parity violation in the background. Unfortunately, this phenomenon also leads to difficulty in matching current observational data. In particular, we show that the original model of Chromo-Natural inflation is in significant tension with current observational constraints from the Planck satellite \cite{Ade:2013uln}.  This is because the tensor-to-scalar ratio is too large in the region where the scalar spectral index matches current constraints. Conversely, in the region where the tensor-to-scalar ratio is acceptable, the spectral index is very red and lies far outside the 2-$\sigma$ contour allowed by Planck \cite{Ade:2013uln}; this result is summarized in Fig. \ref{nsvr}.

Throughout this work, we use natural units where the reduced Planck mass $1/\sqrt{8\pi G} = M_{\rm pl}= c = \hbar = 1$.

%%%%%%%%%%%%%%%%%%%%%%%%%%%%%%%%%
%%%%%%%%%%%%%%%%%%%%%%%%%%%%%%%%%
%
\section{Chromo-Natural Inflation}\label{sec:CNIbackground}
%
%%%%%%%%%%%%%%%%%%%%%%%%%%%%%%%%%
%%%%%%%%%%%%%%%%%%%%%%%%%%%%%%%%%

We consider the theory of Chromo-Natural Inflation \cite{Adshead:2012kp}, which is described by the action
\begin{align}
\label{eqn:CNIaction}
\mathcal{S} =  \int d^{4}x\sqrt{-g}\Bigg[  \frac{1}{2}R -\frac{1}{2}(\partial\axion)^2 -V(\axion)-\frac{1}{2}\tr\[F_{\mu\nu}F^{\mu\nu}\]-\frac{\lambda}{4 f}\axion\tr\[ F\wedge F \]\Bigg] ,
\end{align}
where $\axion$ is a pseudo-scalar (axion) with associated mass scale $f$. We will assume a sinusoidal axion potential with energy scale $\mu$ and decay constant, $f$:
\begin{align}
V(\axion)=\mu^4 \(1+\cos \(\frac{\axion}{f}\)\).
\end{align}
However, we emphasize that the existence of inflationary solutions is not dependent on this choice. We consider a general SU(N) gauge field, $A_\mu$, and adopt the conventions of  Peskin and Schroeder~\cite{Peskin:1995ev}
for its action. In particular, the field-strength tensor and covariant derivative are defined as\footnote{Note that this is opposite to \cite{Adshead:2012kp, Maleknejad:2011jw, Maleknejad:2011sq} who use the opposite sign for the covariant derivative.}
\begin{align}
F_{\mu\nu} = \frac{1}{-ig}\[D_{\mu}, D_\nu\], \quad D_{\mu} = \partial_{\mu} - igA_{\mu},
\end{align}
where $g$ is the gauge field coupling, not to be confused with the determinant of the spacetime metric. We normalize the trace over the SU(N) matrices, which we will denote $J_a$, so that
\begin{align}
\tr\[J_a J_b\] = \frac{1}{2}\delta_{ab}.
\end{align}
Our convention for the antisymmetric tensor is
\begin{align}
\epsilon^{0123} = \frac{1}{\sqrt{-g}}.
\end{align}
while our spacetime metric signature is $(-,+,+,+)$.  Here and throughout, Greek letters will denote spacetime indices, Roman letters from the start of the alphabet denote gauge indices and Roman letters from the middle of the alphabet denote spatial indices.

We will work often work with conformal time, which we define to be a negative quantity during inflation
\begin{align}
\tau = \int^{t}_{0}\frac{dt}{a(t)},
\end{align}
and make use of the near de Sitter expansion to write
\begin{align}
a \approx -\frac{1}{H \tau}.
\end{align}
When we are dealing with fluctuations of the fields, we will work in Fourier space where our convention is
\begin{align}
A({\bf x}) = \int \frac{d^3 k}{(2\pi)^3}A_{\bf k}e^{-i {\bf k}\cdot{\bf x}},
\end{align}
so that we replace spatial derivatives with
\begin{align}
\partial_i A \to -i k_i A_{\bf k}.
\end{align}
We will make extensive use of the fact that the fields satisfy a reality condition, which implies
\begin{align}
A_{-{\bf k}} = \bar{A}_{\bf k}.
\end{align}
It will often prove useful to work with the dimensionless time variable 
\begin{align}
x = -k\tau,
\end{align}
where $k$ is the Fourier space wavenumber; when we match to observations, we will take $k$ to have cosmological units, $h$/Mpc, which will also fix the units for $\tau$. Where necessary, we will match physical length scales to inflationary scales by choosing the scale $k = 0.05$ Mpc$^{-1}$ to leave the horizon 60 efolds before the end of inflation. Throughout we will denote derivatives with respect to cosmic time by an overdot ($\,\dot{}\,$), primes ($\,'\,$) will denote derivatives with respect to $x$, while derivatives with respect to conformal time will be explicit ($\partial_\tau$). Our symmetrization and antisymmetrization conventions throughout will be
\begin{align}
Z_{[ij]} = & \frac{1}{2}(Z_{ij} - Z_{ji}),\nn\\
Z_{(ij)} = & \frac{1}{2}(Z_{ij} + Z_{ji}).
\end{align}

%%%%%%%%%%%%%%%%%%%%%%%%%%%%%%%%%%
%
\subsection{Background solutions}\label{sec:backgroundeqns}
%
%%%%%%%%%%%%%%%%%%%%%%%%%%%%%%%%%%

The background evolution in Chromo-Natural Inflation is found by considering the axion in a classical, homogeneous configuration  $\axion =  \axion(t)$  and the gauge fields in the classical configuration
\begin{align}\label{eqn:gaugevev}
A_{0} = & 0, \quad A_{i} =  \phi\delta^{a}_{i} J_{a} = a\psi \delta^{a}_{i} J_{a},
\end{align}
where $J_a$ is a generator of SU(2) satisfying the commutation relations
\begin{align}
\[J_a, J_b\] = if_{abc}J_c,
\end{align}
and $f_{abc}$ are the structure functions of SU(2). Note that for SU(2), $f_{ijk} = \epsilon_{ijk}$, where $\epsilon_{ijk}$ is the completely antisymmetric tensor in three-dimensions.

On the background field configuration in Eqn.\ (\ref{eqn:gaugevev}), the field strength tensor components are,
\begin{align}
F_{0i} = & \partial_{\tau}\phi \delta^{a}{}_{i}J^a,\quad
F_{ij}  =  g \phi^2 f^{a}_{ij} J^a.
\end{align}
For these degrees of freedom, the mini-superspace action takes the form
\begin{align}
\mathcal{L} =  a^{3}N \Bigg[-3 m_{\rm pl}^2 \frac{\dot a^2}{N^2} + \frac{a^2}{2 N^2} \dot \axion^2 - V(\axion) + \frac{3}{2}\frac{\dot \phi^2}{N^2}  - \frac{3}{2}g^{2}\frac{\phi^4}{a^4} \Bigg]- 3\frac{\lambda}{f}g\axion \dot\phi \phi^2,
\end{align}
where (as mentioned above) an overdot  represents a derivative with respect to cosmic time, and the lapse, $N = a$ on the background solution. This action leads to the equations of motion for the axion and gauge field VEV $\phi$:
\begin{align}
\ddot\axion+3 H \dot\axion+ V'(\axion) = &  -\frac{1}{a^3}\frac{\lambda}{f}g\partial_{t}\(\phi^3\),\\
\frac{\ddot{\phi}}{a}+H\frac{\dot{\phi}}{a} + 2 g^2\frac{\phi^3}{a^3} =  & \frac{\lambda}{f}g\dot\axion \frac{\phi^2}{a^2}.
\end{align}
The equations of motion for the metric are the Friedmann constraint
\begin{align}
3 H^{2} =   \frac{1}{2} \dot \axion^2 + V(\axion) + & \frac{3}{2}\(\frac{\dot\phi^2}{a^2} + g^{2}\frac{\phi^4}{a^4}\) ,
\end{align}
and the equation of motion for the metric
\begin{align}
\dot{H} = -\frac{\dot\axion^2}{ 2} - \frac{\dot{\phi}^2}{ a^2} - g^2\frac{\phi^4}{a^4}.
\end{align}
In \cite{Adshead:2012kp} we showed that this model inflates, with slow-roll provided via a magnetic-drift type force
mediated by the Chern-Simons interaction.  In the large drift force limit ($\lambda \gg 1$), the slow-roll equations for this model are very well approximated by
\begin{align}
\dot \psi & = - H \psi + \frac{f }{3 g \lambda} \frac{V_{,\axion}}{\psi^{2}}  
~,\quad \label{slowrollpsi}\\
\frac{\lambda}{f} \dot \axion %&
 & = 2 g \psi +  \frac{2H^2}{g \psi}  ~\label{slowrollX}\ ,
\end{align}
where $\psi = \phi /a$. To a good approximation, $\psi \approx {\rm const.} $, and Eqn.\ (\ref{slowrollpsi}) is solved by
\begin{align}
\label{psislowrollsoln}
\psi = \(\frac{f V_{,\axion}}{3 g H \lambda}\)^{1/3}.
\end{align}
It will also prove useful to introduce the dimensionless mass parameter
\begin{align}
m_{\psi} = \frac{g\psi}{H},
\end{align}
which, as we will see, characterizes the mass of the gauge field fluctuations in units of the Hubble scale. 

%%%%%%%%%%%%%%%%%%%%%%%%%%%%%%%%%%%%
%%%%%%%%%%%%%%%%%%%%%%%%%%%%%%%%%%%%
%
\section{The quadratic fluctuation action}\label{sec:quadaction}
%
%%%%%%%%%%%%%%%%%%%%%%%%%%%%%%%%%%%%
%%%%%%%%%%%%%%%%%%%%%%%%%%%%%%%%%%%%

In order to find the spectrum of curvature fluctuations in Chromo-Natural Inflation, we will need to understand how the field and metric fluctuations evolve. In this section we derive the action to quadratic order in small fluctuations about the solutions above in Sec.\ \ref{sec:backgroundeqns}. We begin by deriving the action for a the fluctuations of a general SU(N) gauge field about the background field trajectory before we specialize to a two-dimensional representation in Sec.\ \ref{sec:2drep}. The stability of higher dimensional embeddings is considered in Appendix \ref{sec:embeddings}.

To proceed, we write the metric in ADM form \cite{Arnowitt:1962hi},
\begin{align}\label{eqn:adm}
ds^2 = -N^2 d\tau^2 +  \tilde{h}_{ij}(dx^i+N^i d\tau)(dx^j+N^j d\tau),
\end{align}
where $N$ is the lapse, $N^i$ is the shift vector, and $\tilde{h}_{ij}$ is the metric on the spatial hypersurface. At zeroth order in fluctuations, the background Friedmann-Robertson-Walker (FRW) metric in conformal coordinates corresponds to $N = a$ and $N^i = 0$ in our conventions. The metric on the hypersurface, $\tilde{h}$, can be decomposed into scalars, vectors and tensors by writing
\begin{align}\label{eqn:spatialmet}
\tilde{h}_{ij} = a^2\left[(1+A)\delta_{ij}+\partial_{i}\partial_j B + \partial_{(i}C_{j)} + \gamma_{ij}\right],
\end{align}
where $\gamma_{ii} = \partial_i \gamma_{ij} = 0$, and $\partial_iC_i = 0$.
The coordinate invariance of general relativity allows us to impose four conditions on the fields in Eqn.\ (\ref{eqn:spatialmet}).  For this work, we choose spatially flat gauge, where the time threading and spatial coordinates are chosen so that $A = B = 0$. The remaining spatial reparametrizations can then be used to set $C_i = 0$, which completely fixes the coordinates. We further write\footnote{Our summation convention is as follows. Repeated lower Roman indices and all gauge field indices are summed with the Kronecker delta. }
\begin{align}\label{eqn:spatialmet2}
\tilde{h}_{ij} = a^2\[e^{\gamma}\]_{ij} = a^2\left[\delta_{ij} + \gamma_{ij}+\frac{1}{2!}\gamma_{ik}\gamma_{kj}+\ldots\right],
\end{align}
so that $\gamma$ is a spin-2 mode of the metric.

In Appendix \ref{app:gravsector} we demonstrate that the contributions to the action due to integrating out the lapse and the shift (i.e. solving the Einstein constraints) are suppressed relative to the contributions to the action from the non-gravitational terms by small background quantities such as $\dot\axion$ and $\psi$. This means that their contributions are comparable to contributions arising from the evolution of the background itself, and thus we can ignore both self-consistently at leading order.

We denote the fluctuations in the gauge field by 
\begin{align}
\delta\! A_{\mu} = \da_{\mu},
\end{align}
in terms of which, the quadratic Yang-Mills Lagrangian density can be written in the form (see for example \cite{Srednicki:2007qs})
\begin{align}
\delta^2\mathcal{L}_{\rm YM} = &\sqrt{-g}( -\tr\[(\bar{D}_{\mu}\da_{\nu})^2\]+\tr\[(\bar{D}_{\mu}\da_{\nu})(\bar{D}^{\nu}\da^{\mu})\]   +ig\tr\[ \bar{F}^{\mu\nu}\[\da_{\mu},\da_{\nu}\]\]),
\end{align}
where 
\begin{align}
\bar{D}_\mu \equiv \partial_{\mu} - i g \bar{A}_{\mu} = \{\partial_{\tau}, \partial_i - i g a \psi J_i\},
\end{align}
is the background gauge-covariant derivative, $\bar{F}_{\mu\nu}$ is the background field strength tensor, while $J_i$ is an N-dimensional SU(2) matrix. With the background field configuration at Eqn.\ (\ref{eqn:gaugevev}), the quadratic Yang-Mills Lagrangian density can be written,
\begin{align}\label{eqn:actYM}\nn
\delta^2\mathcal{L}_{\rm YM} = &\tr\[(\partial_{i}\da_{0}-i g\phi\[J_i, \da_0\])^2\] -4ig\partial_{\tau}\phi\tr\[\da_0\[\da_i, J_i\]\]
%
%\\ \nn&
-2\tr\[\da_{0}\partial_{\tau}(\partial_{i}\da_{i}-ig\phi \[J_{i}, \da_i\])\]
\\ \nn& +\tr\[\partial_{\tau}\da_i\partial_{\tau}\da_i\] -\tr\[\partial_j \da_{i}\partial_{j}\da_{i}-\partial_i \da_{j}\partial_{j}\da_{i}\]+2g\phi\epsilon_{ijk}\tr\[\partial_i \da_{j}\Omega_k \] \\ &-g^2\phi^2\tr\[(\Omega_k-\da_{k})\Omega_{k}\],
\end{align}
where we have defined
\begin{align}\label{eqn:omegadef}
\Omega_i = i\epsilon_{ijk}\[J_j,\da_k\].
\end{align}
Similarly, the quadratic order Chern-Simons Lagrangian density can be written,
\begin{align}\label{eqn:actCS}\nn
\delta^2\mathcal{L}_{\rm CS} = &  2g\phi^2\frac{\lambda}{f}\daxion\tr\[\partial_i \da_0 J_i\] -\frac{\lambda}{f}\partial_{\tau}\axion\tr\[  g\phi \da_i \Omega_{i}-\epsilon_{ijk}\da_i\partial_{j}\da_{k}\]  + 2g\phi^2\frac{\lambda}{f}\partial_{\tau}\daxion\tr\[ \da_i J_i\]\\ & - 2\frac{\lambda}{f}\epsilon_{ijk}\partial_{\tau}\phi
\daxion \tr\[J_i \partial_j \da_k\].
\end{align}
The axion contribution to the quadratic Lagrangian density is
\begin{align}\label{eqn:actaxion}
\delta^2\mathcal{L}_{\axion} = &   \frac{1}{2}a^2(\partial_{\tau} \daxion )^2- \frac{1}{2}a^2(\partial_{i} \daxion )^2 - a^4\frac{1}{2} \frac{d^2V}{d\axion^2}\daxion^2 .
\end{align}
Finally, the quadratic Lagrangian density for the transverse-traceless components of the metric, and their interactions with the gauge field fluctuations is given by
\begin{align}\nn\label{eqn:actspin2}
\delta^2\mathcal{L}_{\gamma} = & \frac{a^2}{8} \((\partial_{\tau}\gamma)^{2}-(\partial_i \gamma)^{2}+2\(\dot\phi^2-g^{2}\frac{\phi^4}{a^2}\)\gamma^2\)\\  & -a^2 \(\frac{\dot{\phi}}{a} \partial_{\tau}\da_{jl} -g\frac{\phi^2}{a^2}(2\epsilon^{a}_{ij}\partial_{[i}\da^a_{l]}+g\phi \da_{jl}) \)\gamma_{jl},%+ \frac{1}{2}\int dt d^{3}x a^2 \Bigg( \dot{\phi}^2\frac{1}{2}\gamma_{ij}\gamma_{ij}-2\frac{\dot{\phi}}{a} \partial_{\tau}\da^i_j \gamma_{ij}\Bigg)
\end{align}
where $\gamma^2 = \gamma_{ij}\gamma_{ij}$. So the full quadratic Lagrangian is given by
\begin{align}
\delta^2 \mathcal{L}  = \delta^2\mathcal{L}_{\rm YM}+\delta^2\mathcal{L}_{\rm CS}+\delta^2\mathcal{L}_{\axion} + \delta^2\mathcal{L}_{\gamma}+\ldots,
\end{align}
where `$\ldots$' refers to terms that arise from integrating out the gravitational constraints. As discussed above and demonstrated in Appendix \ref{app:gravsector}, these terms are suppressed by additional factors of the small background quantities $\psi$ or $\dot\axion$ relative to the leading order terms.

In order to proceed, we need to choose a specific representation for the gauge field.  As we show in Appendix \ref{sec:embeddings}, the infra-red stability of small fluctuations does not depend on the dimensionality of the embedding, and thus we focus on a two-dimensional representation in what follows for simplicity.

%%%%%%%%%%%%%%%%%%%%%%%%%%%%%%%%%%%%
%%
\subsection{Tensor decomposition}\label{sec:2drep}
%%
%%%%%%%%%%%%%%%%%%%%%%%%%%%%%%%%%%%%

We are able to make a scalar-vector-tensor decomposition of the gauge field perturbations because of the special symmetry of the background solution~\pref{eqn:gaugevev} for the gauge field. The gauge field background identifies spatial indices with $SU(2)$ triplet indices of the internal space.  On this background, the $SO(3)\times SU(2)$ symmetry of global spatial rotations and global gauge transformations is broken down to the diagonal $SO(3)$, which one can use to decompose the spectrum of fluctuations.

Specializing to the case of a two dimensional representation of SU(2), the representation matrices are the Pauli matrices, $J_a = \sigma_a/2$, and we can decompose the gauge field fluctuations into a traceless symmetric tensor, a vector, and a scalar:
\begin{align}\label{eqn:fielddecomp}
\da^a_i \sigma_a = (t^a_{i}+\epsilon^a_{ik}\chi_k + \delta^a_{i}\delta\phi)\sigma_a. 
\end{align}
This represents a total of 9 degrees of freedom.  The temporal components $\da_0^a$ make up another three degrees of freedom that transform as a vector due to the gauge indices.  In terms of the fields defined in Eqn.\ (\ref{eqn:fielddecomp}), the vector in Eqn.\ (\ref{eqn:omegadef}) is given by
\begin{align}
\Omega_i =  (t^a_{i}-\epsilon^a_{ik}\chi_k -2 \delta^a_{i}\delta\phi)\frac{\sigma_a}{2} .
\end{align}
In terms of these degrees of freedom, the Lagrangian derived  in Eqns.\ (\ref{eqn:actYM}) - (\ref{eqn:actspin2})  becomes, 
%\begin{widetext}
\begin{align}\nn\label{eqn:action}
\delta^2 \mathcal{L} = &   \frac{a^2}{2}(\partial_{\tau} \daxion)^2 - \frac{a^2}{2}\partial_{i}\daxion\partial_i \daxion - \frac{a^4}{2}  \frac{d^2V}{d\axion^2}\daxion^2\\ \nn&
 +\frac{1}{2}\da_0^a(-\partial^2 +2g^2\phi^2)\da_0^a+\da_0^a ( \partial_{\tau}\partial_i(t^a_i + \epsilon^a_{ij}\chi_j+\delta^a_i\delta\phi)-2 \partial_{\tau}(g\phi\chi_j)\delta^{a}_{j})\\ \nn&+\da_{0}^a\(4 g \partial_{\tau}\phi  \chi_j  - g\phi^2\frac{\lambda}{ f}\partial_{j}\daxion  \)\delta^a_j\\ \nn&
+\frac{1}{2}\partial_{\tau}t^a_i\partial_{\tau}t^a_i+\partial_\tau \chi_i \partial_\tau \chi_i +\frac{3}{2}\partial_\tau \delta\phi \partial_\tau \delta\phi 
-  \frac{1}{2}\partial_{j}t^a_i\partial_{j}t^a_i-\partial_j\chi_i \partial_j \chi_i -\frac{3}{2}\partial_j\delta\phi \partial_j \delta\phi 
\\ \nn&+\frac{1}{2}\partial_i(t^a_i+\epsilon^a_{ij}\chi_j+\delta^a_{i}\delta\phi)\partial_k(t^a_k+\epsilon^a_{kj}\chi_j+\delta^a_{k}\delta\phi)
+g\phi \big(\epsilon_{ijk}\partial_i t^a_j t^a_k\; - \epsilon_{ijk}\partial_i \chi_j\chi_k +6\partial_i \chi_i\delta\phi\big)\\\nn & 
-2g^2\phi^2( 2 \chi_i \chi_i+9\delta \phi^2)
-\frac{1}{2} g\phi \frac{\lambda}{f}\partial_{\tau}\axion \(t^a_i t^a_i - 2\chi_i \chi_i - 6\delta\phi^2\)\\ \nn&+3 g\phi^2\frac{\lambda}{f}\partial_{\tau}\daxion  \delta\phi+ 2\frac{\lambda}{f}\partial_{\tau}\phi\daxion \partial_{j}\chi_j+\frac{\lambda}{2f}\partial_{\tau}\axion\big(\epsilon_{ijk}\partial_it^a_{j}t^a_k- 2\partial_it_{ij}\chi_j  -\epsilon_{ijk}\partial_i\chi_k\chi_j -4 \delta\phi \partial_i\chi_i \big) \\ \nn&+
 \frac{a^2}{8} \((\partial_{\tau}\gamma)^{2}-(\partial_i \gamma)^{2}+2\(\dot\phi^2-g^{2}\frac{\phi^4}{a^2}\)\gamma^2\)
 \\  & -a^2 \(\frac{\dot{\phi}}{a} \partial_{\tau}t_{jl} -g\frac{\phi^2}{a^2}(2\epsilon^{a}_{ij}\partial_{[i}t^a_{l]}+g\phi t_{jl}) \)\gamma_{jl}.
\end{align}
%\end{widetext}
%Now, 
Analogous to the coordinate invariance of general relativity, SU(2) gauge theories are invariant under local SU(2) gauge transformations. This allows us to eliminate three of the degrees of freedom in the theory by fixing a gauge. Observable quantities, such as the components of the energy-momentum tensor, are by definition gauge invariant. This means as long as the gauge is completely fixed, physical quantities will not  be dependent on the particular choice of gauge. In this work we fix the gauge by imposing the non-Abelian generalization of the Coulomb condition
\begin{align}\label{eqn:NAGcond}
\bar{D}_i \da_i = \partial_{i}\da_i  - i g \phi \[J_i, \da_i\] = 0,
\end{align}
which we refer to as  non-Abelian Coulomb gauge.  A nice property of this gauge choice is that it renders the Gauss's law constraint algebraic, with the solution
\begin{align}\label{eqn:scalargauss}
\da^a_0 = - \frac{1}{-\partial^2+2g^2\phi^2}(4 g \partial_{\tau}\phi  \chi_j- g\phi^2\frac{\lambda}{ f}\partial_{j}\daxion  ) \delta^{a}_j ~.
\end{align}
The non-Abelian Coulomb gauge condition, Eqn.\ (\ref{eqn:NAGcond}), additionally imposes a relationship between the degrees of freedom 
\begin{align}\label{eqn:nacoulombgauge}
\partial_i(t^a_i+\epsilon^a_{ij}\chi_j+\delta^a_{i}\delta\!\phi)= 2 g \phi \chi^a.
\end{align}
It is simple to demonstrate that this gauge condition completely fixes the gauge and can be reached by a finite gauge transformation.

In what follows, we work with explicit components of the fields. Choosing the wavenumber along the $x^3$ direction, we can now explicitly realize a scalar-vector-tensor (SVT) decomposition of the gauge-field modes previously defined in Eqn.\ (\ref{eqn:fielddecomp}) through the use of the remaining diagonal $SO(3)$ symmetry. In this decomposition,
\begin{align}
t^\pm =\frac{1}{\sqrt{2}}\( \frac12(t_{11}-t_{22})\pm i t_{12}  \)
\end{align}
forms the two helicities of a transverse traceless tensor, 
\begin{align}
v^{\pm} = \frac{1}{\sqrt{2}}\(t_{3,1}\pm i t_{3,2}\), \quad u^{\pm} = \frac{1}{\sqrt{2}}\(\chi_{1} \pm i\chi_{2}\)
\end{align}
are helicity states of transverse vectors, and 
\begin{align}
z\equiv\frac16(2t_{33}-t_{11}-t_{22}),
\end{align}
is a scalar along with  $ \chi_3$, and $\delta\!\phi$. Rotational invariance ensures that the particular choice of direction is irrelevant, and thus we drop the `3' subscript on the momenta.

In terms of these fields, the gauge condition as written in Eqn.\ (\ref{eqn:nacoulombgauge}) becomes, 
\begin{align}\label{eqn:vecgaugecon}
-ik(v_{\pm}  \pm iu_{\pm} )= & 2 g \phi u_{\pm},\\
-i k (2z +\delta\phi) = & 2 g \phi \chi_3.\label{eqn:scalargaugecond}
\end{align}
These three conditions remove three degrees of freedom, while Gauss's law removes theree further degrees of freedom, leaving the six physical propagating degrees of freedom of the gauge theory.

We will now analyze the scalar, vector, and tensor fluctuations separately.

%%%%%%%%%%%%%%%%%%%%%%%%%%%%%%%%%%%%
%%%%%%%%%%%%%%%%%%%%%%%%%%%%%%%%%%%%
%%%
\section{Scalar fluctuations}\label{sec:scalars}
%%%
%%%%%%%%%%%%%%%%%%%%%%%%%%%%%%%%%%%%
%%%%%%%%%%%%%%%%%%%%%%%%%%%%%%%%%%%%

There are four scalar fluctuations: $\daxion$, the fluctuation of the axion, and $z$, $\chi_3$ and $\delta\phi$ from the gauge sector. Working in Fourier space, the Lagrangian for these scalars is given by,
%\begin{widetext}
\begin{align}\nn\label{eqn:scalaction2}
\delta^2 \mathcal{L} = &    \frac{a^2}{2}\partial_{\tau} {\daxion} \partial_{\tau} \bar{\daxion} - \frac{a^2}{2}\(k^2 + a^2 \frac{d^2V}{d\axion^2}
+k^2  \frac{g^2\phi^4}{k^2+2g^2\phi^2}\frac{\lambda^2}{ f^2}\)\daxion\bar{\daxion}
\\\nn &
+3\partial_{\tau}z\partial_{\tau}\bar{z}+\partial_\tau \chi_3 \partial_\tau \bar{\chi}_3 +\frac{3}{2}\partial_\tau \delta\phi \partial_\tau \bar{\delta\phi} -  3k^2 z\bar{z}-k^2\chi_3  \bar{\chi}_3 -\frac{3}{2}k^2\delta\phi \bar{\delta\phi} \\\nn
&- \frac{8g^2 (\partial_{\tau}\phi)^2 }{k^2+2g^2\phi^2}\chi_3\bar{\chi}_3-3ikg\phi ( \chi_3 \bar{\delta\phi}- \delta\phi \bar{\chi}_3)
-9g^2\phi^2\delta \phi\bar{\delta \phi}\\ \nn& -\frac{1}{2} g\phi\frac{\lambda}{f}\partial_{\tau}\axion \(6 z\bar{z} - 2\chi_3 \bar{\chi}_3 - 6\delta\phi\bar{\delta\phi}\) + ik \frac{\lambda}{f}\partial_{\tau}\axion\big( (z \bar{\chi}_3 -\bar{z}\chi_3 ) + (\bar{\delta\phi} \chi_3 - \delta\phi \bar{\chi}_3) \big)\\ & +\frac{3}{2} g\phi^2\frac{\lambda}{f}(\partial_{\tau}\bar{\daxion}  \delta\phi +\partial_{\tau}\daxion  \bar{\delta\phi})- \(\frac{i k^3}{k^2+2g^2\phi^2}\)\frac{\lambda}{f}\partial_{\tau}\phi( \bar{\daxion}\chi_3 - \daxion\bar{\chi}_3)
\end{align}
where we have already integrated out the constraint $\da^a_0$. Note that we have written the Lagrangian in terms of fields and their conjugates, so that varying the action with respect to a field yields the equation of motion for its conjugate (e.g. the equation of motion for ${z}$ is found by varying with respect to $\bar{z}$).

We now must eliminate one of the degrees of freedom using the non-Abelian gauge condition at Eqn.\ (\ref{eqn:scalargaugecond}). This unavoidably leads to a kinetic coupling between the remaining degrees of freedom. We choose to  eliminate the field $\chi_3$, but there is nothing particularly special about this choice. Thus, we replace $\chi_3$ using Eqn. (\ref{eqn:scalargaugecond}) and redefine our fields in the following way to diagonalize the kinetic terms:
\begin{align}
\phihat = &\sqrt{2}\(\frac{\delta\phi}{2} + z\),\\
\zhat = & \sqrt{2}(z - \delta\phi).
\end{align}
In terms of these fields, the Lagrangian takes the form
\begin{align}\label{eqn:canscalaraction}
\delta^2 \mathcal{L} = &    \frac{a^2}{2}\partial_{\tau} {\daxion} \partial_{\tau} \bar{\daxion} - \frac{a^2}{2}\(k^2 + a^2 \frac{d^2V}{d\axion^2}
+k^2\frac{g^2\phi^4}{k^2+2g^2\phi^2}\frac{\lambda^2}{ f^2}\)\daxion\bar{\daxion}\\\nn &
+\frac{1}{2}\partial_{\tau}\hz\partial_{\tau}\bar{\hz}  - \frac{1}{2} \(k^2- g\phi\(\frac{\lambda}{f}\partial_{\tau}\axion -4 g\phi\)\) \hz\bar{\hz}
+\frac{1}{2}\(2+\frac{k^2}{g^2\phi^2}\)\partial_{\tau}{\hdp}\partial_{\tau}\bar{\hdp} \\ \nn &  -\frac{1}{2}\(\(6+\frac{k^2}{g^2\phi^2}\) k^2+\frac{k^2}{g^2\phi}\partial^2_{\tau}\(\frac{1}{\phi}\)+4g^{2}\phi^2- k^2\frac{\lambda}{f}\frac{\partial_{\tau}\axion}{g\phi}+\frac{8k^2  }{k^2+2g^2\phi^2}\frac{(\partial_{\tau}\phi)^2}{\phi^2} \)\hdp \bar{\hdp}
\\\nn &
+\frac{1}{2}\Bigg[\(2 k^2  
 -2g\phi\frac{\lambda}{f}\partial_{\tau}\axion -  k^2 \frac{\lambda}{f}\frac{\partial_{\tau}\axion}{g\phi}+4g^2\phi^2\)\big( \hz\bar{\hdp} +\bar{\hz}\hdp \big)\\ \nn & 
  + g\phi^2\frac{\lambda}{f}(\partial_{\tau}\bar{\daxion} (\hdp  - \hz) +\partial_{\tau}\daxion(\bar{\hdp}  - \bar{\hz}))
  -\(\frac{k^4}{k^2+2g^2\phi^2}\)\frac{\lambda}{f}\frac{\partial_{\tau}\phi}{g\phi} ( \bar{\daxion}\hdp +\daxion\bar{\hdp})\Bigg].
\end{align}

Writing the three physical scalars collectively as $\Phi \!=\! (\ada,\phihat , \zhat)$, where 
\be
\ada \!=\! a\, \daxion
\ee
 is the canonically normalized axion fluctuation, we can write the  action  abstractly as
\be
\label{scalaract}
\mathcal{S}_S = \int d^4 x \[\GG_{IJ}\bar\Phi_I'\Phi_J' - \MM_{IJ}\bar\Phi_I\Phi_J + \AA_{IJ}\bar\Phi_I \Phi_J'\].
\ee
The action is a complexified version of that of a charged particle moving in three dimensions in a time-dependent harmonic well, $\MM$, with a magnetic field $\AA$. Making use of the analogy to this dynamical system is a fruitful way to tease apart the rather complicated dynamics that results from Eqn.\ (\ref{eqn:canscalaraction}).

In order to predict the late time behavior of the fields, we will make a series of approximations.  Because we will only be interested in perturbations around slow-roll backgrounds, we can treat the evolution of the background as slow-roll suppressed, giving us the following relationships: 
\be
a \approx -\frac{1}{H\tau} ~,~~
\psi \approx \( \frac{f V_{,\axion}}{3g\lambda H} \)^{1/3},
\ee
together with the slow-roll equation for $\dot \axion$,  Eqn. \pref{slowrollX}, which we use to rewrite $\partial_\tau \axion$ in terms of $\mpsi$.
Apart from the axion mass term, $V_{,\axsub\axsub}\equiv d^2V/d\ada^2$, we find that the action depends only on the combinations 
\be
x = -k\tau ~,~~
\mpsi = \frac{g\psi}{H} ~,~~
\bflam = \frac{\lambda}{f}\psi ~.
\ee
In terms of these, the action can be written
%\begin{widetext}
\bea
\label{orthoscalaract}
\mathcal{S}_S = \half\int  d^4 k \biggl[& & \adab' \ada' + \Bigl(\frac{x^2}{\mpsi^2}+2\Bigr) \phihatb '{\phihat}'+\zhatb'\zhat'
- \frac{\sqrt2 \bflam\mpsi}{x}\(  (\phihatb-\zhatb)\ada'  - \adab(\phihat'-\zhat') \)
\nn\\ & &
-\( 1 - \frac1{x^2}\Bigl(2-\frac{V_{,\axsub\axsub}}{H^2} \Bigr) + \frac{\bflam^2\mpsi^2}{2\mpsi^2+x^2} \) \adab\ada
+\frac{\sqrt2 \bflam(4\mpsi^4+2\mpsi^2 x^2+x^4)}{\mpsi x^2(2\mpsi^2+x^2)} \adab\phihat
\nn\\ & &-\frac{2\sqrt2\bflam\mpsi}{x^2} \adab\zhat
+\frac{\sqrt2\bflam(2\mpsi^4+\mpsi^2x^2+x^4)}{\mpsi x^2(2\mpsi^2+x^2)}\phihatb\ada
%-\frac{8\mpsi^6+4\mpsi^2x^2+12\mpsi^4x^2-2x^4+6\mpsi^2x^4+x^6}{\mpsi^2x^2(2\mpsi^2+x^2)}\phihatb\phihat
\nn\\ & &-\Bigl(\frac{2}{\mpsi^2}+\frac{4}{x^2}\Bigr)\phihatb\zhat
-\frac{\sqrt2\bflam\mpsi}{x^2} \zhatb\ada 
-\Bigl(\frac{2}{\mpsi^2}+\frac{4}{x^2}\Bigr) \zhatb\phihat
-\Bigl(1-\frac{2-2\mpsi^2}{x^2}\Bigr)\zhatb\zhat \nn\\ & & -\Bigl(4 + \frac{8}{2\mpsi^2+x^2} + \frac{x^4-2x^2+4\mpsi^4}{\mpsi^2 x^2} \Bigr) \phihatb\phihat \biggr]~,
\eea
%\end{widetext}
from which we can extract the time-dependent mass matrix, $\MM$,
and analogue magnetic field matrix, $\AA$. To be clear, we will generically refer to terms that involve a single time derivative as `analog magnetic field' terms, while the remaining terms without time derivatives form the mass matrix. 

For the purpose of numerical integration, as well as for a WKB approximation to the equations of motion, it is useful to canonically normalize the kinetic term of $\phihat$ by defining
\be
\label{normphi}
\varphihat = \phihat \, \Bigl(\frac{x^2}{\mpsi^2}+2\Bigr)^{1/2}~,
\ee
but we will not reproduce the action in terms of this new field here.

\subsection{Equations of motion}\label{sec:scalareoms}
The action that results from Eqn. \pref{orthoscalaract}, canonically normalized via Eqn. \pref{normphi}, leads the following system of equations of motion:
\begin{align}
\ada'' + & \( 1 - \frac{2}{x^2} + \frac{V''_\axion}{H^2 x^2} + \frac{\bflam^2 \mpsi^2}{2 \mpsi^2 + x^2} \) \ada + \frac{\bflam \mpsi \sqrt{4 + \frac{2x^2}{\mpsi^2}} \(4 \mpsi^4 + 3 \mpsi^2 x^2 + x^4\)}{\(2 \mpsi^2 x + x^3\)^2} \varphihat \nn \\
&- \frac{2 \sqrt{2} \bflam \mpsi}{x^2} \zhat - \frac{\bflam \mpsi}{x \sqrt{1+ \frac{x^2}{2 \mpsi^2}}} \varphihat' + \frac{\sqrt{2} \bflam \mpsi}{x} \zhat' =0, 
\label{axeom}
\end{align}
%&\mbox{} \nn \\
\begin{align}
\varphihat'' + & \(1 - \frac{2}{2 \mpsi^2 + x^2} + \frac{2 \mpsi^2}{x^2} + \frac{6 \mpsi^2}{\(2 \mpsi^2 + x^2\)^2} \) \varphihat + \frac{2 \sqrt{2 + \frac{x^2}{\mpsi^2}}}{x^2} \zhat + \frac{\bflam \mpsi}{x \sqrt{1 + \frac{x^2}{2 \mpsi^2}}} \ada' \nn \\
& + \frac{\bflam \mpsi \sqrt{4 + \frac{2x^2}{\mpsi^2}} \(2 \mpsi^4 +  \mpsi^2 x^2 + x^4\)}{\(2 \mpsi^2 x + x^3\)^2} \ada =0, \label{phieom}
\end{align}
%&\mbox{} \nn \\
\begin{align}
\label{zeom}
\zhat'' + & \(1 - \frac{2-2 \mpsi^2}{x^2} \) \zhat - \frac{\sqrt{2} \bflam \mpsi}{x^2} \( \ada + x \ada'\) + \frac{2 \sqrt{2 + \frac{x^2}{\mpsi^2}}}{x^2} \varphihat =0. \hspace{0.6in}
\end{align}
We remind the reader that these equations of motion have been written assuming a slowly evolving background, as described above Eqn.~\pref{orthoscalaract}. In particular, we have dropped terms involving the evolution of the gauge field, $\dot\psi$, and the Hubble rate, $\dot{H}$. These terms lead to percent-level corrections to the action and equations of motion, and thus they should not significantly alter the results of our analysis. We will discuss the behavior of this system of equations in the next few sections.

\subsection{Generic features of the scalar dynamics}

\begin{figure}[t] %  figure placement: here, top, bottom, or page
   \centering
   \includegraphics[width=\textwidth]{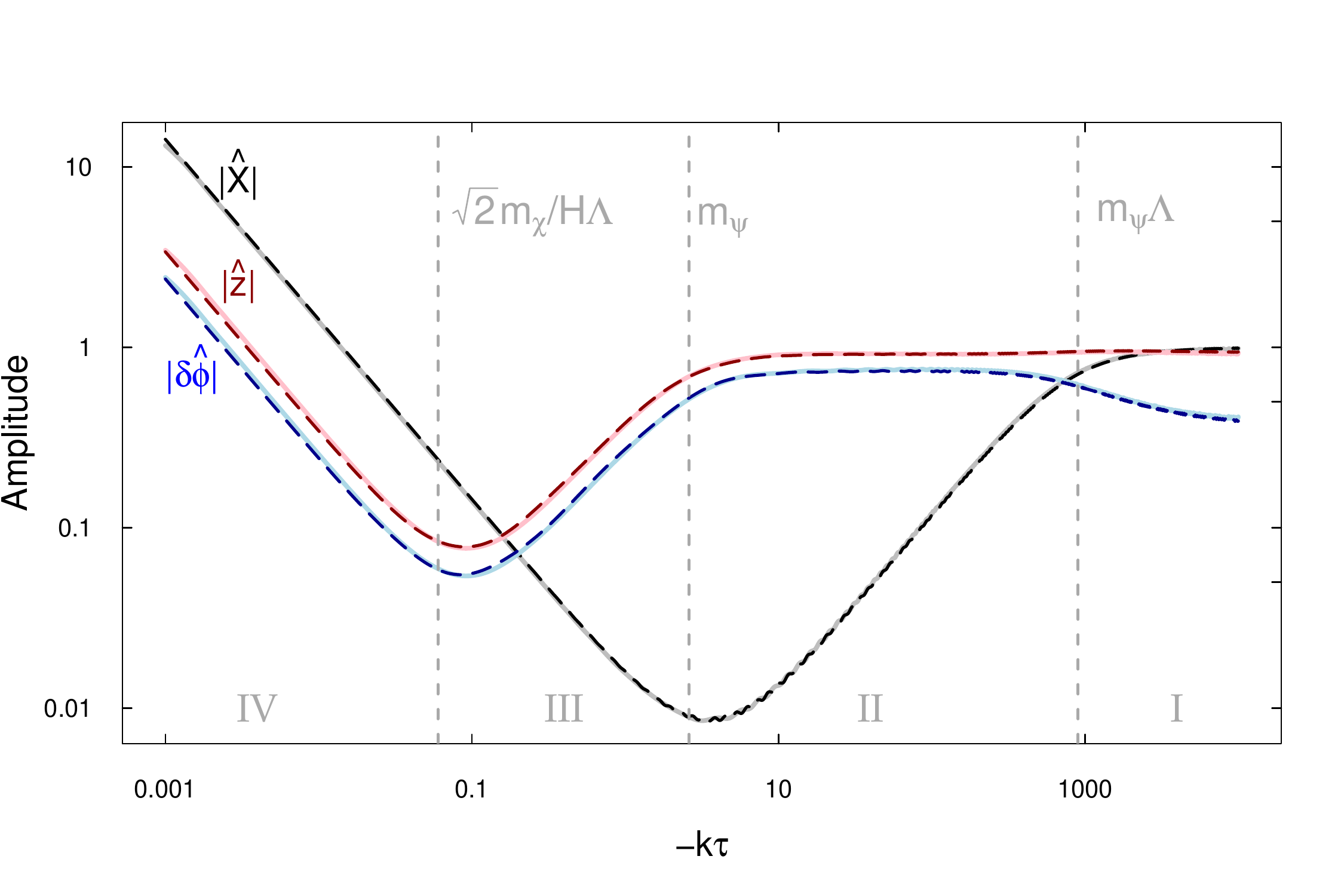} 
   \caption{\it The evolution of the amplitudes of the axion, $|\hat { X}|$ (black), and gauge scalar perturbations, $|\hat z|$ (red) and $|\delta \hat \phi|$ (blue), both allowing the background
   to evolve (solid lines) and with the background fixed at its horizon crossing values (dashed lines). Although there are small differences between the two cases,
   the dynamics, especially at late times, is almost entirely captured by taking the background to be fixed at its horizon crossing values for a given $k$ mode. The model plotted has parameters $g=0.001$, $\mu=0.01$, $\lambda=2000$, $f=0.2$, $m_\psi \simeq 2.6$ and $\bflam\simeq 350$. The transition times between different
   behaviors mentioned in the text are marked with vertical dashed lines and labelled.  Note that in making this plot we have made use of the insights given in Sec. \ref{sec:WKB} to initialize the mode functions in such a way as to isolate the `slow' magnetic drift mode of the system, which has slowly evolving amplitudes. 
   }
   \label{fixVrunning}
\end{figure}
There are four distinct regions over which fluctuations evolve; see Fig. \ref{fixVrunning}:
\begin{enumerate}%[I)]
\item[I.]
At early times, $x \gg \bflam\mpsi$, all the complicated interactions die away as powers of $x$, and the dynamics becomes that of free fields.  The initial conditions are set by canonically quantizing the modes.  The interactions among the fields, while small compared to the kinetic energy, cause energy to gradually oscillate/rotate among them.
\item[II.] 
In the intermediate time window $\bflam\mpsi\gtrsim x \gtrsim \mpsi$, field mixing terms in Eqn.~\pref{orthoscalaract}, such as off-diagonal mass terms and the analogue magnetic field, start to dominate.  We will see that the axion amplitude decreases by a factor of order $\bflam$.
\item[III.]
For $x \lesssim \mpsi$, the axion freezes out and transitions to power law behavior.  The interaction terms proportional to $\bflam$ in the action are still the most important effects.
\item[IV.] When $x \lesssim \sqrt{2} m_\axion/H \bflam$, where $m_\axion^2 \equiv V''(\axion/f)$, the axion's mass term becomes more important than the ${\cal O}(\bflam^2)$ term in the axion dynamics. This feeds back into the gauge field dynamics and leads to a  late time growing solution for the gauge field scalars. 
\end{enumerate}
Note that there is technically a fifth region, $x \lesssim \mpsi/\bflam$, when the ${\cal O}(\bflam^2)$ term in the axion potential becomes less important than the ${\cal O}(\bflam^0)$ terms.  However, by this time the modes are all frozen into their late-time power law behavior, and this change in the dynamics has a negligible effect.

There are six modes.  For early times, these are the positive and negative frequency plane waves for the three fields.  In the intermediate time window, there is one pair of modes orthogonal to the analogue magnetic field that remains relatively unaffected.  Modes oscillating in the plane of the analogue magnetic field split into those whose angular momentum in the harmonic well is aligned with the analogue magnetic field, and those which are anti-aligned.

To discuss the modes that respond to the analogue magnetic field, it is useful to recall the toy model introduced in \cite{Martinec:2012bv}.
Consider two-dimensional motion in the plane $(Y,Z)$ in a constant background magnetic field of magnitude $\lambda$ and harmonic potential well of frequency $\mu$.  The equations of motion are
\bea
\ddot Y + \mu^2 Y &=& \lambda \dot Z \\
\ddot Z + \mu^2 Z &=& -\lambda \dot Y \ .
\eea
There are two normal modes consisting of circular motion about the center of the well.  If the angular momentum is aligned with the magnetic field, the Lorentz force opposes the potential force; in the large field limit, `magnetic drift' balances the potential force against the magnetic force, so that $\mu^2 r_+ \approx \lambda v_+$.  
The orbital frequency of this mode is $\omega_+ = v_+/r_+ \approx \mu^2/\lambda$.  
The other normal mode has the angular momentum anti-aligned with the magnetic field, and in the large field limit the magnetic force balances against inertia, $v_-\approx \lambda r_-$.  
The orbital frequency is just the Larmor frequency $\omega_- = v_-/r_- \approx\lambda$. 

In the problem at hand, we will find that the fast Larmor oscillation matches onto a decaying mode on superhorizon scales; and that the slow magnetic drift mode, which matches onto the growing mode of the inflaton, is the one of interest.  This makes sense~-- as $k\to 0$, the oscillations become the background, and the background dynamics is dominated by magnetic drift.

%%%%%%%%%%%%%%%%%%%%%%%%%%%%%%%%%%%%
%%
\subsection{WKB approximation}\label{sec:WKB}
%%
%%%%%%%%%%%%%%%%%%%%%%%%%%%%%%%%%%%%

To get a rough handle on the dynamics, let us consider a Wentzel-Kramers-Brillouin (WKB) approximation to the equations of motion.  Here and throughout our discussion of approximations to
the scalar dynamics, we will take the background functions $m_\psi$ and $\bflam$ to be fixed during the evolution of perturbations from
their free vacuum through horizon crossing. This is an excellent approximation to the full dynamics, as we illustrated in Fig. 
\ref{fixVrunning}.
The WKB approximation sets 
\be
\label{wkbapprox}
\ada(x)		= A(x) e^{i S(x)},~
\varphihat(x)	= B(x) e^{i S(x)},~
\zhat(x) 		= C(x) e^{i S(x)} .
\ee
For region I and most of region II it is a good approximation to retain only the leading terms in $\bflam/x$ in the mass matrix and analogue magnetic field.
The leading order equations of motion are then, inserting Eqn.~\pref{wkbapprox} into Eqns.~\pref{axeom}, \pref{phieom}, and \pref{zeom}, 
\bea
0 &=& \Bigl(\bfkap^2+\frac {\bfgam^2}{x^2} - (S')^2\Bigr) A + \frac{\bfnu_{\axsub \varphihat}}{x} B  - \frac{\bfxi_{\axsub\zhat}}{x^2} C 
- i\frac{\bfbeta_{\axsub\varphihat}}{x^2} B S'  + i\frac{\bfbeta_{\axsub\zhat}}{x} C S' 
\nn\\ %\label{axionwkb} \\
0 &=& \Bigl(\bfkap^2 - (S')^2 \Bigr) B + \frac{\bfnu_{\varphihat\axsub }}{x} A  + i\frac{\bfbeta_{\axsub\varphihat}}{x^2} A S'  
\label{scalarwkb}\nn \\ %\label{Awkb} \\
0&=& \Bigl(\bfkap^2 - (S')^2 \Bigr) C - \frac{\bfxi_{\zhat\axsub}}{2x^2} A  - i\frac{\bfbeta_{\axsub\zhat}}{x} A S'  
 %\label{zwkb}
\eea
where we have
\begin{align}
\bfkap=1, \hspace*{0.12in}\qquad  \qquad \quad &\bfgam = \bflam \mpsi ,\qquad \hspace{0.6in} \bfbeta_{\axsub\varphihat} = \sqrt{2}\bflam\mpsi, \nn\\
\quad\bfbeta_{\axsub\zhat}  = \sqrt{2}\bflam\mpsi, \qquad  &\bfnu_{\axsub \varphihat} = \bfnu_{\varphihat \axsub} = \sqrt{2}\bflam \hspace{0.25in} \quad \bfxi_{\axsub\zhat} = \bfxi_{\zhat\axsub} = 2\sqrt{2}\bflam\mpsi.
\end{align}
For compactness in what follows, we define two more combinations:
\be
\bfbeta=\sqrt{2}\bflam\mpsi, \qquad \bfnu \equiv \sqrt{2} \bflam.
\ee
The system of equations given in Eqs.~\pref{scalarwkb} has a trivial solution 
\be
\label{boring}
S' = \pm\bfkap ~,~~
A = 0 ~,
\ee
with the first equation given in Eqs.~\pref{scalarwkb} determining the relative amplitudes of $B$ and $C$.
This mode oscillates entirely between the two gauge scalars and does not affect the amplitude of the axion at late times; we will largely ignore it in what follows.

Modes with $A\ne 0$ are more interesting.
Solving for the eikonal $S(x)$ in the region $x\gg 1$, one finds the relation
\be
\Bigl(\bfkap^2-(S')^2\Bigr)\Bigl(\bfkap^2+\frac{\bfgam^2}{x^2}-(S')^2\Bigr) - \frac{\bfbeta^2}{x^2}(S')^2 - \frac{\bfnu^2}{x^2} \approx 0
\ee
so that
\bea
\label{eikeq}
(S')^2 = \frac{1}{2x^2}\Bigl(\bfbeta^2+\bfgam^2 + 2\bfkap^2 x^2  \pm \sqrt{(\bfbeta^2 + \bfgam^2)^2 + 4x^2(\bfkap^2\bfbeta^2 + \bfnu^2)}\Bigr) ~.
\eea
The upper sign is the fast mode of the toy model, whose frequency grows as the rms combination of the harmonic well frequency $\bfgam/x$ and the Larmor frequency $\bfbeta/x$ as one enters region II where these terms dominate; the lower sign is the magnetic drift mode where these two effects oppose one another, and the oscillation frequency is approximately constant in region II.

In the asymptotic past of region I, where $\bfbeta,\bfgam,\bfnu \ll x$, both roots approach the free field eikonal $S \sim \pm \bfkap x$:
\be
\label{wasymp}
\pm S' \sim  \bfkap \pm \frac{\sqrt{\bfbeta^2\bfkap^2+\bfnu^2}}{2\bfkap x}
\ee
where the upper (lower) sign on the RHS is the fast (slow) mode; one also has the relations among the asymptotically constant amplitude functions
\bea
B &\sim& \pm \frac{\bfnu A}{\sqrt{\bfbeta^2\bfkap^2+\bfnu^2}}
\nn\\
C &\sim& \mp \frac{i\bfbeta\bfkap A}{\sqrt{\bfbeta^2\bfkap^2+\bfnu^2}}
\label{asympamps}
\eea
where again the upper (lower) sign is the fast (slow) mode.

\begin{figure}
\begin{center}
\includegraphics[width= \textwidth]{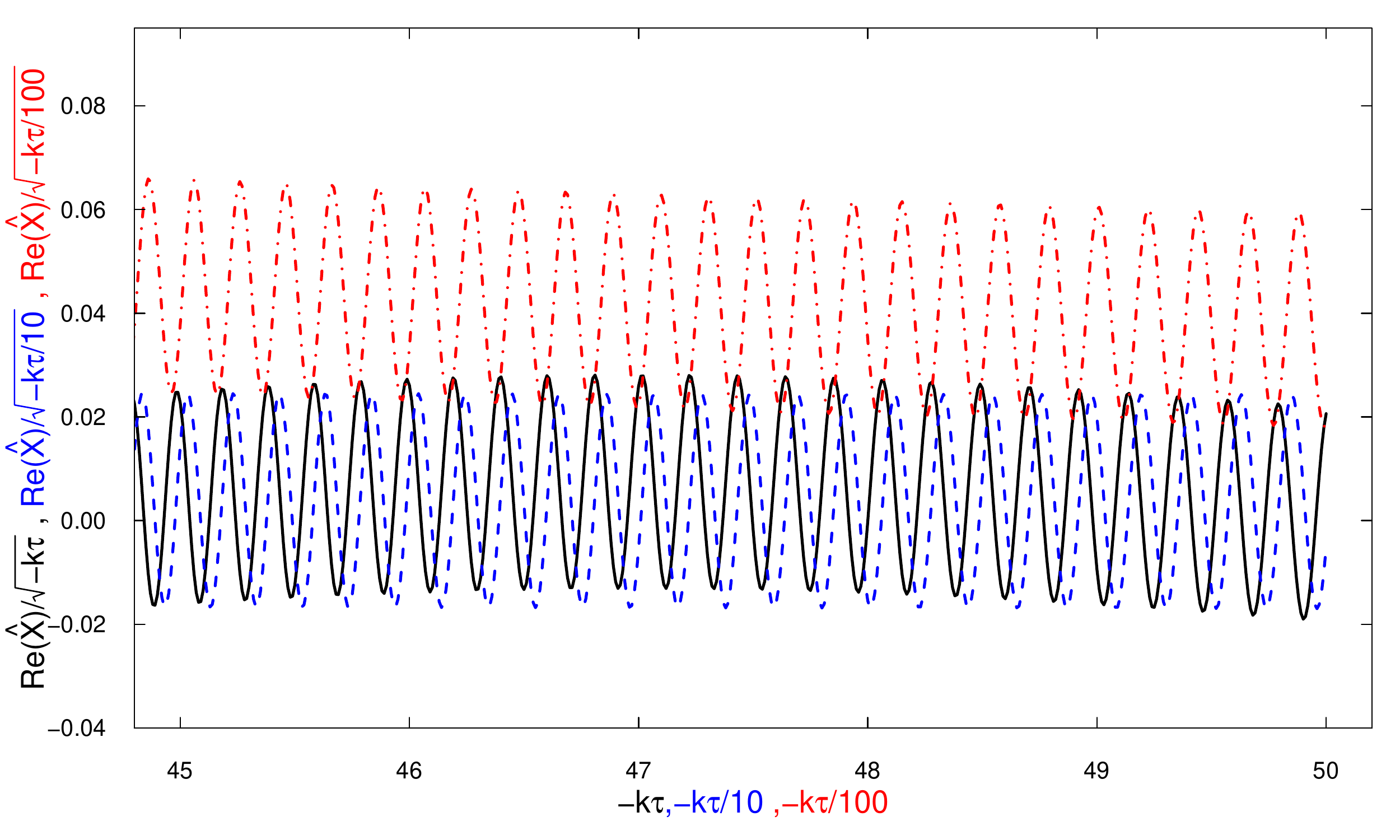}
\caption{\it Example of fast mode scaling: Numerical evaluation of the solution for the axion, with generic initial conditions, showing the Real part of
$\ada(x)/\sqrt{x}$ (black solid curve),
$\ada(x/10)/\sqrt{x/10}$ (blue dashed curve),
$\ada(x/100)/\sqrt{x/100}$ (red dot-dashed curve).
The amplitudes and the frequencies match, apart from an overall drift due to the slow mode oscillation 
(the red curve shows the start of the growth in the axion just after horizon crossing).  The parameters used here are $\{\bflam=350,\mpsi=2.6\}$. }\label{fig:fastmodescaling}
\end{center}
\end{figure}

Entering region II, one has $\bfbeta,\bfgam,\bfnu \gtrsim x$; this region has a large analogue magnetic field and a steepening harmonic potential.  The frequency of the fast mode is well approximated by 
\be
\label{omegafast}
\omega_{\rm fast}^2\approx\frac{\bfbeta^2+\bfgam^2}{x^2}~;
\ee
thus the WKB approximation for this mode becomes more accurate as $x$ decreases.  The fast mode amplitude $A_{\rm fast}(x)$ scales as $\sqrt{x}$;
this scaling can be understood from the adiabatic approximation to the rapid harmonic motion, which keeps fixed the action integral $J=\int pdx$ over a period.  Then the energy $\omega J = \omega^2 \ada^2$ has $\omega \ada^2$ approximately constant; since the frequency $\omega_{\rm fast}\propto 1/x$, we have $A_{\rm fast}\propto \sqrt x$.  This scaling is confirmed in numerical approximations to the solution, as we illustrate in Fig. ~\ref{fig:fastmodescaling}.

The analogue magnetic field in the action, Eqn.~\pref{orthoscalaract}, continues to increase through the horizon crossing transition from region II into regions III and IV; therefore, we conclude that the fast mode matches onto a decaying mode at late time, and is not relevant in determining the scalar curvature perturbation spectrum.

\begin{figure}
\begin{center}
\includegraphics[width=0.9 \textwidth]{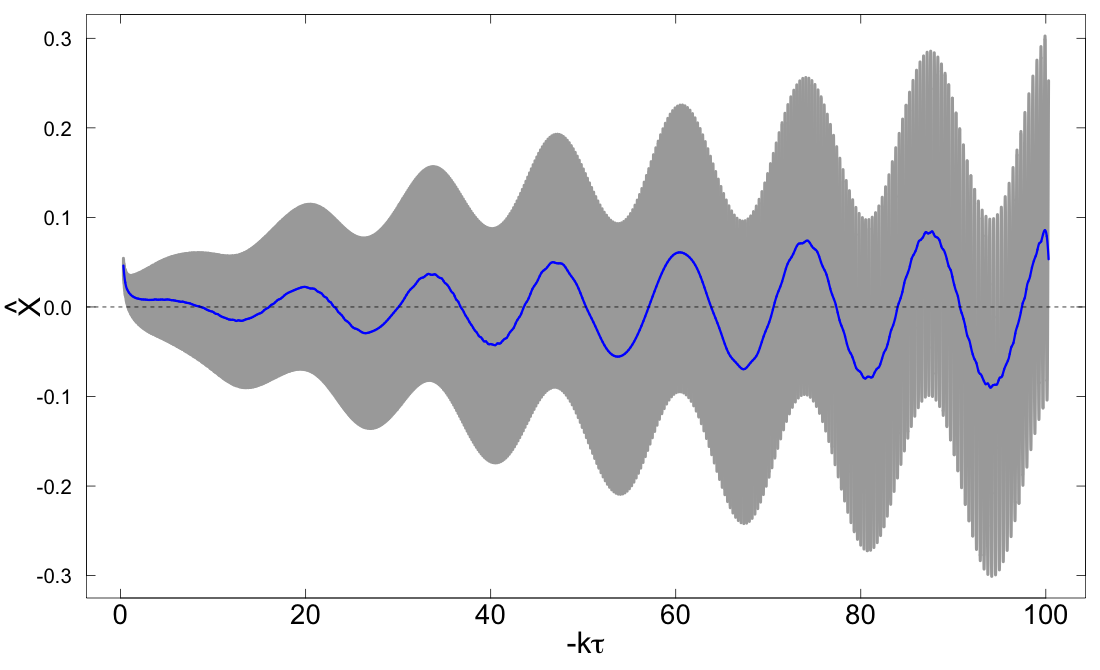}
\caption{\it The gray shaded plot shows the numerical solution for the axion over the range $0.25<x<100$ with generic initial conditions; the plot is dense due to the rapid oscillation of the fast mode.  A running average reveals that the slow magnetic drift mode has approximately constant frequency, and amplitude decreasing linearly in conformal time, down to $x\sim \mpsi$.}\label{fig:slowmoderevealed}
\end{center}
\end{figure}

However, the fast Larmor mode, whose high frequency only grows as time goes on, can make it hard to discern the behavior of the slower ``magnetic drift'' mode for generic initial conditions; this is illustrated in Fig.~\ref{fig:slowmoderevealed}.
For generic initial conditions, if we observe the behavior of the axion perturbation over a long time window, it appears that the amplitude is decaying. However, it is hard to see the precise manner of decay because of the superposed fast motion. If we instead average over a time window, we find a residual slow mode. This is nothing but the magnetic drift mode that the WKB predicts should exist. Looking at the mode's behavior in this figure suggests that it has approximately constant frequency, but that its amplitude seems to decrease linearly with conformal time in region II.

Now, we expect the transition region II to extend from $x\approx \bflam \mpsi$ to $x\approx\mpsi$. At early times (large $x$), the magnetic field and large off-diagonal mass terms shut off and the fields transition to their early-time free field asymptotics.  Since the slope of the slow mode axion amplitude $A(x)$ appears to be of order $\bflam^{-1}$ and the oscillation period is $\delta x\sim 1$, WKB should be a good approximation to the behavior of both the slow and the fast modes when $\bflam$ is large. Hence, we should be able to see these two behaviors -- constant frequency and decaying amplitude -- from the WKB approximation.

To this end, let us unwind our various WKB parameter   
definitions and write the slow mode solution to the eikonal in terms of the background parameters. Doing so, we find that the
slow mode frequency is indeed approximately constant over region II, as suggested by the averaging procedure in Fig. \ref{fig:slowmoderevealed}:
\be
\label{leadingeikonal}
S_{\slow} \approx \omega_\slow x   ~, ~~ 
\omega_\slow^2 = \Bigl(\frac{\bfkap^2\bfgam^2-\bfnu^2}{\bfbeta^2+\bfgam^2}\Bigr) = \Bigl(\frac{\mpsi^2-2}{3\mpsi^2} \Bigr).
\ee
Note that the frequency goes imaginary, leading to an instability, when $\mpsi^2<2$. The same instability was noted in the closely related analysis of \cite{Dimastrogiovanni:2012ew}.

Approximating the eikonal using Eqn.~\pref{leadingeikonal}, the gauge mode equations of motion, Eqn.~\pref{scalarwkb}, have the property that $B(x)\propto A(x)/x$, and $C(x)\propto A(x)/x)$.
Let us assume a regime of power law behavior of the amplitude functions,
\be
A(x) \sim A_0 x^b~,~~
B(x) \sim B_0 x^{b-1} ~,~~
C(x) \sim C_0 x^{b-1} ~ ,
\ee
which we take to form part of a series expansion of these functions in powers of $1/x$.
Inserting these expansions into the equations of motion, Eqn.~\pref{scalarwkb}, one finds relations between the amplitudes of the gauge scalars and the axion; at leading order, these are
\bea
B_0 &=& \frac{\bfnu A_0}{\bfkap^2-\omega_\slow^2} \nn\\
C_0 &=& \frac{i \bfbeta\,\omega_\slow \, A_0}{\bfkap^2-\omega_\slow^2} ~.
\label{regionIIamps}
\eea
When these relations are substituted into the remaining equation of motion for the axion, the expansion at large $x$ 
(\ie\ consider $x$ large enough that an expansion in inverse powers of $x$ is valid, but bearing in mind that we want to keep $x<\bflam\mpsi$ to remain in region II)
has a leading term which recovers the expression, Eqn.~\pref{leadingeikonal}, for $\omega_\slow$, and a subleading term which determines 
\be
b =1~,
\ee
thus reproducing the observed linear decrease in the slow mode amplitude in Fig.~\ref{fig:slowmoderevealed}. 

To summarize, the WKB approximation to the axion mode function reveals that in the intermediate time region II, underneath a rapidly oscillating fast mode which behaves as 
\be
\ada_{\rm fast} \propto x^{1/2\pm i\sqrt3 \bflam\mpsi} ~,
\ee
there is a slow mode with an approximately constant frequency $\omega_{\slow}$, Eqn.~\pref{leadingeikonal}, which has an amplitude decreasing linearly in $x$.  We will see that it is this slow mode that matches onto the growing late-time behavior of the axion after horizon crossing; since the intermediate time region II extends over $\mpsi<x<\bflam\mpsi$, at horizon crossing this mode has decreased in amplitude by a factor of order $\bflam$; therefore we expect the amplitude of the late-time fluctuation spectrum to be reduced by this factor.  Unfortunately, the WKB approximation breaks down when $x$ approaches the transition point $x\approx\mpsi$ between regions II and III; a better approximation is needed to extract an accurate estimate of the amplitude of the axion perturbations after horizon crossing.

%%%%%%%%%%%%%%%%%%%%%%%%%%%%%%%%%%%%
%%
\subsection{Reduced action}\label{sec:reduced}
%%
%%%%%%%%%%%%%%%%%%%%%%%%%%%%%%%%%%%%

Now that we have an indication from the WKB analysis that the mode of interest for the axion is the slow ``magnetic drift'' mode. Starting from the action given in Eqn.~\pref{orthoscalaract}, we can analyze the dynamics of this mode more precisely.  In the magnetic drift regime, forces from the potential balance against the large analogue magnetic field. This implies that the kinetic energy term for the axion is largely irrelevant in both the action and equations of motion.  Therefore, when the analogue magnetic field is sufficiently large (\ie\ for $x\ll \bflam\mpsi$), we can approximate the dynamics
of the axion perturbation well by dropping its kinetic entirely.  We can also make further simplifications. 
The diagonal mass term for the axion perturbation, $\ada$, is given by
\be
1 - \frac1{x^2}\Bigl(2-\frac{V_{,\axsub\axsub}}{H^2} \Bigr) + \frac{\bflam^2\mpsi^2}{2\mpsi^2+x^2}. 
\ee
For large $\bflam$, the last term is the dominant one over the entirety of regions II and III, which encompasses horizon crossing.  Therefore, we can make a further approximation and retain only this last term in the $\adab \ada$ potential term in the action.

The equation of motion for $\ada$ that results from this pair of approximations is
\be
\label{Xdrift}
\ada \approx 
\frac{\sqrt{2} \bigl((4 \mpsi^4+2 \mpsi^2 x^2+x^4) \phihat
-\mpsi^2(2 \mpsi^2+x^2)(x(\phihat'-\zhat')+2 \zhat)\bigr)}{{\bflam} \mpsi^3 x^2}.
\ee
Plugging this solution back into the action gives a reduced action for the gauge scalars
\bea
\mathcal{S}_{\rm red} &=& \half \int \biggl[
\Bigl(4+\frac{4\mpsi^2}{x^2}+\frac{x^2}{\mpsi^2}\Bigr)\phihatb'\phihat'
+ \Bigl( 3+\frac{4\mpsi^2}{x^2} \Bigr) \zhatb' \zhat'
- \Bigl(2+\frac{4\mpsi^2}{x^2} \Bigr) (\phihatb' \zhat' + \zhatb' \phihat' ) 
\\
& &\hskip 1cm - \Bigl(4-\frac{4}{x^2}-\frac{2x^2}{\mpsi^4}-\frac{8-x^2}{\mpsi^2}+\frac{4\mpsi^2(2+x^2)}{x^4} \Bigr)\phihatb\phihat
- \Bigl(\frac{6}{\mpsi^2}-\frac{8\mpsi^2}{x^4}+\frac{8}{x^2}\Bigr) \phihatb\zhat
\nn\\
& & \hskip 1cm+ \frac{2x}{\mpsi^2} \bigl(\phihatb\zhat' - \zhatb \phihat'\bigr)
- \Bigl(\frac{8}{\mpsi^2}-\frac{8\mpsi^2}{x^4}+\frac{8}{x^2}\Bigr) \zhatb\phihat
- \Bigl(\frac{x^2(x^2-6)+2\mpsi^2(x^2+4)}{x^4}\Bigr) \zhatb\zhat
\biggr]\nn
\eea
whose solution will determine the axion dynamics through the relation given in  Eqn.~\pref{Xdrift}.  Note that in the reduced dynamics $\bflam$ has dropped out; the only dependence of the late-time axion amplitude on $\bflam$ is through the overall factor of $1/\bflam$ in Eqn.~\pref{Xdrift}.

Switching to new fields $\phitil$ and $\ztil$ via the field redefinition
\be
\label{nicevars}
\phihat \to \frac{\mpsi (\sqrt{3} x \phitil + 2\mpsi \ztil)}{\sqrt{6}(2 \mpsi^2 + x^2)} ~,~~
\zhat \to \frac{\ztil}{\sqrt{6}},
\ee
simplifies the reduced action considerably:
\bea
\label{Sred}
{\cal S}_{red} &=& 
\half \int  
\Bigl[ {\phitilb}'_k {\phitil}'_k+{\ztilb}'_k {\ztil}'_k
 -\frac{2}{\sqrt{3} {\mpsi}}( {\ztilb_k} {\phitil}'_k+ {\phitilb}_k {\ztil}'_k)
 \\
& & \hskip 2cm 
- \frac{(\mpsi^2 \!-\! 2)}{{\mpsi}^2} {\phitilb_k} {\phitil_k}
-\frac{2(\mpsi^2 \!+\! 1) }{\sqrt{3} {\mpsi} x} ( {\ztilb_k}{\phitil_k}+{\phitilb_k} {\ztil_k})
-\frac{6 \mpsi^2 \!+\! x^2 \!+\! 6 }{3 x^2} {\ztilb_k} {\ztil_k} \Bigr] .
\nn
\eea
Note that the $\phitil$ dynamics is unstable when $\mpsi^2 < 2$, in accord with the instability observed in the WKB analysis.

Tracking the ``magnetic drift'' mode, Eqn.~\pref{Xdrift}, through the field redefinition, Eqn.~\pref{nicevars}, 
\be
\label{xsoln}
X \approx \frac{\sqrt{2}\bigl(3(\mpsi^2+x^2)\phitil - \mpsi x(3\mpsi\phitil'  - \sqrt{3}x\ztil')\bigr)}{3\bflam\mpsi^2 x} ~,
\ee
the amplitude of the axion on super-horizon scales is determined (up to order one corrections) by the behavior of $\phitil$ near $x=0$:
\be
\ada \sim \frac{\sqrt{2}\, \phitil(0)}{\bflam\, x} + O(x)~.
\ee
Therefore, we need to analyze the dynamics of the reduced system to determine the late-time amplitude of the axion perturbations.

The reduced system has a constant analogue magnetic field of strength $2/(\mpsi\sqrt{3})$.
At large $x$, the mass term of the reduced system goes to a constant
\be
\frac{(\mpsi^2 \!-\! 2)}{\mpsi^2} {\phitilb_k} {\phitil_k} + \frac13 {\ztilb_k} {\ztil_k}
\ee
The system therefore exhibits plane wave behavior 
\be
\label{planewavered}
\phitil \sim \phitil_\infsub e^{i\omega x} ~,~~
\ztil \sim \ztil_\infsub e^{i\omega x}
\ee
at large $x$; there are two modes,
\bea
\omega &=& \pm\omega_\slow ~,~~ \ztil_\infsub = \mp i\phitil_\infsub \sqrt{\mpsi^2-2} ~;
\\
\omega &=& \pm 1 ~,~~ \ztil_\infsub = \pm i\phitil_\infsub \sqrt{3}/\mpsi ~.
\nn
\eea
This plane wave asymptotic matches onto the WKB solution in region II, in which both WKB and the reduced system dynamics are good approximations. 
At small $x$, we look for power law solutions
\be
\phitil \sim x^\alpha(\phitil_0+\phitil_1 x) ~,~~ 
\ztil \sim x^\alpha(\ztil_0+\ztil_1 x) ~.
\ee
and find two solutions
\bea
\label{powred}
\alpha &=& 0 ~,~~ \ztil_0=0 ~,~~ \ztil_1=-\frac{\phitil_0}{\mpsi\sqrt{3}}
\\
\alpha &=& \half \pm i\sqrt{7+8\mpsi^2} ~,~~ \phitil_0=0 ~,~~ \ztil_1=0 ~,~~ \phitil_1=\frac{\ztil_0}{\mpsi\sqrt{3}}
\nn
\eea
The second solution in each of Eqns.~\pref{planewavered} and \pref{powred} is the ``trivial" solution that we previously found
in the WKB analysis, Eqn.~\pref{boring}, that consists mostly of gauge fluctuations. It damps away at late times, so we will again ignore it in what follows. The first solution is the slow mode of interest.
Note that in the slow mode solution, $\ztil$ decays linearly to zero at late times with a slope set by the constant term $\varphi(0)$, while $\varphi$ has undetermined constant and linear terms.

\begin{figure}
\begin{center}
\includegraphics[width= \textwidth]{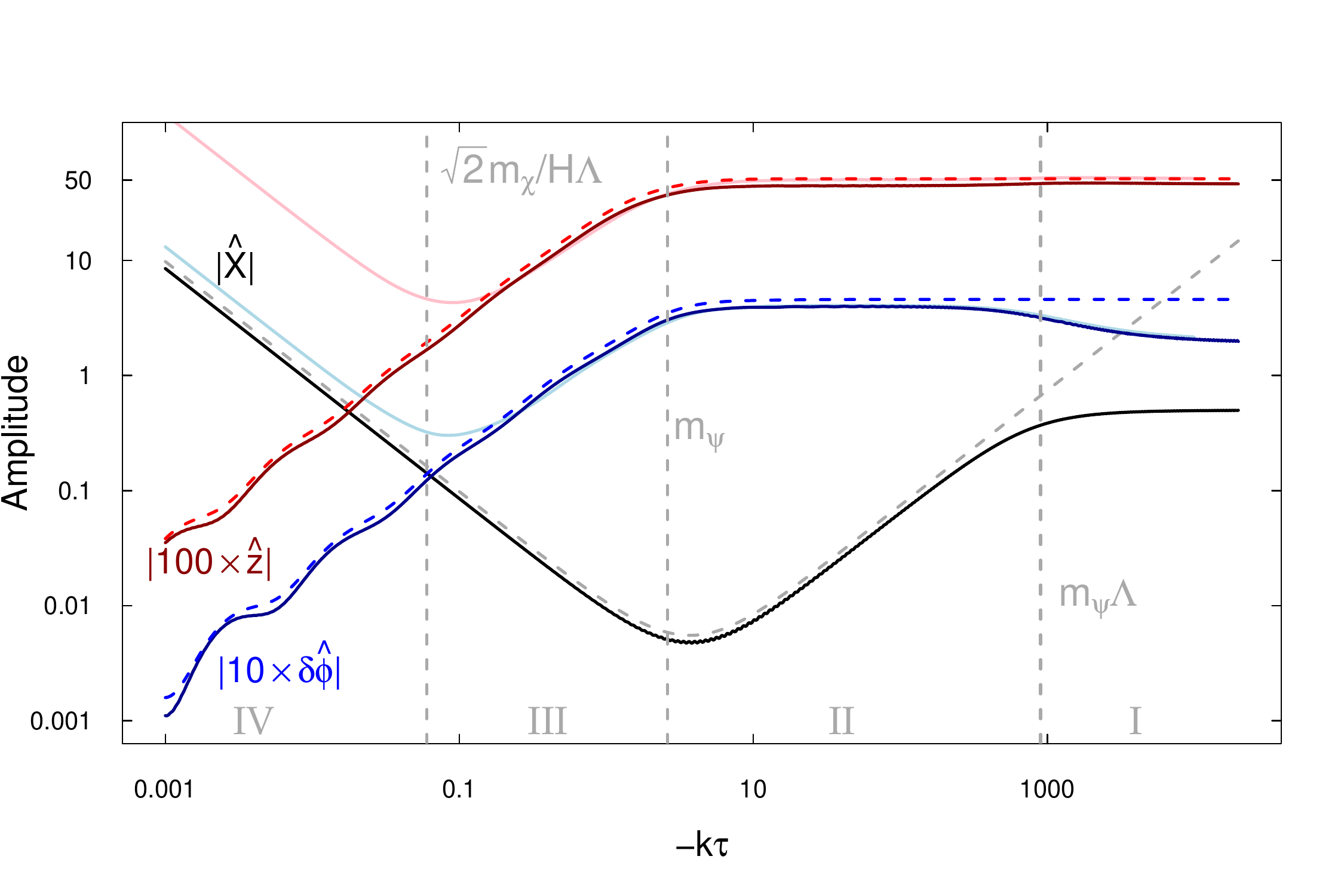}
\caption{\it Comparison of numerical integration of the full dynamics (solid lines) and reduced dynamics (dashed lines).  We show the absolute values of the fields, with boundary conditions set by the WKB solution for the slow mode in the far past, and the relative amplitudes fixed by e.g. Eqn.~\pref{fudgefactor}; parameters are chosen so that $\bflam=350$, $\mpsi=2.6$.  The value of $\ada$ in the reduced system is determined from Eqn.~\pref{xsoln}.  $\phihat$ has been multiplied by 10, and $\zhat$ by 100, 
so that the curves don't lie on top of one another.  The agreement persists until $x\gtrsim \bflam\mpsi$, where the terms dropped in the reduced system contribute significantly to the dynamics. The difference between the light and dark solid lines is that the dark solid lines represent solutions that omit, and the light solid lines solutions that include, the contribution to the dynamics
coming from the axion's mass term, $m_\axion^2 = V''(\axion/f)$. 
In the reduced system, the gauge perturbations do not have a late-time growing solution. Their late time growth as $1/x$ is entirely due to their coupling to the axion
when it has a mass term; this term is dropped in the reduced dynamics. As we show, the deviation between the two solutions commences around $x \simeq \sqrt{2} m_\axion/ H\bflam$.  }
\label{fig:FullvsRed}
\end{center}
\end{figure}

The reduced system holds over intermediate time regions II and III; in particular it provides a good approximation to the horizon crossing transition for the scalar modes.  A comparison of the full and reduced dynamics is shown in Fig.~\ref{fig:FullvsRed}.
WKB is a good approximation to the dynamics in regions I and II.  There is thus an overlap of validity of both approximation schemes in the intermediate time region II.  We can therefore use WKB to determine the intermediate time fields in region II starting from free field normalizations in the asymptotic past of region I, and then use these intermediate time values to set initial conditions for the reduced system which then carries the system through horizon crossing.
We have not succeeded in solving for the overall amplitude of the WKB motion progressing between regions I and II; however, a reasonable numerical approximation over the range of $\mpsi$ of interest relates the initial conditions in the asymptotic past of the full system to those of the reduced system as follows:
\be
\label{fudgefactor}
2( |\phitil_{\rm red}  |^2 + |\ztil_{\rm red}  |^2) \approx | \ada_{\rm full} |^2 + |\varphihat_{\rm full}  |^2 + |\zhat_{\rm full}  |^2
\ee
This approximation is accurate to about 10\% for large $\mpsi$, and to about 30\% for $\mpsi\approx\sqrt{2}$.

Qualitative features of the reduced dynamics can be deduced from the action, Eqn.~\pref{Sred}.  The transition between the two asymptotics, Eqn.~\pref{planewavered} and Eqn.~\pref{powred} takes places around $x\approx\mpsi$.  At this point, $\ztil$ starts to decrease toward zero, and $\phitil$ rises towards its final value.  For large $\mpsi$, there are several oscillation periods of the fields, and an adiabatic approximation is reasonable.  The WKB frequency is approximately constant: $\omega_\slow^2\approx \frac13-\frac{2}{3\mpsi^2}$.  As in the discussion below Eqn.~\pref{omegafast}, in the adiabatic approximation the action over a period is conserved; with constant frequency, the amplitude is approximately conserved, and so the final value $\varphi(0)$ is the same as the amplitude of the initial oscillation.

\begin{figure}
\begin{center}
\includegraphics[width=0.75 \textwidth]{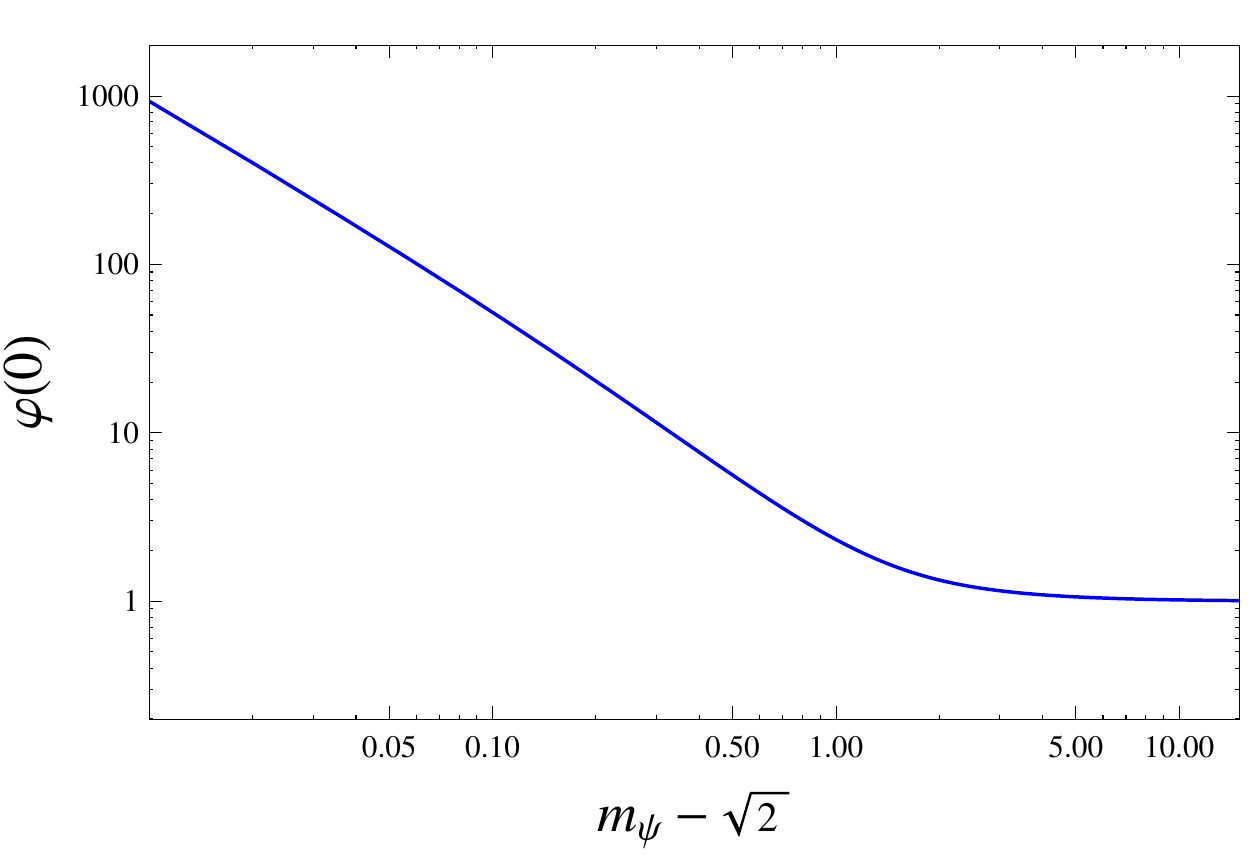}
\caption{\it In this figure, we show the late-time value of $\phitil$ obtained from exact numerical integration of the reduced system as a function of $(\mpsi-\sqrt{2})$.  We take initial conditions that match the unit amplitude WKB slow mode. }
\label{fig:phizero}
\end{center}
\end{figure}

On the other hand, as $\mpsi\to\sqrt{2}$, $\varphi(0)$ diverges.  The local WKB frequency goes imaginary at 
\be
x_{\rm unstable} \approx \biggl[\frac{2(8+7\mpsi^2-\mpsi^4)}{\mpsi^2-2}\biggr]^{1/2} ~.
\ee
Hence, whenever we have $\mpsi<2\sqrt{2}$, there will be a period of instability for $x<x_{\rm unstable}$ where $\varphi$ will grow.
As $\mpsi\to\sqrt{2}$ this era of instability lasts for an increasingly long period, and $\varphi(0)$ grows larger and larger with decreasing $\mpsi$.  One can estimate in this (admittedly crude) WKB approximation that 
\be
\varphi'(x) \approx i\omega_{\rm loc}(x) \varphi(x) ~,~~ \omega_{\rm loc}(x) \approx i\frac{\sqrt{6}}{x} + O(\mpsi-\sqrt{2}).
\ee
This approximation predicts an $\mpsi$ dependence in the growth of $\varphi(0)$
\be
\label{vphi}
\varphi(0) \sim (\mpsi-\sqrt{2})^{-\sqrt{3/2}}
\ee
which is in reasonable agreement with numerics for~$\mpsi\lesssim2$; the exact numerically-determined
behavior of $\phitil(0)$ is plotted in Fig.~\ref{fig:phizero}.
%

%%%%%%%%%%%%%%%%%%%%%%%%%%%%%%%%%%%%
%%%%%%%%%%%%%%%%%%%%%%%%%%%%%%%%%%%%
%%%
\section{Vector fluctuations}\label{sec:vectors}
%%%
%%%%%%%%%%%%%%%%%%%%%%%%%%%%%%%%%%%%
%%%%%%%%%%%%%%%%%%%%%%%%%%%%%%%%%%%%

\begin{figure}[t]
\includegraphics[width = 3. in]{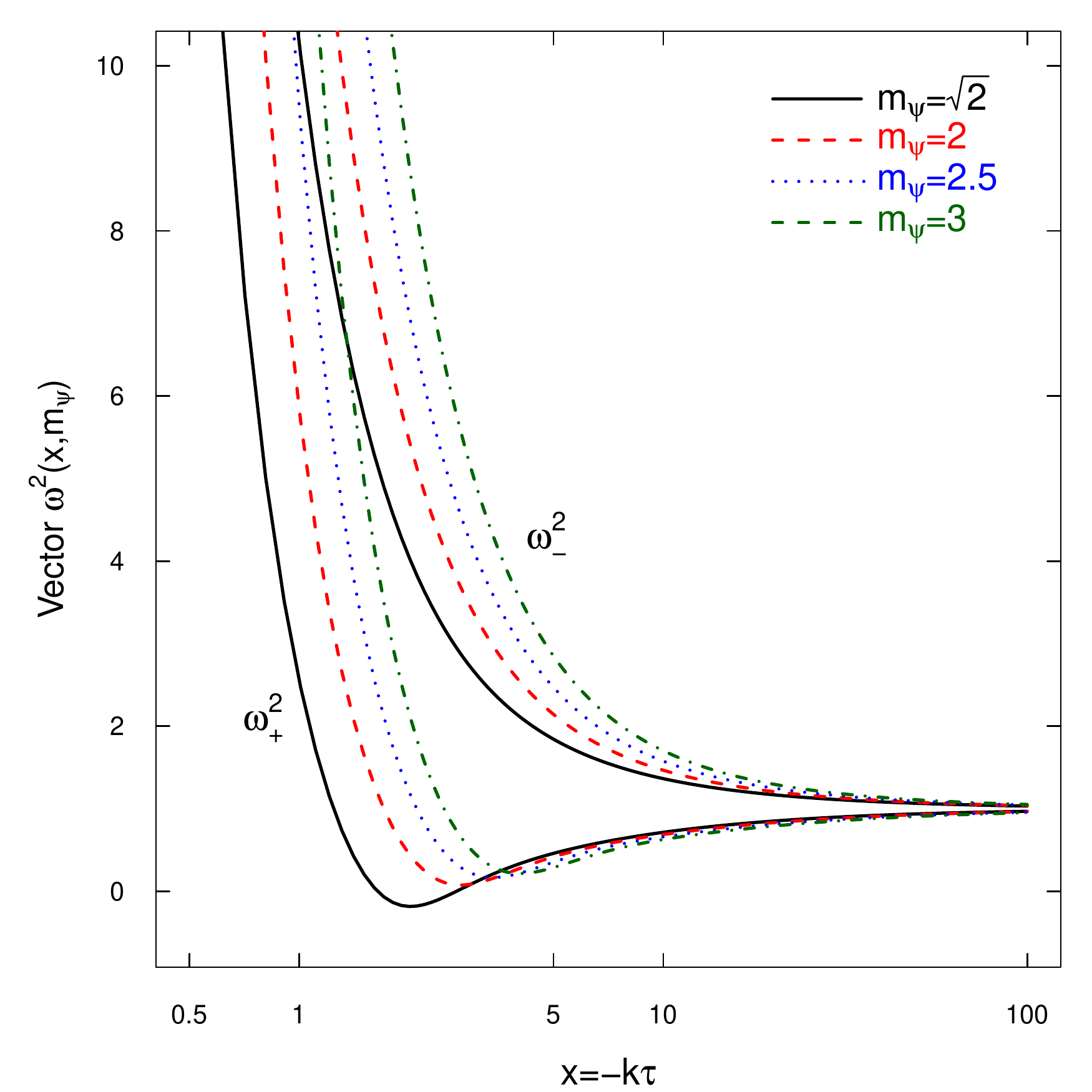}\includegraphics[width = 3. in]{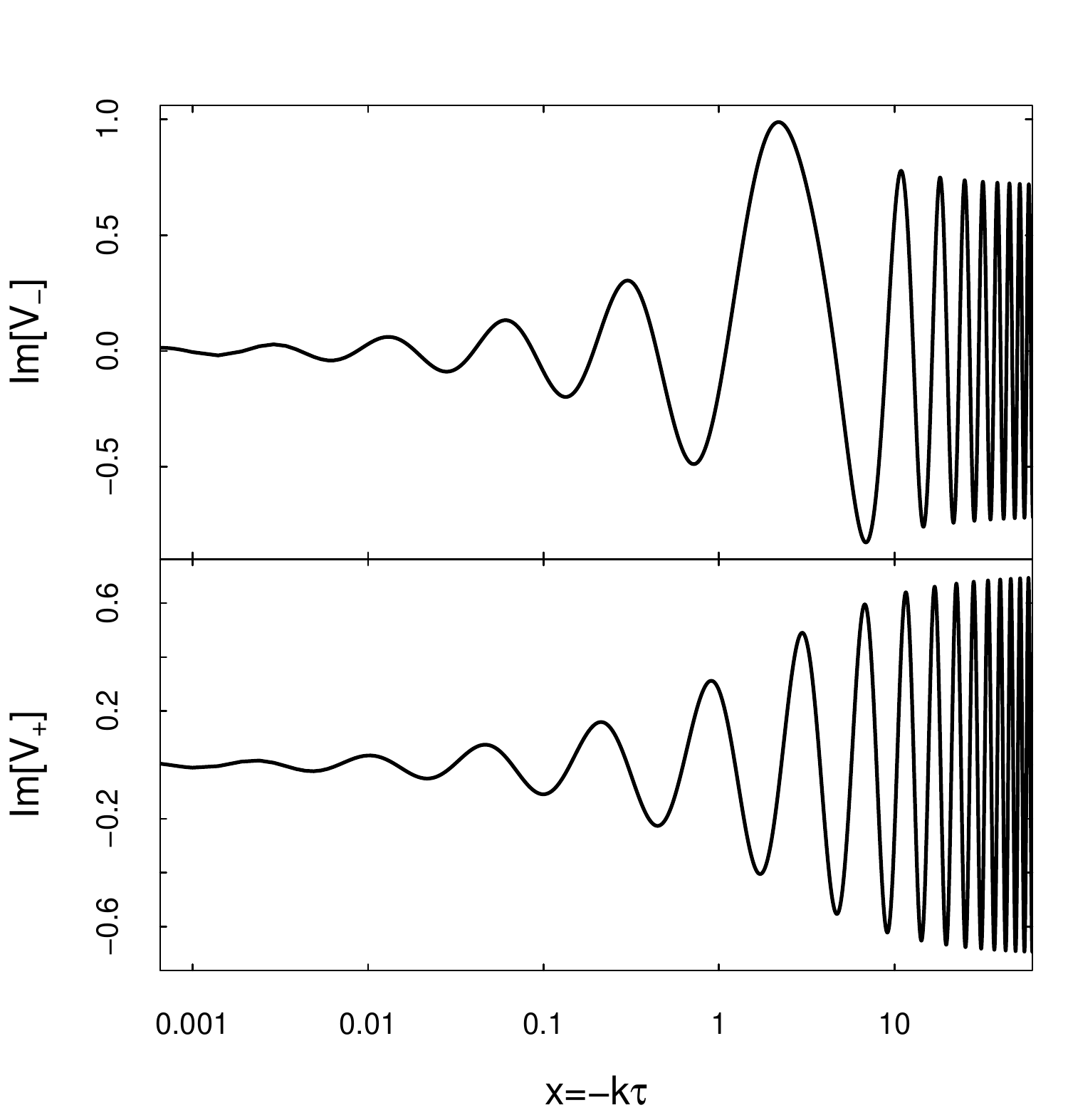}
\caption{\it (Left)Vector field effective frequencies. We plot the effective frequencies $\omega^2_{\pm}(x, \mpsi)$ as a function of $x$ for various $\mpsi$. The lower curves correspond to $\omega^2_+$ while the upper curves correspond to $\omega^2_-$. (Right) Vector field mode functions.  Parameters chosen: $\{\lambda = 500, \, g= 3\times10^{-4},\, f = 0.02,\, \mu= 10^{-3} \}$, which give $\mpsi \approx 2.64$.}\label{fig:vecmasses}
\end{figure}

For the  vector degrees of freedom, $u^{\pm}$ and $v^{\pm}$,  the Fourier space action reads,
\begin{align}\nn\label{eqn:helvecaction}
\delta^2 \mathcal{L}_{V} =  
&\partial_{\tau}v_{\pm}\partial_{\tau}\bar{v}_{\pm}+\partial_\tau u_\pm \partial_\tau \bar{u}_\pm-k^2(v_{\pm}\bar{v}_{\pm} + u_\pm \bar{u}_\pm) +\(2g^2\phi^2-8 \frac{g^2(\partial_{\tau}\phi)^2}{k^2+2g^2\phi^2}\)u_\pm\bar{u}_\pm\\ \nn&
\pm g\phi k\big(v_{\pm}\bar{v}_{\pm} - u_\pm\bar{u}_\pm\big) 
-g^2\phi^2(v_{\pm}\bar{v}_{\pm}+ u_\pm\bar{u}_\pm )-g\phi\(\frac{\lambda}{f}\partial_{\tau}\axion-g\phi\) \(v_{\pm}\bar{v}_{\pm}- u_\pm\bar{u}_\pm\)\\
& +k\frac{\lambda}{2f}\partial_{\tau}\axion  \big(\pm v_{+}\bar{v}_{+} +i\(v_+ \bar{u}_{+}+v_-\bar{u}_- -{\rm c.c.}\)  \pm  u_+\bar{u}_+\big),
\end{align}
where the Gauss's law constraint has been integrated out. That is, we solve the vector part of Gauss's law for $\da_0$ and substitute it back into the action.  Note that the action for the various vector modes is polarization dependent. As above, these four degrees of freedom are not independent, but are again related by the gauge condition Eqn.\ (\ref{eqn:vecgaugecon}). For the vectors, we can write this condition as,
\begin{align}
v_{\pm}
%%
%= &  i\(\frac{2 g \phi}{ k}  \mp 1 \)u_{\pm} 
= 
i\(\frac{2 m_{\psi }}{x}  \mp 1 \)u_{\pm} =  i \cF_{\pm}u_{\pm},
\end{align}
and substituting to eliminate the $v_{\pm}$ modes in favor of the $u_{\pm}$ modes. We finally introduce the canonically normalized fields
\begin{align}\label{eqn:canonnorm}
\v_{\pm} = \sqrt{2(1+\cF_{\pm}^2)}u_{\pm}.
\end{align}
Making use of the background equations of motion Eqns.\ (\ref{slowrollpsi}) and (\ref{slowrollX}), the action for the rescaled field is
%\begin{widetext}
\begin{align}\nn
\delta^{2}\mathcal{L}_{\rm V} 
= & 
  \frac{1}{2}\Bigg[\v_{\pm}'\bar{\v}_{\pm}'-\v_\pm \bar{\v}_{\pm}  -\(2\frac{\mpsi^2}{x^2}\(1 + \frac{1}{\mpsi^2}\)\cF_{\pm}^2+\frac{\mpsi^2}{x^2}\( \frac{8}{x^2+2\mpsi^2}\)-\frac{\cF'_{\pm}{}^2}{(1+\cF_{\pm}^2)}\) \frac{\v_{\pm}\bar{\v}_{\pm}}{1+\cF_{\pm}^2}\\
&
\pm \frac{1 }{x}\( \( 2\mpsi + \frac{1}{\mpsi} \)\cF_{\pm}^2+\( \mpsi + \frac{2}{\mpsi} \)\)\frac{\v_\pm \bar{\v}_{\pm}}{1+\cF_{\pm}^2}\Bigg]
\end{align}
Varying the action and working in the slow roll limit we find the equations of motion for the vector modes,
\begin{align}\nn
\v^{\pm}{}'' +\omega^2_\pm(\mpsi, x)\v_{\pm} = 0
\end{align}
where we have written
\begin{align}
\omega^2_\pm(\mpsi, x) = & 1+\(2\frac{\mpsi^2}{x^2}\(1 + \frac{1}{\mpsi^2}\)\cF_{\pm}^2+\frac{\mpsi^2}{x^2}\( \frac{8}{x^2+2\mpsi^2}\)-\frac{\cF'_{\pm}{}^2}{(1+\cF_{\pm}^2)}\) \frac{1}{1+\cF_{\pm}^2}\\\nn
&
\mp \frac{1 }{x}\( \( 2\mpsi + \frac{1}{\mpsi} \) \cF_{\pm}^2+\( \mpsi + \frac{2}{\mpsi} \)\)\frac{1}{1+\cF_{\pm}^2}.
\end{align}
In the left panel of  Fig.\ \ref{fig:vecmasses} we plot the effective frequencies. Aside from a brief period near $x = \mpsi$ for the $\v_+$ modes at  low values of $\mpsi$, the frequencies are positive, and thus these degrees of freedom are stable. 
In  the right panel of Fig.\ \ref{fig:vecmasses} we show the behaviour of the canonically normalized field $\v_{\pm}$.  We have plotted the imaginary part of the field; the real part is qualitatively similar.  The physical field is related to the canonically normalized field by Eqn.\ \ref{eqn:canonnorm}. Note that the physical fields are vanishing even faster at late times than $\v_{\pm}$, since for $x \ll \mpsi$, $u_{\pm} \approx \v_{\pm} x/(2\sqrt{2}\mpsi)$. There does not appear to be any observable consequence of these modes.

%%%%%%%%%%%%%%%%%%%%%%%%%%%%%%%%%%%%
%%%%%%%%%%%%%%%%%%%%%%%%%%%%%%%%%%%%
%%%
\section{Tensor fluctuations}\label{sec:tensors}
%%%
%%%%%%%%%%%%%%%%%%%%%%%%%%%%%%%%%%%%
%%%%%%%%%%%%%%%%%%%%%%%%%%%%%%%%%%%%

We now consider transverse traceless excitations of the metric and gauge field configuration~-- the tensor modes
\begin{align}
t^{\pm} = & \frac{1}{\sqrt{2}}\( \frac12(t_{11}-t_{22})\pm i t_{12}  \),\\
\gamma^{\pm} = &\frac{1}{\sqrt{2}}\( \frac12(\gamma_{11}-\gamma_{22})\pm i \gamma_{12}  \).
\end{align} 
These excitations are gauge invariant at linear order. That is, the gravitational waves and spin-2 fluctuations of the gauge fields are invariant to linear order in infinitesimal coordinate transformations and SU(2) gauge transformations, respectively. Furthermore, they are not subject to the constraints from the Einstein equations, or Gauss's law at this order. Both of these properties are simple consequences of the fact that a scalar, vector and tensor decomposition results in modes that are independent at linear order in perturbation theory. After rescaling the amplitude of the tensor modes via
\begin{align}
\hg^{\pm} = \frac{a\gamma^{\pm}}{\sqrt{2}},\quad \hy^{\pm} = \sqrt{2}t^{\pm}
\end{align}
to canonically normalize the fields, and making use of the fact that the fields satisfy a reality condition, which allows us to write $\gamma^{\pm}(-{\bf k}) = \bar{\gamma}^{\mp}({\bf k})$, the action at Eqn.\ (\ref{eqn:actspin2}) for the canonically normalized fields becomes
\begin{align}\label{act:spin2hel}
\mathcal{S}_T  = & \frac{1}{2}\int \frac{d^3 k}{(2\pi)^3} d\tau \Bigg[  \partial_{\tau}\hg_k^{\pm}\partial_{\tau}\bar{\hg}_k^{\pm}-\(k^2-\frac{1}{a}\frac{\partial^2 a}{\partial\tau^2}-2\dot{\phi}^2 + 2 g^{2}\frac{\phi^4}{a^2}\)\hg_k^{\pm}\bar{\hg}_k^{\pm}\\ \nn & \hskip 2 cm 
\partial_{\tau}\hy_k^{\pm}\partial_{\tau}\bar{\hy}_k^{\pm}-\(k^2 +  g\phi\frac{\lambda}{f}\partial_{\tau}\axion  \)\hy_k^{\pm}\bar{\hy}_k^{\pm}  \pm k\(\frac{\lambda}{f}\partial_{\tau}\axion + 2g\phi\) \hy_k^{\pm}\bar{\hy}_k^{\pm}
\\\nn &  \hskip 2cm -  2\dot{\phi}( \partial_{\tau} \hy^{\pm}_k\bar{\hg}_k^{\pm}+\partial_{\tau} \bar{\hy}_k^{\pm}{\hg}_k^{\pm}) \mp 2k g\frac{\phi^2}{a} (\hy_k^{\mp}\bar{\hg}_k^{\pm}+\bar{\hy}_k^{\pm}{\hg}_k^{\pm}) +2g^2\frac{\phi^3}{a} (\hy_k^{\pm}\bar{\hg}_k^{\pm}+\bar{\hy}_k^{\pm}{\hg}_k^{\pm}) \Bigg],
\end{align}
where we remind the reader that overdots are derivatives with respect to cosmic time. There are two important features of this action. Firstly, notice that the action depends on the particular helicity of the modes in two places. The most important is the last term on the second line. This term behaves as a negative, or tachyonic, mass for a period of time near horizon crossing for the $t^{+}$ modes. As we will see in the next section, this will lead to the (exponential) growth of this mode near horizon crossing. The second dependence, which is less important, appears in the coupling between the modes. The second feature we wish to highlight is that this chiral splitting of the polarizations of the tensor modes is \emph{not} entirely due to the Chern-Simons interaction. The helicities are still split even in the limit $\lambda \to 0$. In this limit, the tensor fluctuations of the Yang-Mill field alone appears to depend on polarization. At first, the appearance of this dependence may seem puzzling since the Yang-Mills action preserves parity. This apparent contradiction is resolved when we recall that the background gauge field texture itself violates parity (and CP). Under parity ($P$) the gauge field configuration changes sign but there is no compensating automorphism of SU(2) with which to preserve $P$. The perturbations hence inherit their parity
violating nature from the background. We should further note that the gauge field texture on its own would quickly decay
if it were not being renewed by the parity-violating Chern-Simons interaction; in that sense, all of the parity violation of this theory is sourced by the Chern-Simons interaction.

The equations of motion for the fields follows from the variation of the action. We take the slow roll limit where
$$
H\approx {\rm const}, \quad a \approx -\frac{1}{H\tau} , \quad \dot{\phi} \approx H\phi , 
$$ 
and use the dimensionless time variable $x = - k \tau$. The equations of motion for the left and right-handed gravitational wave amplitudes are given by
\begin{align}\label{eqn:metricfull}
\hg_k^{\pm}{}''+\Big(1-\frac{2}{x^2}-\frac{2}{x^2}(1-\mpsi^2 )\psi^2\Big)\hg_{k}^{\pm}=  2 \frac{ \psi}{x} t^{\pm}_k{}'
+ 2\mpsi(\mpsi \mp x )\frac{\psi}{x^2}t_{k}^{\pm}, %\mp 2\mpsi \frac{\psi}{x}t^{\pm}_k ,
\end{align}
while the left and right-handed spin-2 fluctuations of the gauge field satisfy
\begin{align}\label{eqn:gaugefull}
t^{\pm}_{k}{}''+\(1 + \mpsi \frac{\lambda}{f}\frac{\dot{\axion}}{H}\frac{1}{x^2}\)t^{\pm}_k  \mp \(\frac{\lambda}{f}\frac{\dot{\axion}}{H}+ 2\mpsi\)\frac{1}{x} t^{\pm}_k=  -2\psi\partial_{x}\(\frac{ \hg^{\pm}_k}{x}\)+ 2\mpsi(\mpsi \mp x )\frac{\psi}{x^2}\hg_{k}^{\pm}.%\mp 2\frac{g\psi^2}{H x}\hg^{\pm}_k
\end{align}

\begin{figure}[t!] %  figure placement: here, top, bottom, or page
   \centering
   \includegraphics[width=0.85\textwidth]{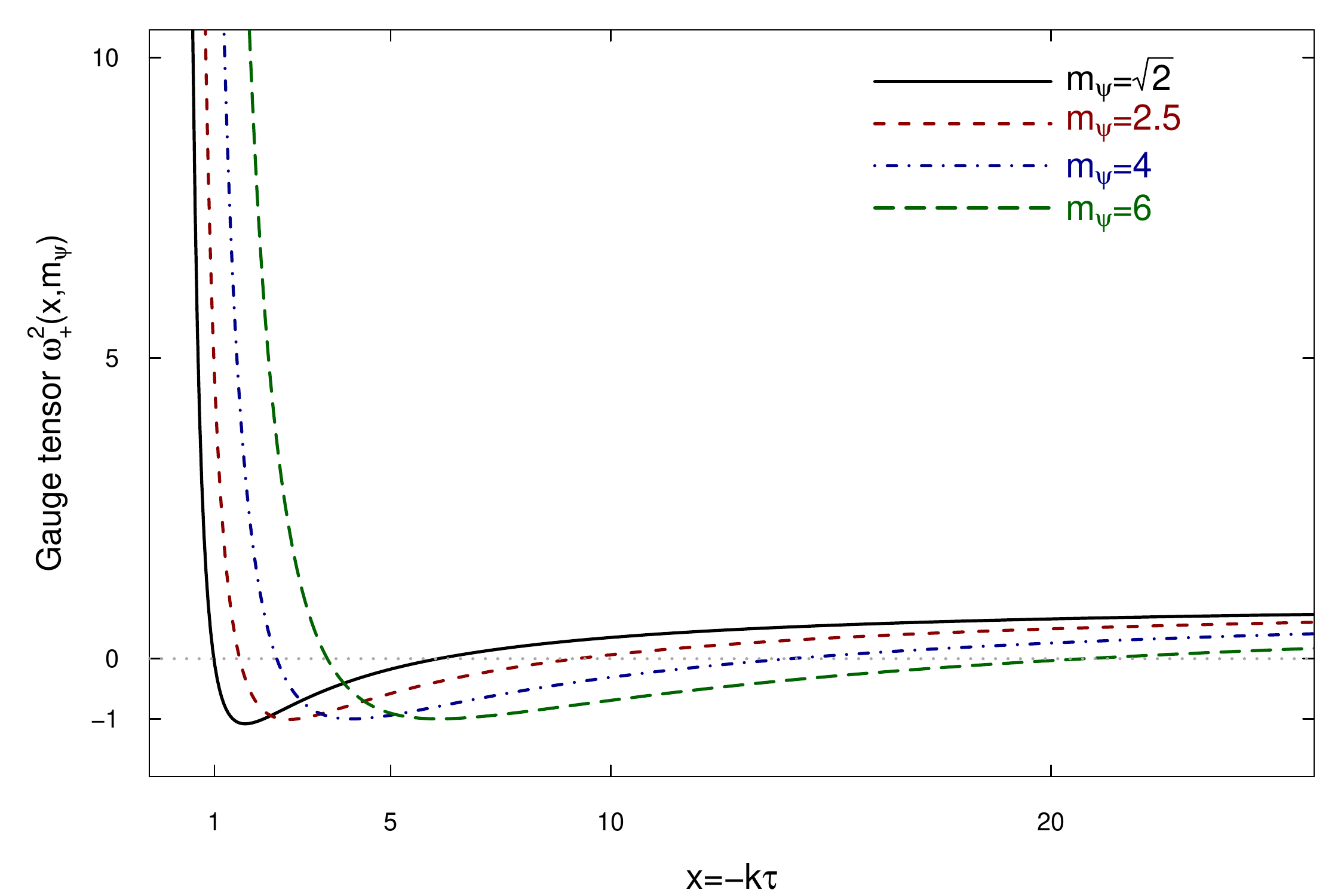} 
   \caption{The effective frequencies for the unstable left-handed tensor gauge modes as a function of the parameter, $\mpsi$, showing the tachyonic behavior over the range
   described in Eqn. \pref{eqn:tensinstab} in the main text.}
   \label{tensoromega}
\end{figure}

Making use of the background equations of motion, we note that this equation has a tachyonic mass term for a range of time (see Fig. \ref{tensoromega})
\begin{align}\label{eqn:tensinstab}
2\mpsi +\frac{1}{\mpsi}- \sqrt{2\mpsi^2 +2+\frac{1}{\mpsi^2} } < x < 2\mpsi +\frac{1}{\mpsi}+ \sqrt{2\mpsi^2 +2+\frac{1}{\mpsi^2} } ,
\end{align}
which means the instability persists for a longer and longer region of time as one increases $\mpsi$. As we will see, this presents a severe problem in providing a viable spectrum of fluctuations. A similar instability occurs for low masses. However, since the scalar fluctuations are (disastrously) unstable when $\mpsi < \sqrt{2}$, we focus only on the region of parameter space for which $\mpsi >\sqrt{2}$.

%
%%%%%%%%%%%%%%%%%%%%%%%%%%%%%%%%%%%%
%%%
\subsection{Approximate solutions}\label{sec:gravapprox}
%%%
%%%%%%%%%%%%%%%%%%%%%%%%%%%%%%%%%%%%
%

Upon first encounter, one is likely to suspect that the coupled system of equations (\ref{eqn:metricfull}) and (\ref{eqn:gaugefull}) governing the evolution of the tensor fluctuations are not liable to admit analytic solutions. However, there is an approximation within which we can find an exact solution which agrees remarkably well with the full dynamics. The first thing to notice is that the evolution of the gauge field modes is dominated by its own mass terms. Hence, 
to a very good approximation, we can approximate the equation of motion for the gauge field by its free field form. The negative helicity modes remain heavy, and the corresponding gravitational wave is unaffected. We thus focus on the positive helicity modes $t^+$.

Working in the slow roll limit, we treat the gauge field background, $\psi$, as constant, and write the equation of motion for the gauge tensor as
\begin{align}\label{eqn:gaugeapprox}
t^{\pm}_{k}{}''+\(1 + \frac{m}{ x^2}  \mp \frac{m_t}{x} \)t^{\pm}_k =  0.
\end{align}
We have made use of the background equations of motion, Eqs.\ (\ref{slowrollpsi}) and \ (\ref{slowrollX}), to define
\begin{align}
m = & %\mpsi\frac{\lambda}{f}\frac{\dot\axion}{H} = 
2(1+\mpsi^2) = \frac{1}{4}-\beta^2,\\% \quad
m_t = & % \frac{\lambda}{f}\frac{\dot\axion}{H} +2\mpsi = 
2\(2\mpsi+\frac{1}{\mpsi}\) = -2 i \alpha,
\end{align}
and have further defined $\alpha$ and $\beta$ for later convenience.
%%%%%%%%%%%%%%%%%%%%%%%%%%%%%%%%%%%%%%%%%%%%%%%%%%%%
%%%%%%%%%%%%%%%%%%%%%%%%%%%%%%%%%%%%%%%%%%%%%%%%%%%%%
\begin{figure}[t]
\center
\includegraphics[width = 0.9 \textwidth]{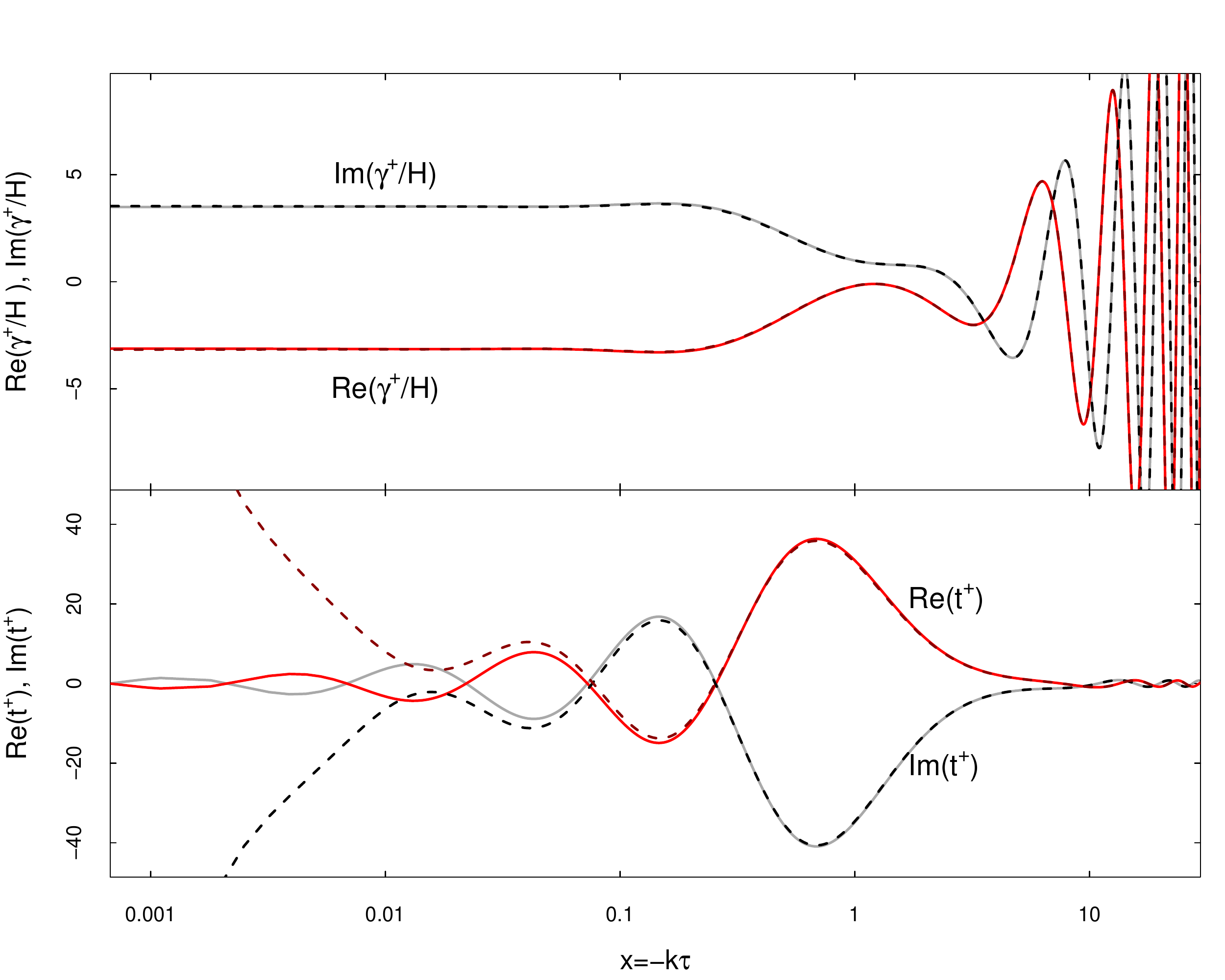}
\caption{\it We show the comparison of a full numerical evaluation of Eqns.\ (\ref{eqn:metricfull}) and (\ref{eqn:gaugefull}) in solid curves versus the result of numerically evaluating the system with Eqn. (\ref{eqn:gaugefull}) replaced by the free equation at Eqn. (\ref{eqn:gaugeapprox}), shown in dashed curves. The upper panel show the evolution of the metric perturbation in units of the Hubble rate, while the lower panel show the evolution of the transverse traceless perturbation to the gauge field. Although it is evident in the lower panel that the approximate solution for the gauge tensor is breaking down for $x \lesssim 0.1$, this has no impact on the gravitational wave amplitude
since the breakdown of the approximation occurs after horizon crossing ($x=1$). Concretely, this can be seen by observing that the approximate and exact solutions for the gravitational waves in the upper panel are in very good agreement throughout. Since the gravitational waves are the observable quantity, this implies that the approximation made in Eqn. \pref{eqn:gaugeapprox} is adequate for our purposes. In making this plot, we chose parameters such that $m_t = 6.61$, $m = 3.77 $ and $\psi \approx 0.035$.}\label{fig:numapprox}
\end{figure}
%%%%%%%%%%%%%%%%%%%%%%%%%%%%%%%%%%%%%%%%%%%%%%%%%%%%%%
%%%%%%%%%%%%%%%%%%%%%%%%%%%%%%%%%%%%%%%%%%%%%%%%%%%%%%
In Fig.\ \ref{fig:numapprox}, we show the accuracy of this approximation. Notice that the solution of Eqn.\  (\ref{eqn:gaugeapprox}) is an excellent approximation to Eqn.\ (\ref{eqn:gaugefull}) until well after horizon crossing; in this region, 
the amplitude of $\gamma^{+}$ is frozen. At late times the solutions for the comoving gauge field perturbation, $t^{+}$, begin to grow since they couple to the exponentially growing, canonically normalized field $\hat{\gamma}$. However, the physical gauge field perturbation, $t^{+}/a$, freezes out with a small amplitude at late times. 
In this limit, the equation of motion for $t$, Eqn.\ (\ref{eqn:gaugeapprox}), has an exact solution:
\begin{align}
\hy_k(x) = &  A_k M_{\alpha,\beta}(2 i x)+B_k W_{\alpha,\beta}(2 i x),
\end{align}
where $M_{\alpha,\beta}(2 i x)$ and $W_{\alpha,\beta}(2 i x)$ are the Whittaker M and W functions. %, and
We set the values of the constants $A_k$ and $B_k$, by imposing the Bunch-Davies vacuum conditions in the asymptotic past. That is, we demand that the solutions approach canonically normalized positive frequency free plane waves as $x = -k\tau \to \infty$,
\begin{align}
\hg^{\pm}, \hy^{\pm} \to \frac{1}{\sqrt{2 k}}e^{i x}.
\end{align}
In this large $x$ limit, the Whittaker functions have asymptotic expansions,
\begin{align}\nn
M_{\alpha,\beta}(2 i x) \to &\frac{(2 i)^{-\alpha }  \Gamma (2 \beta +1) }{\Gamma \left(-\alpha +\beta +\frac{1}{2}\right)}e^{i x - \alpha\ln x} +\frac{i 2^{\alpha } i^{\beta } (-i)^{\alpha -\beta } \Gamma (2
   \beta +1) }{\Gamma \left(\alpha +\beta +\frac{1}{2}\right)}e^{-i x+\alpha\ln x}, \\
W_{\alpha,\beta}(2 i x) \to & (2 i)^{\alpha } e^{-i x+ \alpha\ln x}.
\end{align} 
Using these, we find that the constants are given by
\begin{align}
A_k = & \frac{1}{\sqrt{2k}}\frac{\Gamma \left(-\alpha +\beta +\frac{1}{2}\right)}{(2 i)^{-\alpha }  \Gamma (2 \beta +1) },\\
B_k = &  \frac{1}{\sqrt{2k}}\frac{\Gamma \left(-\alpha +\beta +\frac{1}{2}\right)}{\Gamma \left(\alpha +\beta +\frac{1}{2}\right)}2^{\alpha } i^{\beta+1 } (-i)^{\alpha -\beta }.
\end{align}
We can now solve for the  resulting positive helicity gravitational wave mode by inverting Eqn.\ (\ref{eqn:metricfull}) using the Greens function for the free gravitational wave equation,
\begin{align}
G(x, x') %= & -\frac{(u_1(x)u_{2}(x') - u_{1}(x')u_2(x))}{W[u_1(x), u_2(x)]} \Theta(x'-x)\\
= & \frac{\Im\[u_1(x)u_{2}(x')\]}{W[u_1(x), u_2(x)]} \Theta(x'-x),
\end{align} 
where $u_1$ and $u_2$ are solutions of the free gravitational wave equation and $W[u_1(x), u_2(x)]$ is the Wronskian.
The formal solution for the  mode function is then given by
\begin{align}\label{eqn:gwhplus}
\hg^{+}(x) = & \hg_0(x)+ 2\int^{\infty}_{0} dx' G(x, x')\(\frac{\psi}{x'} \partial_{x'}  -  \mpsi(\mpsi - x') \frac{\psi}{x'^2}\) \hy_0^{+}(x')+\ldots,
\end{align}
where `$\ldots$' refers to terms dropped in neglecting the backreation of $\hg^+$ on $\hy^+$, and the subscript $0$ here refers to the solutions of the respective free wave equations, Eqns.\ (\ref{eqn:gaugeapprox}) and the left hand side of Eqn.\ (\ref{eqn:metricfull}).  We are interested in the deflection of the modes near where the gauge field modefunction becomes large. In this region, the Whittaker $M$-function is decaying and thus  irrelevant for our considerations. Hence we can approximate
\begin{align}
\hy_{0}^{+} \approx B_k  W_{\alpha,\beta}(2 i x).
\end{align}
Furthermore, to a very good approximation we can ignore the mass terms and terms that arise from the deviation from de Sitter expansion in the free equation of motion for the gravitational wave. In this limit, the equations admit the solutions
\begin{align}
u_{1}(x) = \(1+\frac{i}{x}\)e^{ix}, \quad u_2(x) = u_{1}^*(x),
\end{align}
yielding a correspondingly simple Green's function. Now, fortunately for us, the integrals that result from Eqn.\ (\ref{eqn:gwhplus}) can be performed by making use of the identities
\begin{align}\nn
\int dx \,& x^{n} e^{i x} W_{\alpha, \beta}(2i x)= \frac{x^{n+1} G_{2,3}^{2,2}\left(2 i x\left|
\begin{array}{c}
 -n,\alpha +1 \\
 \frac{1}{2}-\beta ,\beta +\frac{1}{2},-n-1
\end{array}
\right.\right)}{\Gamma \left(-\alpha -\beta +\frac{1}{2}\right) \Gamma \left(-\alpha +\beta +\frac{1}{2}\right)}
\end{align}
and
\begin{align}\nn
\int dx \, & x^{n} e^{-i x} W_{\alpha, \beta}(2i x) =  x^{n+1} G_{2,3}^{2,1}\left(2 i x\left|
\begin{array}{c}
 -n,1-\alpha  \\
 \frac{1}{2}-\beta ,\beta +\frac{1}{2},-n-1
\end{array}
\right.\right)
\end{align}
where $G$ is the Meijer G-function. Thus a closed form analytic solution is available for the evolution of the metric modes. The final expression is long and easily reproduced from the preceding discussion, so we will not write it down it explicitly in the text. At late times, the solution is well approximated by
\begin{align}\label{eqn:gammaplus}
\gamma^{+}(x) = &   \frac{H x}{\sqrt{k^3}} u_1 (x)+2\sqrt{2}\frac{H}{k} B_k \psi  \(I_1+\mpsi I_2 -\mpsi^2 I_3\),
\end{align}
where
\begin{align}\nn
I_1 = & \frac{\left(m^2-2 i m m_t+2 m-2 m_t^2\right) \sec \left( \pi \beta \right) \sinh \left(-i\pi\alpha\right) \Gamma \left(\alpha\right)}{2 m (m+2)} \\ \nn&-\frac{\pi ^2 \left(m^2+2 i m m_t+2 m-2 m_t^2\right) \sec \left( \pi \beta \right) \text{csch}\left(-i\pi\alpha\right)}{2 m (m+2) \Gamma \left(\alpha+1\right) \Gamma \left(-\alpha-\beta+\frac{1}{2}\right) \Gamma \left(-\alpha+\beta+\frac{1}{2}\right)}~,\\
\nn
I_2 = & \frac{\pi  \sec \left(\pi \beta\right) \Gamma \left(-\alpha\right)}{2 \Gamma \left(-\alpha-\beta+\frac{1}{2}\right)
   \Gamma \left(-\alpha+\beta+\frac{1}{2}\right)}
   -\frac{\pi  \sec \left(\pi \beta\right) \Gamma \left(1-\alpha\right)}{m \Gamma \left(-\alpha-\beta+\frac{1}{2}\right) \Gamma \left(-\alpha+\beta+\frac{1}{2}\right)} \\ \nn&
    +\frac{\pi  m \sec \left( \pi  \beta\right)-i \pi  m_t \sec \left( \pi  \beta\right)}{2 m \Gamma \left(1-\alpha\right)} ~,
\nn\\
I_3 = &\frac{\pi ^2 (m+i m_t) \text{sec} \left(\pi 
   \beta\right) \text{csch}\left(-i\pi\alpha \right)}{m (m+2) \Gamma \left(i\alpha \right) \Gamma \left(-\alpha-\beta+\frac{1}{2}\right) \Gamma \left(-\alpha+\beta+\frac{1}{2}\right)} +  \frac{\pi  (m_t+i m) \text{sec} \left( \pi\beta\right)}{m (m+2) \Gamma \left(-i\alpha\right)}~.
\end{align}
\begin{figure}[t]
\center
\includegraphics[width =  \textwidth]{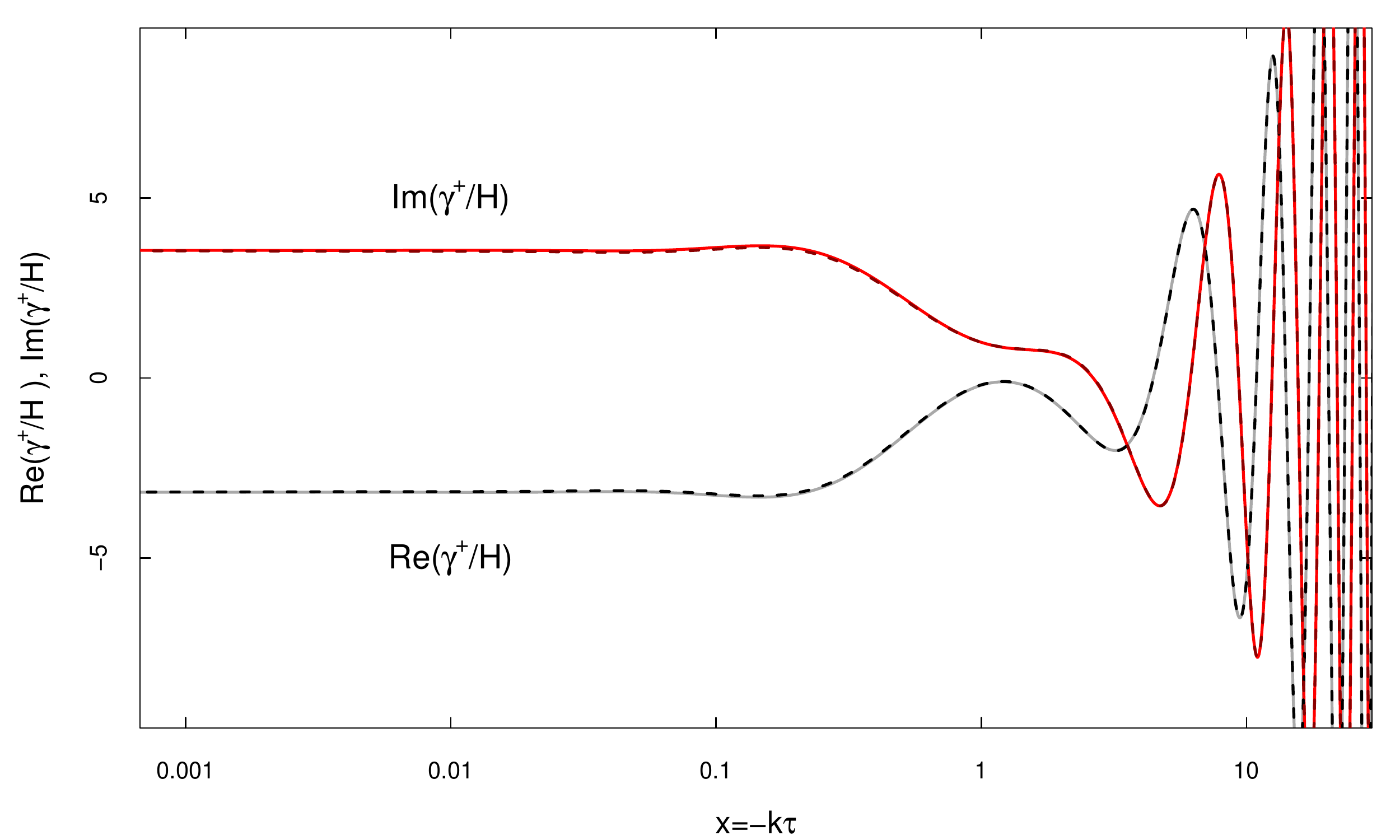}
\caption{\it 
In this plot, we compare our analytic expressions to a full numerical analysis. The solid lines are a plot of the full analytic version of Eqn.\ (\ref{eqn:gwhplus}), where the solution of the gauge field equations are taken to be given only by the Whittaker $W$ function. The results from the exact numerical evaluation of the equations are plotted dashed lines, demonstrating the accuracy of this approximation. The parameters are the same as those in Fig.\ \ref{fig:numapprox}.}\label{fig:analsol}
\end{figure}
The total gravitational wave power spectrum at late times is given by
\begin{align}
\Delta^{2}_{\gamma}(k) =2 \Delta^{2}_{\gamma^+}(k)+2\Delta^2_{\gamma^-}(k) 
\end{align}
where the spectra of left and right-handed modes are defined by
\begin{align}
\langle \gamma^{\pm}_{\bf k}(\tau_*)\gamma^{\pm}_{\bf k'}(\tau_*) \rangle= (2\pi)^3\delta^{3}({\bf k}+{\bf k'})\frac{2\pi^2}{k^{3}}\Delta^{2}_{\gamma^{\pm}}(k) .
\end{align}
Now, the right handed modes $\gamma^{-}$ are, to a very good approximation, unaffected by their interactions with the spin-2 fluctuations of the gauge fields.  Their spectrum is given by the usual result,
\begin{align}\label{eqn:rhgravspec}
\Delta^{2}_{\gamma^-}(k) = \frac{ H^2}{2\pi^2}.
\end{align}
\begin{figure}%[htbp] %  figure placement: here, top, bottom, or page
   \centering
   \includegraphics[width=0.8 \textwidth]{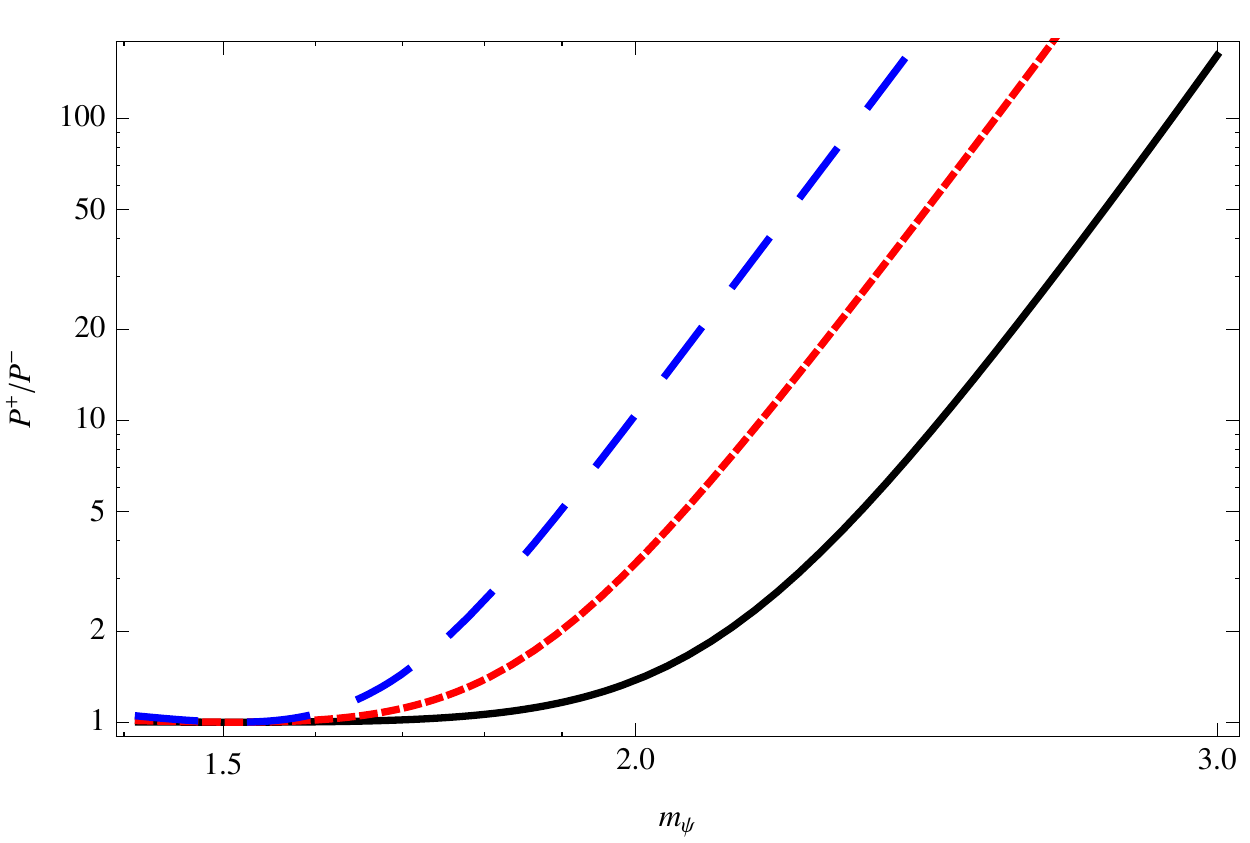} 
   \caption{\it We show the behaviour of the ratio of the power in left handed gravitational waves to the power in right handed gravitational waves as a function of the gauge field mass $\mpsi$ at various values of the gauge field vev $\psi = 0.01$ (black, solid), $\psi = 0.025$ (red, short dashed), and $\psi = 0.05$ (blue, long dashed).}
   \label{fig:powerratio}
\end{figure}
For the left handed modes, $\gamma^+$, the vacuum fluctuations are uncorrelated with the contribution due to their interaction with the gauge field fluctuations, and thus to a good approximation (See Fig. \ref{fig:analsol}),
\begin{align}\label{eqn:lefthandedgw}
\Delta^{2}_{\gamma^+}(k) = \frac{ H^2}{2\pi^2}+4k \frac{H^2}{\pi^2}\psi^2  |B_k|^2  |I_1+\mpsi I_2 -\mpsi^2 I_3|^2.
\end{align}
In Fig.\ \ref{fig:powerratio}, we plot the ratio of the power in left-handed gravitational  waves, Eqn.\ (\ref{eqn:lefthandedgw}), to the power in right handed gravitational waves, Eqn.\ (\ref{eqn:rhgravspec}). For $\mpsi \gtrsim 2$ the gravitational wave power spectrum is dominated by the contribution that is sourced by the gauge field instability, the second term in Eqn.\ (\ref{eqn:lefthandedgw}). 

\begin{figure}%[htbp] %  figure placement: here, top, bottom, or page
   \centering
   \includegraphics[width=0.8 \textwidth]{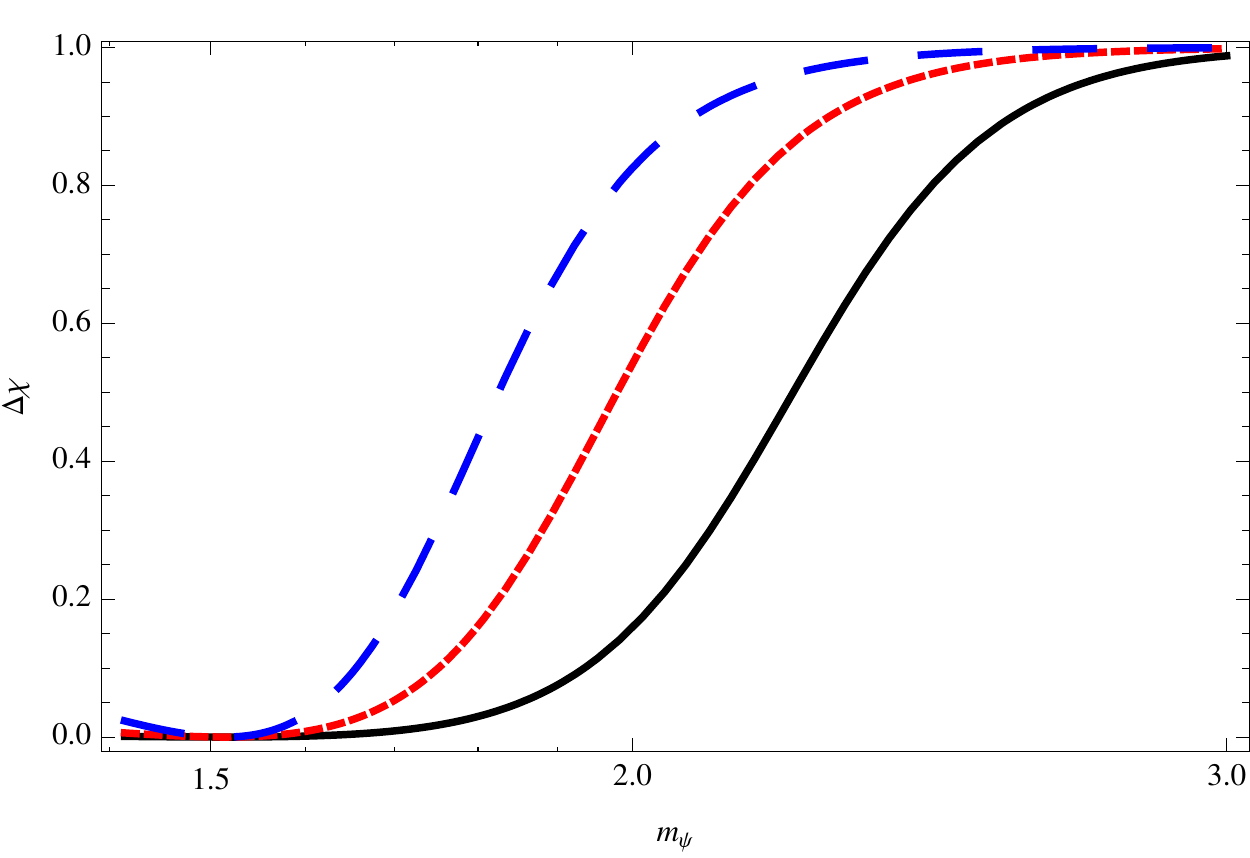} 
   \caption{\it We show the behavior of $\Delta \chi$, which measures the chirality of the gravitational wave power, as a function of the gauge field mass $\mpsi$ at various values of the gauge field VEV $\psi = 0.01$ (black, solid), $\psi = 0.025$ (red, short dashed), and $\psi = 0.05$ (blue, long dashed).}
   \label{fig:chiparam}
\end{figure}

We can also compute the chirality parameter
\begin{align}
\Delta\chi = \frac{\Delta^{2}_{\gamma^+} - \Delta^{2}_{\gamma^-}}{\Delta^{2}_{\gamma^+} + \Delta^{2}_{\gamma^-}}.
\end{align}
We plot the behaviour of this quantity as a function of the mass of the gauge field fluctuations for various values of the gauge field vev $\psi$ in Fig. \ref{fig:chiparam}. This function very quickly approaches unity as the mass $\mpsi \gtrsim 2$.

Finally, we note that the spectral index of the tensor modes takes on a variety of values, depending on the value of the parameter $\mpsi$.  When $\mpsi$ is small, it returns to the usual value for inflation ( $n_t \approx -2 \epsilon_H$) because the second term of Eqn.\ (\ref{eqn:lefthandedgw}) is negligible. On the other hand, it becomes strongly blue at larger values of $\mpsi$; numerically, we find $n_t \sim 0.1$ for $\mpsi \gtrsim 2.5$. This blue spectrum can be easily understood. During inflation, $\mpsi$ slowly increases primarily due to the decreasing hubble rate. This means that modes that leave the horizon later in inflation experience a larger value of $\mpsi$ and consequently a larger instability leading to a larger tensor amplitude.

%%%%%%%%%%%%%%%%%%%%%%%%%%%%%%%%%%%%
%%%%%%%%%%%%%%%%%%%%%%%%%%%%%%%%%%%%
%%
\section{Curvature fluctuations and related observables}\label{sec:curvature}
%%
%%%%%%%%%%%%%%%%%%%%%%%%%%%%%%%%%%%%
%%%%%%%%%%%%%%%%%%%%%%%%%%%%%%%%%%%%

In order to determine how the fluctuations in the fields are imprinted as curvature fluctuations, we need to calculate the gauge invariant quantity \cite{Bardeen:1980kt}
\begin{align}
\mathcal{R} =  \frac{A}{2} + H\delta u
\end{align}
which coincides with the curvature perturbation in comoving gauge. The function $A$, defined in Eqn.\ (\ref{eqn:spatialmet}), characterizes the curvature of spatial hypersurfaces, and $\delta u$ is found by computing the velocity potential, which in turn is defined by (see e.g.\  \cite{Weinberg:2008zzc})
\begin{align}
%\delta T_{ij} = & \bar{p}h_{ij} + a^{2}\[\delta_{ij}\delta p + \partial_{i}\partial_{j}\pi^{s} + \partial_{i}\pi_{j}^{V}+\pi_{ij}^{T}\]\\
%
\delta T_{0i} = & \bar{p}\, \tilde{h}_{i0} - a (\bar\rho+\bar p)(\partial_i \delta u + \delta u_{V}^i).%\\
%
%\delta T_{00} = & -\bar{\rho}h_{00} +  a^2\delta\rho
\end{align}
In this expression, we see that %
$\delta u$ and $\delta u^i_V$ are the scalar and vector perturbations to the velocity potential respectively, while $\tilde{h}_{0i} = a^{2}\delta_{ij}N^j$ is the perturbation to the space-time components of the metric $g_{\mu\nu}$ and $\bar\rho$ and $\bar p$ are the background energy density and pressure respectively. In spatially flat gauge, $A = 0$, and
\begin{align}
\mathcal{R} = H\delta u.
\end{align}
The stress tensor for the theory defined at Eqn.\ (\ref{eqn:CNIaction}) is given by
\begin{align}
T_{\mu\nu} = & 2\tr\[F_{\mu\alpha}F_{\nu\beta}\]g^{\alpha\beta}-\frac{g_{\mu\nu}}{2}\tr\[F_{\alpha\beta}F^{\alpha\beta}\]+ \partial_{\mu}\axion\partial_{\nu}\axion -g_{\mu\nu}\left[\frac{1}{2}g^{\rho\sigma}\partial_{\rho}\axion\partial_{\sigma}\axion+V(\axion)\right].
\end{align}
We can calculate the momentum flux from this expression; to linear order in fluctuations, it is
\begin{align}
T_{0i} 
= & (\bar{p}_{\axion}+\bar{p}_{YM})a^{2}\delta_{ij}N^j+a\dot\axion\partial_{i}\daxion
 + \frac{\dot{\phi}}{a}(2\partial_{[k}\da^k_{i]} + g \phi \epsilon_{aik}   \da^a_k)\\\nn
 &+g\frac{\phi^2}{a^2}(  \epsilon^{a}_{ki}\partial_\tau \da^a_k -  \epsilon^{a}_{ik}\partial_{k}\da^a_0 -2 g \phi \da^i_{0}) -2g^2\frac{\phi^4}{a^2}\delta_{ij}N^j .
\end{align}
Inserting our field decomposition, we find
\begin{align}
T_{0i} 
\approx & (\bar{p}_{\axion}+\bar{p}_{YM})a^{2}\delta_{ij}N^j+a\dot\axion\partial_{i}\daxion
  -\(H\psi \partial_{i}(2 z+4\delta\phi) - g a \psi^3\frac{\lambda}{f}\partial_{i}\daxion\)-2g^2\psi^4a^{2}\delta_{ij}N^j \,,
  \end{align}
where the `$\approx$' indicates that we have dropped terms that decay at late times and worked in the slow-roll approximation. We have also made use of the non-Abelian gauge condition and imposed the Gauss's law constraint, Eqn. (\ref{eqn:scalargauss}), in the long wavelength limit ($k \to0$). In this expression, $\bar{p}_{\axion}$ and $\bar{p}_{\rm YM}$ are the background isotropic pressures due to the axion and gauge fields respectively.

Ignoring the term term proportional to $N^j$, since it is of slow-roll order (see Appendix \ref{app:gravsector}), and noting that $g\psi^3 \lambda/f \gg H\psi$, we find the curvature perturbation to be
\begin{align}
\curv \approx \frac{H}{\bar{\rho}+\bar{p}}g\psi^3 \frac{\lambda}{f}\daxion = \frac{g \psi^3}{2 H \epsilon_H}\frac{\lambda}{f}\daxion% =  \frac{V'}{6\dot{H}}\delta\axion
\end{align}
where 
\begin{align}
\epsilon_{H} = -\frac{\dot{H}}{H^2} = \frac{1}{2 H^2}(\bar{\rho}+\bar{p}),
\end{align}
and $\bar{\rho}$ and $\bar{p}$ are the total background energy density and pressure respectively.
Now recall that on the background solution, to a very good approximation,
\begin{align}
\psi^3 = \frac{ f}{3 g H \lambda}\frac{d V}{d\axion}~.
\end{align}
Since the Hubble rate is dominated by the potential,
$
H^2 \approx V/3
$
and we can write
\begin{align}
\dot{H} \approx \frac{1}{6}\frac{d V}{d\axion} \frac{\dot{\axion}}{H}.
\end{align}
Denoting by $\Delta \axion = \dot{\axion}/H$, which represents the classical field evolution of the axion in a hubble time, we can write the curvature perturbation in the form
\begin{align}
\curv \approx \frac{\delta \axion}{\Delta\axion}.
\end{align}
Somewhat remarkably, we recover the familiar result for single clock inflation. That is, that the curvature fluctuation can be thought of as a fluctuation in the time inflation ends from place to place.  This result suggests that an alternate choice of gauge, where one chooses spatial coordinates so that the axion is unperturbed, might have been advantageous. In the superhorizon limit, such a gauge choice would correspond to comoving gauge.

Using the results above, we find the scalar curvature fluctuation is given by 
\be
\label{PsubR}
{\cal R} \simeq 
\frac{1}{\sqrt{2 k^3}}\cdot
\frac{H \mpsi  \phitil(0)}{\sqrt{2} (1+ \mpsi^2)} \Bigl(\frac{\lambda \mpsi V}{-V_{,\theta}}\Bigr)^{\!1/2}  ~;
\ee
where the subscript $,\theta$ represents a derivative with respect to $\theta = \axion/f$
(and note that in the reduced system, the scalar d.o.f.s are normalized to unit amplitude in the far past); 
expressed in terms of the slow roll parameters
\bea
\label{slowrollparams}
\epsilon_H &=&- \frac{\dot{H}}{H^2}= -\frac{(1+\mpsi^2)V_{,\theta}}{\lambda\mpsi V}
\\
\nonumber
\eta_H &=& \epsilon_H+\frac1{2}\frac{ \dot\epsilon_H}{H \epsilon_H}
=  - \frac{(\mpsi^2+1)V_{,\theta\theta}}{\lambda\mpsi V_{,\theta}}  - \frac{(\mpsi^2-1)V_{,\theta}}{\lambda\mpsi V}  ~,
\eea
one has
\be
\label{Reps}
{\cal R} \simeq
\frac{1}{\sqrt{2 k^3}}\cdot
\frac{H}{\sqrt{2\epsilon_H}}\cdot \frac{\mpsi \phitil(0)}{(1+\mpsi^2)^{1/2}} ~,
\ee
and we define the dimensionless power spectrum of curvature fluctuations in the usual way
\begin{align}
\langle \curv_{\bf k}\curv_{\bf k'}\rangle = (2\pi)^3\delta^{3}({\bf k}+{\bf k'})\frac{2\pi^2}{k^3}\Delta^2_{\curv}(k).
\end{align}
In these expressions, the subscript $\theta$ represents a derivative with respect to $\theta = \axion/f$.
Relative to the corresponding expressions for single-field inflation, the slow-roll parameters, Eqn.~\pref{slowrollparams}, directly exhibit the additional suppression of motion along the inflaton potential by the coupling $\lambda$ to the gauge sector; 
and in the perturbations, there is an additional enhancement from $\phitil(0)$ when $\mpsi$ is small enough.
Differentiating Eqn.~\pref{Reps} with respect to $dN = -H dt$, one finds the spectral index
\be
\label{nseq}
n_s -1 \simeq 
-2\epsilon_H + \eta_H + 2\frac{d\log\phitil(0)}{dN} ~.
\ee
Combining Eqn.\ (\ref{Reps}) with Eqn.\ (\ref{eqn:rhgravspec}) we find the tensor-to-scalar ratio for the unenhanced (right-handed) graviton is given by
\be
\label{runenhance}
r_- \simeq 
 \frac{8(1+\mpsi^2)\epsilon_H}{\mpsi^2\phitil(0)^2} ~,
\ee
while the tensor-to-scalar ratio of the left-handed graviton $r_+$ is enhanced by the effect discussed above, determined from Eqn.\ (\ref{eqn:lefthandedgw}). For low values of $\mpsi \gtrsim \sqrt{2}$, the total tensor to scalar ratio is simply twice the result in Eqn.\ (\ref{runenhance}). However, as demonstrated in the previous section, as $\mpsi$ increases the left handed gravitational waves have a significantly larger amplitude than the right handed modes and the dominant contribution to the tensor to scalar ratio arises from Eqn.\ (\ref{eqn:lefthandedgw}). As we will show in the next subsection, it appears to be impossible to satisfy current observational bounds with this model. 

Let us emphasize here that these formulae are approximate, and differ from exact numerical evaluation of the full 
system of equations by factors of ${\cal O}(1)$ (for $\cal R$), which leads to ${\cal O}(10)$ disagreements in the value
of $\Delta_{\cal R}^2$. These disagreements come mostly from the fact that we have not been able to derive $\phitil(0)$
analytically, and are instead compelled to use a qualitatively suggestive but imperfect fitting function for its
value, Eqn. \pref{vphi}. This inaccuracy affects our ability analytically to evaluate the overall amplitude of fluctuations, but
the formulas mostly do give a good approximation to the relative dependence of the amplitude on the model's parameters. 
In practice, this means that the formula we derived for the spectral tilt, Eqn. \pref{nseq}, is much more accurate than the formula 
for the (unenhanced) tensor-to-scalar amplitude, Eqn. \pref{runenhance}. To compare with data we will need to use numerics, as we
will describe below.

\subsection{Parameter dependence and observational constraints}

\begin{figure}[htbp] %  figure placement: here, top, bottom, or page
   \centering
   \includegraphics[width=0.85 \textwidth]{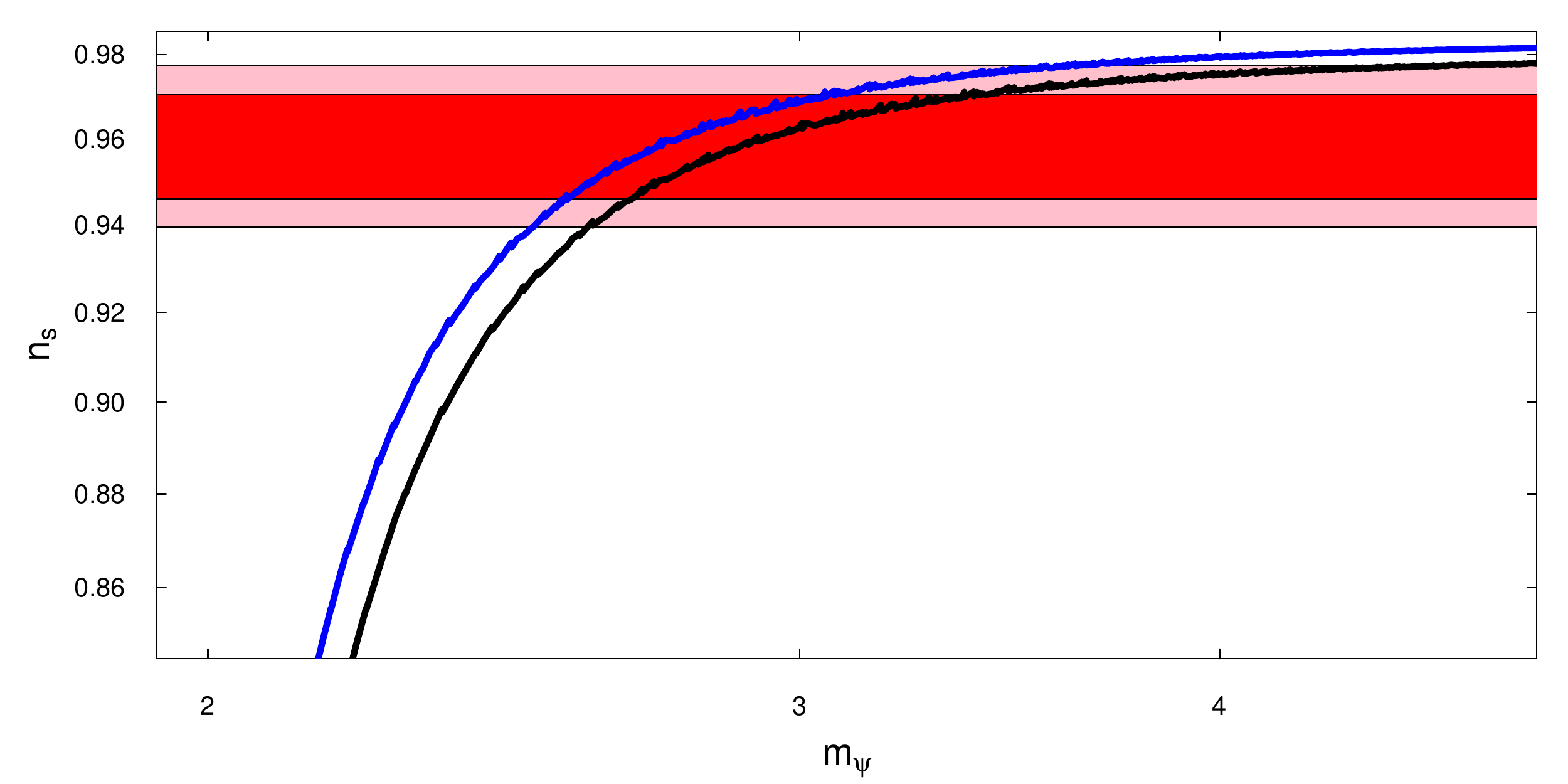} 
   \caption{\it The black (blue) lines show the predicted spectral index, $n_s$ as a function of the gauge field ``mass" parameter, $m_\psi$ evaluated   
    for a mode crossing the horizon 50 (60) efoldings before the end of inflation. The Planck one (two) sigma error bands on $n_s$ are
    shown in red (pink). }
   \label{nsVmass}
\end{figure}

\begin{figure}[htbp] %  figure placement: here, top, bottom, or page
   \centering
   \includegraphics[width=0.85 \textwidth]{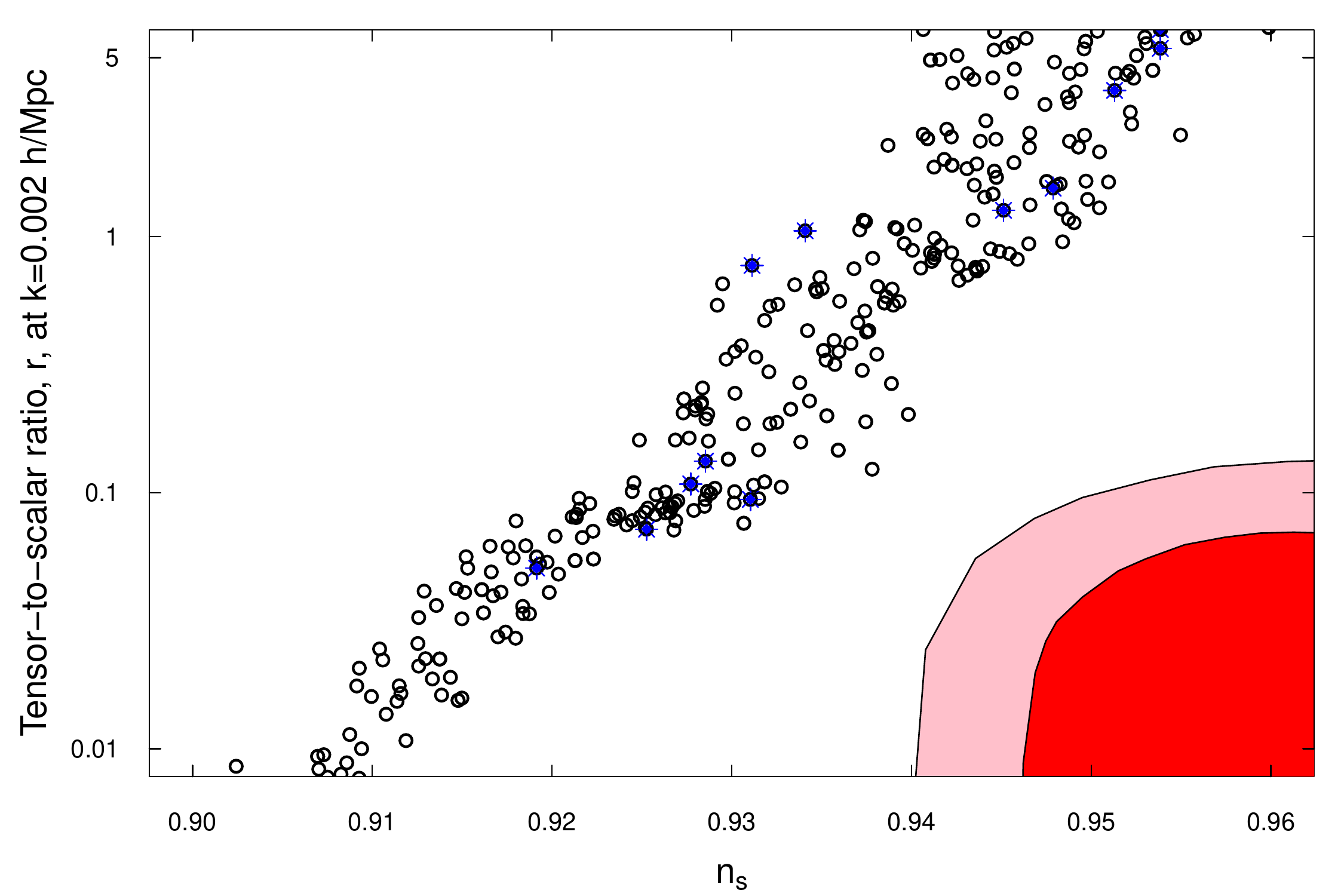} 
   \caption{\it Comparison of the tensor-to-scalar ratio, r, evaluated at $k=0.002$ h/Mpc and the spectral tilt, $n_s$ (evaluated at $k=0.05$ h/Mpc) for models
    drawn from a numerical exploration of the $g, f, \mu, \lambda$ parameter space with the data from Planck; for this plot,
    we have assumed that $k=0.05\,h$/Mpc maps to the point 60 efoldings before the end of inflation. The value for $r$
    presented here includes the contributions from both gravitational wave helicities and is computed numerically using the gravitational
    wave mode functions. The open
   black circles represent parameter combinations whose scalar power spectrum amplitudes are outside of the Planck error bars; blue stars represent models
   with acceptable power spectrum amplitudes. A finer-grained exploration of parameter space would more thoroughly fill in the 
   band with points of acceptable amplitude. The Planck one and two sigma contours are plotted in red and pink, respectively.
   Note that the y-axis is logarithmic, and that in this model it is possible to have $r>1$ due to the chirally enhanced gravitational wave spectrum.}
   \label{nsvr}
\end{figure}

The model as written has four free parameters: $g, \mu, f,$ and $\lambda$. Our previous study of the background solutions \cite{Adshead:2012qe} (see also \cite{Maleknejad:2012dt}) revealed that the background evolution depends only on $\lambda$ and the combination $g/\mu^2$ (and not on $f$ at all), suggesting that the ``magnetic drift" limit has less freedom than is implied by the four free parameters. This is even more true at the level of perturbations, where we will see that the single dimensionless
quantity $\mpsi$ plays a decisive role.
 
We first studied the parameter dependence of the spectral tilt, $n_s$, because preliminary numerical work had suggested that the equation for it in Sec.\ \ref{sec:curvature}, Eqn. (\ref{nseq}), was reasonably accurate if we used the approximate formula given in Eqn. \pref{vphi} to evaluate $\phitil(0)$. This formula in principle gives us the value for $n_s - 1$ at any time during inflation. To compare with observational constraints, these expressions need to be evaluated at the appropriate point in the inflationary history so that the modes that are leaving the horizon at that point are correctly mapped onto the physical modes we observe in the CMB. This point is conventionally taken to be either 50 or 60 e-foldings before the end of inflation for a ``pivot" momentum of $k=0.05 h$/Mpc.
% This required us to solve numerically for the background value of the axion, $\axion$ that corresponded to the point 50 (or 60) e-foldings before inflation ended.
 As was previously demonstrated in \cite{Adshead:2012qe}, to a very good approximation, the background evolution of the axion can be found by using the slow-roll approximation for the evolution of the axion, Eqn. \pref{slowrollX}, and solving %That is, the number of e-foldings of inflation between any two points along the axion potential can be computed by calculating
\be
\Delta N = \int_{\axion_i}^{\axion_f} d\axion \frac{H}{\dot \axion}.
\ee 
The remaining background parameters can then be found accurately by making use of the slow roll solutions. In summary,  the following procedure was used:
\begin{enumerate}
\item For a grid of values for $g/\mu^2$ and $\lambda$, we calculated the value of $\axion$ that corresponded to the point along the potential 50 (or 60) e-folds before the end of inflation, where we took the end of inflation to be when the axion reached $\axion_f = f \pi$. 
\item Using these values, we computed $\mpsi$ and $n_s$ at the point 50 (or 60) e-folds before the end of inflation for this broad range of parameters.
\end{enumerate}
The results of this procedure are plotted in Fig. \ref{nsVmass}. In this plot, we show the effect of varying the dimensionless gauge field mass parameter, $\mpsi$, on the value of the scalar spectral index. As described above, each point in this plot corresponds to 50 or 60 efoldings before the end of inflation for the given set of parameters. As the mass is decreased below $\mpsi\sim 3$, the tilt is dominated by the  $d \log \phitil(0)/dN$ contribution.\footnote{Note that the results of the analysis of  \cite{Dimastrogiovanni:2012ew} also found spectra that tended to be very red for low values of the parameter $\mpsi$, leading those authors to speculate that this may pose an issue for this model. }  Let us emphasize that these lines only vary with the parameters of the model insofar as the parameters alter the value of $m_\psi$ at
horizon crossing. We also note that this figure has been made using the approximate formulae presented in Sec. \ref{sec:curvature}, which overestimate the value of $n_s-1$ for
large $m_\psi$ because that the fitting formulae at Eqn.\  \pref{vphi} breaks down at large $\mpsi$. Although they are quite narrow, these lines in fact represent the full range of variation over a very broad range in 
$\{g/\mu^2, \lambda\}$ parameter space. In particular, although it would appear from Eqs. \pref{slowrollparams} that the values of $\epsilon_H$ and $\eta_H$ can be lowered arbitrarily by taking a large value for $\lambda$, we have found that enforcing the 
evaluation of the functions at the point 50 (or 60) e-folds before the end of inflation in fact removes this possibility. This should not surprise us, though. A large value for
$\lambda$ will give a very long period of inflation during which the slow roll parameters are very small. However, when inflation is approaching its termination,
the slow roll parameters must necessarily begin to grow. This is because fact that our model's slow-roll solutions mimics the behavior of large field inflation models in that
$\epsilon_H \sim \mathcal{O}(N^{-1})$, where $N$ is the number of e-foldings before the end of inflation at which it is evaluated, while $\ddot{\axion}/(H\dot\axion) = 2(\eta_H - \epsilon_H) \sim \mathcal{O}(N^{-2})$.

Studying the overlap of the curve in Fig. \ref{nsVmass} with the constraints from the Planck results, we see immediately that we are driven to adopt parameters for the model that will yield $m_\psi \gtrsim 2.5$ at horizon crossing. This is the first sign of trouble for the model. As was demonstrated in Sec. \ref{sec:tensors},  for $m_\psi \gtrsim 2$ one of the gravitational wave helicity modes is exponentially enhanced.  To discover whether this enhancement leads to a disagreement with data, we needed to compute the tensor-to-scalar ratio. 
Although we have very good approximate results for the amplitude of the gravitational waves from Sec. \ref{sec:tensors}, our analytic estimates for the amplitude of scalar curvature fluctuations are not accurate enough for doing comparisons with data.

To proceed further, we used numerical solutions of the full set of equations of motion for the scalar perturbations to compute the tensor-to-scalar ratio as a function of the model parameters. 
To this end,  code that solves both the background and perturbation equations simultaneously was used to perform an extensive numerical exploration of the parameter space. Due to the magnetic, or velocity couplings between the fields in this model, generic initial conditions lead to a mixture of ``slow" and ``fast" modes in the
numerical solutions, which make precise calculations difficult to perform. In order to produce stable numerical results, our numerical solutions were initialized at early times ($x_i \sim 10^4$) such that only the ``slow" mode was excited by  making use of the WKB analysis presented in Sec. \ref{sec:WKB}. 
In particular, using the language of the WKB analysis, we set the axion amplitude, $A=1$ at early times, and then used Eqns.\ (\ref{asympamps}) to initialize the gauge scalars, and Eqn.\ (\ref{wasymp}) to set the initial velocities for the perturbations. This procedure produced very clean results, with slowly varying field amplitudes and final results that were independent of both the precise time at which initial conditions were set and the final time at which the observable quantities were calculated. 

The scalar and tensor power spectrum amplitudes were numerically evaluated at $x=-k\tau=0.02$, though the results were independent of this precise choice. The amplitude and tilt of the scalar perturbations were calculated at $k=0.05\,h$/Mpc,  corresponding to the wavelength that crossed the horizon at the point 60 e-foldings before the end of inflation. We chose 60 rather than 50 efoldings from the end of inflation to set this condition for our parameter exploration because the results from our semi-analytic computation of $n_s$ suggested
that evaluating the observable quantities earlier would give values for $n_s$ that were closer to the results from the Planck experiment. The tilt was evaluated numerically by calculating the logarithmic change in the amplitude (calculated at fixed $x = -k\tau = 0.02$) over a range of $k$ values near $k=0.05\,h$/Mpc. Following the Planck satellite  analyses \cite{Ade:2013uln},  the tensor-to-scalar index, $r$, was calculated at $k=0.002\,h$/Mpc. For that computation, we similarly evolved the gauge tensor and gravitational wave modes to $x=-k\tau=0.02$ and directly evaluated their power spectrum, comparing the amplitude of its dimensionless power spectrum to that of the scalar perturbations.

We summarize the result of our parameter exploration in Fig.\ \ref{nsvr}. This figure demonstrates that seems impossible to simultaneously satisfy the observational constraints on $r$, $n_s$, and the power spectrum amplitude using this model. 
When $m_\psi$ is large enough to produce an acceptable spectral tilt, the chirally enhanced gravitational wave spectrum is too large, pushing 
up the value of $r$ beyond the observably permitted region.  

%It will be important to revisit the results of  \cite{Maleknejad:2011jw,Maleknejad:2011sq} for the closely related model of Gauge-flation in light of our negative conclusions.

%%%%%%%%%%%%%%%%%%%%%%%%%%%%%%%%%%%%
%%%%%%%%%%%%%%%%%%%%%%%%%%%%%%%%%%%%
\section{The massive gauge field limit and effective field theory}\label{sec:mgggh}
%%%%%%%%%%%%%%%%%%%%%%%%%%%%%%%%%%%%
%%%%%%%%%%%%%%%%%%%%%%%%%%%%%%%%%%%%

In the superhorizon limit, the only growing mode is the axion, and the gauge fields are massive and decaying.  One may then ask whether there is an effective field theory that integrates out these massive modes, and captures the low-energy axion dynamics.  This strategy is the opposite of the one taken in Sec. ~\ref{sec:reduced}, where the axion was integrated out and the gauge fields retained.  However, the `magnetic drift' mode of interest is one which involves a competition between potential forces and the analogue magnetic field arising from the Chern-Simons term; the gauge scalars and the axion are in effect locked together, and either can be eliminated to generate an effective action for the other.

In order to compare to the analysis of \cite{Dimastrogiovanni:2012st}, we work (in this section only) in the same gauge as they do -- the symmetric gauge
\be
\chi^j = 0
\ee
that turns $\Psi_i^a$ into a symmetric matrix.  Solving the Gauss's law constraint, one finds an action for the scalar modes similar to Eqn.~\pref{orthoscalaract} where again the coefficient functions are rational functions of $x$ and $\mpsi$.  In particular, the mass terms quadratic in the gauge scalar fluctuations are
\be
2\biggl(\Bigl(\frac{3\mpsi^2 \!-\! 6 \!+\! x^2}{x^2}\Bigr)\dphib\dphi + \Bigl(\frac{6\mpsi^2 \!+\! 6 \!+\!x^2}{x^2}\Bigr) \bar z z - \dphib z - \bar z\dphi\biggr).
\ee
This is to be compared with the axion, for which the diagonal mass term is
\be
1-\frac{1}{x^2}\Bigl(2-\frac{V_{,\axsub\axsub}}{H^2}\Bigr) + \frac{\bflam^2\mpsi^2}{2\mpsi^2+x^2}.
\ee
The gauge scalars are thus much more massive than the axion in the limit $\mpsi/x \gg \bflam \gg 1$, and integrating out the gauge scalar modes is justified in this limit.  Expanding the full scalar mode action in powers of $x/\mpsi$, one finds
\bea
\mathcal{S}_{\rm S} \approx &&\half\int \biggl[ 
\bar\ada'\ada' + 3 \dphib'\dphi' + 6 \bar z' z'
+ \frac{3\bflam\mpsi}{x} ( \dphib \ada' - \bar \ada \dphi' ) 
+ \frac{3\mpsi}{x^2} ( 2\bar \ada\dphi + \dphib \ada )
\nn\\
& & - \frac{6\mpsi^2}{x^2} \dphib\dphi - \frac{12\mpsi^2}{x^2} \bar z z + 2(\dphib z + \bar z \dphi)
- \biggl(\frac{1}{x^2}\Bigl(2-\frac{V_{,\axsub\axsub}}{H^2}\Bigr) + \frac{\bflam^2}{2}\biggr) \bar \ada \ada\biggr].
\eea
The equation of motion for $z$ is a massive wave equation
\be
z'' + \Bigl(\frac{\mpsi^2}{x^2}\Bigr) z -\frac13 \dphi = 0
\ee
whose source is another massive field; thus we are justified in dropping $z$ from the dynamics.  In the equation of motion for $\dphi$,
\be
\dphi'' + \frac{2\mpsi^2}{x^2}\dphi +\frac{\bflam \mpsi}{x^2} (\ada + x \ada') = 0 ~,
\ee
the kinetic term is much smaller than the remaining terms; the solution is well approximated by
\be
\dphi \approx \frac{\bflam (\ada +x \ada')}{2\mpsi} ~.
\ee
We then arrive at an effective equation of motion for the canonically normalized axion fluctuation $\ada = a\daxion$,
\be
\label{axioneffeom}
\ada'' + \Bigl( \frac13 -\frac{2}{x^2}\Bigr) \ada = 0 ~.
\ee
This equation exhibits the reduced sound speed, $c_s^2 = 1/3$, observed in~\cite{Dimastrogiovanni:2012st}; otherwise, it is the standard wave equation a massless scalar in de~Sitter space. Hence, as described in \cite{Dimastrogiovanni:2012st}, in the large
$\mpsi$ limit, the model of Chromo-Natural Inflation provides a concrete realization of the so-called gelaton mechanism, which was introduced in \cite{Tolley:2009fg}.

We now come to the question of the regime of validity of Eqn.~\pref{axioneffeom}.  The gauge fields are heavier than the axion for $\mpsi/x \gg \bflam$, and so the above procedure is certainly justified in this limit.  In fact, the solutions to the gauge scalar equations of motion are well approximated by the above so long as $\mpsi/x \gg 1$, regardless of whether it is larger than $\bflam$.  In effect, there is only one light (magnetic drift) mode, and it has a projection onto both the gauge scalar and axion motions; one can keep track of its dynamics by retaining any of the fields involved.

The condition $x \ll \mpsi$ is always true at sufficiently late times -- the expansion will always stretch modes beyond the scale set by $\mpsi$.   However, for the effective action to be useful, it must cover a long enough period before horizon crossing that we trust it to describe the transition of modes from oscillation to freeze-out; and we must have the expectation that modes are delivered to the effective action in something approximating their ground state.  If we allow ourselves to ignore the tensor modes, these conditions are met when $\mpsi \gg 1$.  The perturbative vacuum of the full theory evolves to the perturbative vacuum of the effective theory; the state handed to the effective theory is the vacuum state prior to horizon crossing, and then we can use the effective theory to describe its subsequent evolution.  All this remains approximately true for scales of interest in a realistic inflationary trajectory.  Inflation lasts a finite time, but the inflating state approaches the de~Sitter vacuum below some scale that is well outside our current horizon.

The tensor mode instability changes this picture.  For large $\mpsi$, tensor modes grow exponentially over a time window $(2-\sqrt{2})\mpsi \lesssim x \lesssim (2+\sqrt{2})\mpsi$ (see Eqn.\ (\ref{eqn:tensinstab})).   At some point these gauge modes grow so large that they go non-linear and become strongly coupled. The non-linear interactions of the fields most likely cuts off the exponential growth, leading to a bath of fluctuations with energy density of the same order of magnitude as the background gauge fields. It thus seems that these modes will not end inflation. However, in this regime linear perturbation theory has broken down, and thus the fluctuations that leave the horizon during this epoch are likely to be highly non-gaussian. One may, however, imagine a scenario where the masses of the gauge modes is sufficiently large that the modes of interest that are observed in large-scale structure and the CMB leave the horizon when the theory is well described by the low-energy effective field theory \cite{Tolley2013}.   A consistent value of $r$ would seem to require that the modes seen in the CMB are all described by this theory rather than the full dynamics, from the onset of inflation.  This would imply an $\mpsi$ of order $e^{10}$ or larger.

%%%%%%%%%%%%%%%%%%%%%%%%%%%%%%%%%%%%
%%%%%%%%%%%%%%%%%%%%%%%%%%%%%%%%%%%%
%%
\section{Relation to Gauge-flation}\label{sec:gaugeflation}
%%
%%%%%%%%%%%%%%%%%%%%%%%%%%%%%%%%%%%%
%%%%%%%%%%%%%%%%%%%%%%%%%%%%%%%%%%%%

Chromo-Natural Inflation is very closely related to a class of models of inflation where the axion has been integrated out, known as Gauge-flation \cite{Maleknejad:2011jw,Maleknejad:2011sq} (see \cite{Adshead:2012qe, SheikhJabbari:2012qf}).  Gauge-flation is described by the action
\begin{align}\label{eqn:GFact}
\mathcal{S} =  \int d^{4}x\sqrt{-g}\Bigg[  \frac{1}{2}R -\frac{1}{2}\tr\[F_{\mu\nu}F^{\mu\nu}\]+\frac{\kappa}{96}\tr\[ F\wedge F \]^2\Bigg] ~,
\end{align}
which may be put in a form suitable for applying our analysis by the introduction of a pseudo-scalar auxiliary field $\axion$
\begin{align}\nn\label{eqn:CNIform}
\mathcal{S} =  \int d^{4}x\sqrt{-g}\Bigg[  \frac{1}{2}R -\frac{1}{2}\mu^4\( \frac{\axion}{f}\)^2-\frac{1}{2}\tr\[F_{\mu\nu}F^{\mu\nu}\]-\frac{\lambda}{4 f}\axion\tr\[ F\wedge F \]\Bigg] ~.
\end{align}
Integrating out this auxiliary field yields the Gauge-flation action with the identification of parameters
\begin{align}
\kappa = 3\frac{\lambda^2}{\mu^4} ~.
\end{align}

The analysis of fluctuations arising from~\pref{eqn:GFact} parallels closely that leading to the reduced action of section~\ref{sec:reduced}.  The approximation leading to the reduced action amounts to dropping the kinetic energy of the axion, as well as the mass term for fluctuations $\frac{1}{2}\mu^4\({\daxion}/{f}\)^2$, which is subdominant to the quadratic term in $\daxion$ arising from the Gauss law constraint until well after horizon crossing.  The background equations of motion in the slow-roll approximation are the same, namely eqs.~\pref{slowrollX}-\pref{psislowrollsoln}.  The transition between region I and II of the axion evolution (the onset of the intermediate time region) takes place when the dynamics switches between being KE dominated and magnetic drift dominated.  Here, since there is no kinetics for the axion, region II of the scalar dynamics extends to $x\to \infty$ and so the reduced model is valid from the beginning of time until well after horizon crossing.  We can still use the formulae for the scalar curvature perturbations in terms of the axion amplitude; now the axion just codes some particular combination of the gauge modes via the auxiliary field equation of motion~\pref{Xdrift}.  We thus expect the fluctuation spectrum of this model to exhibit exactly the same features found above for Chromo-Natural Inflation.

In light of our analysis, the results presented in e.g. \cite{Maleknejad:2011jw,Maleknejad:2011sq} regarding the observable consequences of Gauge-flation are puzzling. Our analysis of Chromo-Natural Inflation does not appear to approach the findings presented in \cite{Maleknejad:2011jw,Maleknejad:2011sq}, even in the limit where the axion is very massive and could be integrated out.  Indeed, the conclusions of those works are the opposite of what we find: They seem to find a spectral index which is too close to scale invariant to match the data. The region of of parameter space in which these models overlap is found by taking very large values for our parameter $\lambda$ (e.g. $\lambda \gtrsim {\rm few}\times 10^{4}$, together with $g \sim 10^{-3}$), which leaves the axion very near the bottom of its potential for the entire duration of inflation. In our analysis, we have explored this very region of parameter space, yet doing so does not ameliorate the tension of our findings with the data: the tilt of the scalar spectrum remains very red when the tensor-to-scalar ratio is small enough.  These findings are not expected to change much with the modification~\pref{eqn:CNIform}; lower values of $\mpsi$ are needed to avoid a large tensor-to-scalar ratio, in which case the redness of the spectrum is driven by the decrease in $\phitil(0)$ with increasing $\mpsi$ (recall that $\mpsi$ is inversely dependent on $H$) indicated in figure~\ref {fig:phizero}.

%%%%%%%%%%%%%%%%%%%%%%%%%%%%%%%%%%%%
%%%%%%%%%%%%%%%%%%%%%%%%%%%%%%%%%%%%
%%
\section{Discussion}\label{sec:conclusions}
%%
%%%%%%%%%%%%%%%%%%%%%%%%%%%%%%%%%%%%
%%%%%%%%%%%%%%%%%%%%%%%%%%%%%%%%%%%%

In this work we have performed a detailed analysis of all of the linear fluctuations in Chromo-Natural Inflation. We find that the theory, in its present form, appears to be in significant tension with current constraints on the scalar spectral index and tensor-to-scalar ratio. Let us, however, emphasize that our analysis has been confined to linear perturbation theory. The estimates of the tensor amplitude of Sec.\ \ref{sec:tensors} clearly cannot increase without bound as $\mpsi$ increases as depicted in Fig.\ \ref{fig:powerratio}. As the spin-2 modes of the gauge field become more unstable and grow ever larger, linear perturbation theory will break down as the non-linear interactions of the gauge fields become increasingly important. These interactions will tend to cut off the exponential growth of these modes. However, in this region, the scalar-vector-tensor decomposition will have broken down, and there will be a contribution to the scalar curvature perturbation  due to the non-linear tensors. Such an analysis is beyond the scope of this work, and we leave it for future study.

Within the context of linear perturbation theory, our exploration of the parameter space of this model has revealed that it is only possible to generate a spectrum of scalar perturbations with the correct amplitude and tilt in a region of parameter space where the parameter that characterizes the mass of the gauge field fluctuations, $\mpsi$, is large. However, we have also shown that one of the polarizations of the spin-2 modes of the gauge field becomes unstable and undergoes exponential growth for a length of time proportional to the same mass parameter. These fluctuations become increasingly large as the mass of the gauge field fluctuations is increased which in turn source large chiral gravitational waves with an amplitude that is ruled out by current data.  Whether a region exists where the non-linear theory is compatible with data remains to be seen; such an analysis would require extending the methods presented here and in, e.g., \cite{Anber:2009ua, Anber:2012du} to model
 the strong coupling regime of non-Abelian gauge theory.

We note that the expected phenomenology of the model is significantly different to the na\"ive estimates presented in \cite{Adshead:2012kp}. While the form of the curvature perturbation was correct in that work, the axion fluctuation was assumed to freeze out with an amplitude of order the Hubble scale. However, as we demonstrated in Sec.\ \ref{sec:scalars}, the late-time amplitude of the axion, is suppressed relative to this naive estimate by, among other factors, a factor of the large parameter $\lambda$. The effect of this suppression is that, contrary to the conclusions of \cite{Adshead:2012kp}, one is actually forced to inflate at a reasonably high scale in order to produce curvature fluctuations at a level sufficient to source the observed level of temperature fluctuations in the cosmic microwave background. This means that while the axion only rolls a very short distance in field space, since the energy scale is high, it produces gravitational radiation at an observable level. This violates simple formulations of the Lyth bound \cite{Lyth:1996im}.

While the original model of Chromo-Natural Inflation appears to be ruled out by the analysis presented here, given the rich phenomenology afforded by the background SU(2) gauge fields, it will be important to determine if variants of the scenario can be made workable by considering e.g.\ higher dimensional embeddings of the gauge group, or considering more than one axion field coupled to the classical gauge field configuration.  If such a model could be made compatible with observations, there are many important future directions to pursue. Important among these are the question of 
non-adiabatic, or so-called iso-curvature, perturbations. Although the negative conclusion we reached concerning the adiabatic modes means that we have not pursued calculation of iso-curvature observables in detail, we believe that they would not be very large in this model. This can be argued in two ways. Most simply, a late-time power law analysis of our system reveals that only one scalar mode survives on super-horizon scales; the rest decay. Hence, any iso-curvature mode would be decaying outside the horizon. Relatedly, the partial success of the large gauge mass EFT first described by \cite{Dimastrogiovanni:2012st} and outlined in Sec.\ \ref{sec:mgggh} again points to the fact that,  in the magnetic drift limit, there is only one important scalar degree of freedom on superhorizon scales at late times in our system.

Perhaps the most important novel feature revealed by our analysis is a new mechanism for producing chiral gravitational waves. Their presence can provide important new observational handles for constraining or observing any scenario in which they appear. In fairness, we should mention that chiral gravitational waves are not unique to this scenario, and have been noticed in related contexts \cite{Contaldi:2008yz, Alexander:2004wk}. In a particularly closely related example, inflation driven by an axion that is coupled to U(1) gauge fields is known to produce chiral gravitational radiation \cite{Sorbo:2011rz, Anber:2012du, Barnaby:2012xt}. The violation of parity by these fluctuations allows for a non-zero correlation between the gravitational-wave-sourced temperature fluctuations and the B-mode spectrum \cite{Gluscevic:2010vv}. Furthermore, given that the chiral gravitational wave spectrum is produced with a strongly blue spectrum, interesting constraints on this scenario may also be possible from direct detection, or lack thereof, of gravitational waves   \cite{Barnaby:2012xt}. Furthermore one may even hope to measure the polarization of a detection \cite{Crowder:2012ik} to determine whether such radiation was parity violating or not.

Another possible new feature of these models is the fact that they end inflation in a state of radiation domination, filled with decaying non-Abelian gauge fields. Hence, it will be interesting ask whether new possibilities for reheating could be found that go beyond the standard story of parametric resonance. The fact that the background violates CP also provides intriguing new possibilities for leptogenesis or baryogenesis following inflation. In particular, it seems that the scenario of gravitational leptogenesis presented by \cite{Alexander:2004us, Alexander:2007qe} that makes use of the gravitational chiral anomaly in the standard model of particle physics to generates lepton number violation could readily be adapted to this scenario. Some work along this direction has been present by \cite{Noorbala:2012fh}.

{\bf Acknowledgements:}
We thank Andrew Tolley and Wayne Hu for useful discussions. We also acknowledge David Seery for early collaboration on this project. This work was supported in part by DOE grant DE-FG02-90ER-40560 and by the Kavli Institute for Cosmological Physics at the University of Chicago through grant NSF PHY-1125897 and an endowment from the Kavli Foundation and its founder Fred Kavli. P.A. thanks the Kavli Institute for Theoretical Physics for hospitality and support through National Science Foundation Grant No. NSF PHY11-25915 as this work was nearing completion.

%%%%%%%%%%%%%%%%%%%%%%%%%%%%%%%%%%%%
%%%%%%%%%%%%%%%%%%%%%%%%%%%%%%%%%%%%
%%%%%%%%%%%%%%%%%%%%%%%%%%%%%%%%%%%%
\vskip 2cm

%\newpage
\bibliographystyle{JHEP}
\bibliography{PertsinCNI}

%%%%%%%%%%%%%%%%%%%%%%%%%%%%%%%%%%%%
%%%%%%%%%%%%%%%%%%%%%%%%%%%%%%%%%%%%
%%%%%%%%%%%%%%%%%%%%%%%%%%%%%%%%%%%%

\appendix

%%%%%%%%%%%%%%%%%%%%%%%%%%%%%%%%%%%%
%%%%%%%%%%%%%%%%%%%%%%%%%%%%%%%%%%%%
%
\section{Gravitational sector}\label{app:gravsector}
%
%%%%%%%%%%%%%%%%%%%%%%%%%%%%%%%%%%%%
%%%%%%%%%%%%%%%%%%%%%%%%%%%%%%%%%%%%

In the main body of this paper, we have ignored the contribution of gravitational perturbations to the dynamics.
In this Appendix, we will justify this approximation by explicitly showing that the contribution of the gravitational degrees of freedom to the action in our choice of spatial coordinates is suppressed relative to the quadratic action we have considered above.

As in the main text, we will work in the ADM formulation of general relativity \cite{Arnowitt:1962hi}, for which the line element is written in the form
\begin{align}\label{eqn:ADMmetric}
ds^2 = -N^2 d\tau^2 +  \tilde{h}_{ij}(dx^i+N^i d\tau)(dx^j+N^j d\tau),
\end{align}
where $N$ and $N^j$ are the lapse and the shift, and $\tilde{h}_{ij}$ is the metric on the spatial hypersurface. 
Inserting the ADM metric into the action in Eq.\ (\ref{eqn:CNIaction}), we find,\footnote{Our summation convention in this section is the same as the above, repeated lower indices are summed with the Kronecker delta, while upper indices paired with lower indices are summed with the metric $\tilde h_{ij}$ and its reciprocal $\tilde h^{ij}$.}
\begin{align}\nn
S = & \int d^4 x  \sqrt{\tilde{h}}\[N R{}^{(3)} + \frac{1}{N}(E^{ij}E_{ij} - E^2)\]\\ \nn& + \int d^4x \sqrt{\tilde{h}} N\[\frac{1}{2N^2}\(\partial_{\tau}\axion - N^j \partial_j \axion\)^2  - \frac{\tilde{h}^{ij}}{a^2} \partial_{i}\axion\partial_j\axion- V(\axion)\]\\ \nn & +
\int d^{4}x \frac{\sqrt{\tilde h}}{a^2N}\tr\[(F_{0i}+N^{k}F_{ik})\tilde{h}^{ij}(F_{0j}+N^{l}F_{jl})\]-\frac{1}{2}\int d^{4}x \frac{\sqrt{\tilde h}}{a^4}N\tr\[ \tilde{h}^{ik}\tilde{h}^{jl}F_{ij}F_{kl}\]\\ & -\lambda\int d^{4}x\axion\;\epsilon_{ijk}\tr\[F_{0i}F_{jk}\].
\end{align}
In this expression, $E_{ij}$ is related to the extrinsic curvature of the spatial slices
\begin{align}
E_{ij}   = & \frac{1}{2}\left(\partial_{\tau}\tilde{h}_{ij} -
\nabla_{i}N_{j}-\nabla_{j}N_{i}\right),\quad 
E  =  E^{i}_{\;i},
\end{align}
and $\nabla_i$ is the covariant derivative constructed from $\tilde{h}$.  Note that spatial indices are raised and lowered using $\tilde h_{ij}$ and its reciprocal $\tilde h^{ij}$.  As indicated in the main text, we work in spatially flat gauge where our coordinates are chosen so that $\tilde{h}_{ij} = a^2 \[e^{\gamma}\]_{ij}$. In what follows, we will drop the transverse traceless fluctuations of the metric. Since these are gauge invariant and decoupled from the other degrees of freedom in the quadratic action, it is self-consistent
to ignore them for the present analysis.

We proceed by expanding the lapse and shift about a Friedmann-Robertson-Walker spacetime
\begin{align}
N = & a(1+\dN_{(1)} +\dN_{(2)}),\quad 
N^{i} = N^{i}_{(1)}+N^{i}_{(2)}
\end{align}
where $N_{(1)}$ and $N_{(2)}$ are, respectively, first and second order in fluctuations.

In spatially flat gauge, neglecting gravitational waves, the curvature of the spatial slices ${}^{3}R = 0$ and the connection
for $\nabla_{i}$ (the covariant derivative compatible with the metric $\tilde{h}_{ij}$ on the hypersurface) vanishes and thus $\nabla_{i}\rightarrow\partial_{i}$.  The Einstein-Hilbert action to quadratic order in scalar and vector fluctuations is given by
\begin{align}\nn\label{eqn:EHaction}
\delta^2 S_{EH} = & \frac{1}{2}\int d^4 x a^2 \Bigg[-(4a^{-2} \h \partial_{i}N_{i}\dN+6\h^{2}\dN^2)  +a^{-4}(\partial_{(i}N_{j)}\partial_{(i}N_{j)} - \partial_{i}N_{i}\partial_{j}N_{j})
\Bigg]
\end{align}
where we have defined for convenience in this section,
\begin{align}
\h = \frac{d\ln a}{d\tau} = a H.
\end{align}
The gravitationally coupled action for the axion is straightforward to derive, and reads,
\begin{align}\label{eqn:axionact}
\delta^2  S_{\axion} = \int d^4x \Bigg[\delta^2\!\mathcal{L}_{\axion}-\partial_{\tau}\axion\partial_{\tau}\daxion\dN+\frac{1}{2}(\partial_{\tau}\axion)^2\dN^2-a^{-2}\partial_{\tau}\axion N_j \partial_j \daxion  -a^2 V'\daxion \dN\Bigg],
\end{align}
where $\delta^2 \mathcal{L}_{\axion}$ is the quadratic action from Eqn.\ (\ref{eqn:actaxion}).  Finally, the gauge field action to quadratic order in field and metric fluctuations is given by
\begin{align}\label{eqn:GFaction}
\delta^2  S_{A} = & \int d^4 x \[\delta^{2}\!\mathcal{L}_{\rm YM} +{\delta^2}\! \mathcal{L}_{\rm CS}\] - \frac{1}{2}\int d^{4}x a^4 \dN\(g\frac{\phi^2}{a^4}(2\epsilon_{a ij}\partial_{i}\da_{aj}+4 g\phi\da_{ii} )\)\\ \nn&+ \frac{1}{2}\int d^{4}x a^2 \Bigg[3 \dot{\phi}^2\dN^2 -2\frac{\dot{\phi}}{a} (\partial_{\tau}\da_{ii} - \partial_i \da_{i 0} )\dN +2g^{2}\frac{\phi^4}{a^6} N_{i}N_{i}\\ \nn & 
+\frac{\dot{\phi}}{a^3}2 N_k((\partial_{i}\da_{ik}-\partial_{k}\da_{ii})+ g \phi \epsilon_{aki}   \da_{ai})+   g\frac{\phi^2}{a^4}2 N_k(  \epsilon_{aik}\partial_\tau \da_{ai} -  \epsilon_{aik}\partial_{i}\da_{a0} -2 g \phi \da_{k0}) \Bigg],
\end{align}
where $\delta^{2}\!\mathcal{L}_{\rm YM}$ and  $\delta^2\! \mathcal{L}_{\rm CS}$ are the quadratic actions from Eqn.\ (\ref{eqn:actYM}) and Eqn.\ (\ref{eqn:actCS}) in the main text. Note that we have  made use of the background equations of motion to eliminate the second order terms $\dN_{(2)}$ and $N^{i}_{(2)}$, and hence have dropped the order-of-perturbation labels on the remaining gravitational perturbations: $\dN_{(1)} \equiv \dN$ and $N^i_{(1)} \equiv N^i$. 

Working with the $N = 2$ representation above,  the part of the action containing gravitational and gauge constraints can be written,
%\begin{widetext}
\begin{align}\label{eqn:constractionN2}\nn
\delta^2 S_{\rm constr.} = & \frac{1}{2}\int d^4 x a^2 \Bigg[\((\partial_{\tau}\axion)^2 +3 \dot{\phi}^2 -6\h^{2}\)\dN^2-(4a^{-2}\h \partial_{i}N_{i}+2\partial_{\tau}\axion\partial_{\tau}\daxion +2a^2V'\daxion)\dN \Bigg]\\\nn
& 
 - \frac{1}{2}\int d^{4}x a^2 \dN\[\frac{\dot{\phi}}{a} (6\partial_{\tau}\delta\phi- 2\partial_i \da_{a 0} \delta_{ai})+g\frac{\phi^2}{a^2}(-4\partial_{i}\chi_i+12 g\phi\delta\phi)\]\\ \nn&
+  \frac{1}{2} \int d^4 x \Big[\da_{i0}(-\partial^2 +2g^2\phi^2)\da_{i0} +2\da_{i0}\(4 g \partial_{\tau}\phi  \chi_i - g\phi^2\frac{\lambda}{ f}\partial_{i}\daxion  \) \\ \nn& \hskip 2.5 cm +   2g\frac{\phi^2}{a^2}  N_k(  2\partial_\tau \chi_k -  \epsilon_{aik}\partial_{i}\da_{a0} -2 g \phi \da_{k0}) \Big]\\ \nn&
+ \frac{1}{2}\int d^{4}x a^2 \Bigg[a^{-4}(\partial_{(i}N_{j)}\partial_{(i}N_{j)} - \partial_{i}N_{i}\partial_{j}N_{j}) +2g^{2}\frac{\phi^4}{a^6}N_{i}N_{i}
\\ & \hskip 2.5 cm+\frac{\dot{\phi}}{a^3}2 N_k(\partial_i t_{ik}-2\partial_k \delta\phi -\epsilon_{ikj}\partial_i \chi_j - 2 g\phi \chi_k) -2a^{-2}\partial_{\tau}\axion N_j \partial_j \daxion ,
\Bigg].
\end{align}

The key advantage of the ADM formulation of general relativity is that lapse $N$ and shift $N^i$ are non-dynamical. Hence, we can eliminate them from the action
by solving their equations of motion, then substituting the result back into the action. However, it will be important to solve the gauge constraint simultaneously with the
gravitational constraints. This is the calculation we perform now.

Varying the action with respect to $\da_0$ yields the Gauss's Law constraint,
\begin{align}\label{eqn:gaussfull}
(-\partial^2 &+2g^2\phi^2)\da_{i0} +\frac{1}{2}a\dot\phi \partial_i \delta N+4 g \partial_{\tau}\phi  \chi_i+g\frac{\phi^2}{a^2}(\epsilon_{ijk}\partial_{j}N_k -2 g \phi N_i)  = g\phi^2\frac{\lambda}{ f}\partial_{i}\daxion .
\end{align}
while varying with respect to $\dN$ gives the linear order Hamiltonian (or energy) constraint,
\begin{align}\label{eqn:hamconstr}
\(-6\h^{2}+(\partial_\tau\axion)^2+3\dot\phi^2\)\dN & + a^2 V'(\axion)\daxion
 -\partial_\tau\axion\partial_{\tau}\daxion- 2\frac{\h}{a^2} \partial_i N_i
\\ \nn &  + g\frac{\phi^2}{a^2} (2\partial_{i}\chi_{i}-6 g\phi\delta\phi ) -\frac{\dot{\phi}}{a} \(3\partial_{\tau}\delta\phi  -\partial_i\da_{i0}\) = 0.
\end{align}
Finally, the linear order momentum constraint from varying $N^i$ is
\begin{align}\label{eqn:momconstr}
&\(-\partial^2+2g^{2}\frac{\phi^4}{a^2}\)\frac{N_{i}}{a^2}+a^{-2}\partial_i\partial_j N_j+2\h a^{-2} \partial_{i}\delta N-\partial_{\tau}\axion \partial_i\daxion
\\\nn & \hskip 1cm+ g\frac{\phi^2}{a^2}(  2\partial_\tau\chi_i-\epsilon_{aik} \partial_i \da_{a0} -2 g \phi \da_{i0})-\frac{\dot{\phi}}{a}\(\partial_i t_{ik}-2\partial_k \delta\phi -\epsilon_{ikj}\partial_i \chi_j - 2 g\phi \chi_k\)  = 0.
\end{align}
In order to proceed, we decompose the fields into scalars and vectors by introducing the following helicity decomposition for the shift vector and temporal part of the gauge fields,
\begin{align}
N^{\pm} = &\frac{1}{\sqrt{2}}\( N_1 \pm i N_2\),\quad \tilde N_i = N_3 \delta_{i3},\\
\da^{\pm}_0 = & \frac{1}{\sqrt{2}}\(\da_{10}\pm i \da_{20}\), \quad \tilde \da_{a0} = \da_{30} \delta_{a3},
\end{align}
where $\tilde N_i $ and $\tilde \da_{a0}$ are the scalar valued longitudinal modes. 

We can solve Eqns.\ (\ref{eqn:gaussfull}) - (\ref{eqn:momconstr}) exactly; however, the resulting expressions are long and not particularly illuminating. For our purposes,
we simply need to demonstrate that the solutions for the lapse and shift do not yield important corrections to the action, justifying our decision to neglect them in the analyses
performed in the main text.
  To this end, we recall from above that
\begin{align}
g\phi  =  \h \mpsi.
\end{align}
For the background solutions of interest, we have $\mpsi \sim 2 -3 $, while $\psi \sim 10^{-2}$. Thus, $\psi$ can be thought a small parameter for this expansion. Note also that
self-consistency of the analysis always requires $\psi$ to be small: for inflation to proceed, the potential energy of the axion must always be dominant, while the gauge field energy scales scales as $g^2 \psi^4$. Hence, we are bound to consider only parameters and solutions for which $g^2 \psi^4/H^2 = \mpsi^2 \psi^2 \ll 1$.  In the main text, the terms we kept were all $\mathcal{O}(\psi^{0})$ in this expansion. Below,  we will show that all contributions from the gravitational constraints are at least $\mathcal{O}(\psi)$ or smaller, and thus can be consistently ignored relative to the terms we have previously included.

%%%%%%%%%%%%%%%%%%%%%%
\subsection{Scalar constraints}
%%%%%%%%%%%%%%%%%%%%%%

Let us write out the scalar part of the constraint equations. Gauss's law gives us
\begin{align}\label{eqn:gaussscal}
(-\partial^2 +2g^2\phi^2)& \tilde\da_{i0} + \frac{1}{2} a\dot\phi  \partial_i\delta N+4 g \partial_{\tau}\phi   \chi_i -2 g^2\frac{\phi^3}{a^2}\tilde N_i  =   g\phi^2\frac{\lambda}{ f} \partial_i \daxion,
\end{align}
the linear order Hamiltonian constraint gives us
\begin{align}\label{eqn:hamconstrscal}
\(-6\h^{2}+(a\dot\axion)^2+3\dot\phi^2\)\dN & + a^2 V'(\axion)\daxion -\partial_\tau\axion\partial_{\tau}\daxion \\ 
& \nn-  2\frac{\h}{a^2} \partial_i \tilde N_i
 + g\frac{\phi^2}{a^2} (2 \partial_i \chi_i-6 g\phi\delta\phi )-\frac{\dot{\phi}}{a} \(3 \partial_{\tau}\delta\phi  - \partial_i \tilde\da_{i0}\) = 0,
\end{align}
and the scalar part of the linear-order momentum constraint is
\begin{align}\label{eqn:momconstrscal}
2g^{2}\frac{\phi^4}{a^4}\tilde N_{i}&+2 \h  \partial_i \delta N- \partial_{\tau}\axion \partial_i \daxion
 + g\frac{\phi^2}{a^2}(  2\partial_\tau\chi_i-2 g \phi \tilde \da_{i0})+  \frac{\dot{\phi}}{a}(2\partial_i z-2\partial_i \delta\phi  - 2 g\phi \chi_i) = 0.
\end{align}
Note that the Gauss's law constraint  Eqn.\ (\ref{eqn:gaussscal}) receives its largest contribution by far from the term on the right hand side (since $\lambda/f \gg 1$), so we have the approximate solution
\begin{align}\label{eqn:largelamgauss}
\tilde \da_{i0} \sim  \frac{g\phi^2}{(-\partial^2 +2g^2\phi^2)}\frac{\lambda}{ f}\partial_{i}\daxion
\end{align}
where `$\sim$' indicates we are solving only approximately for the leading behavior. The leading contribution to the solution of Gauss's law is thus unaffected by the inclusion of the
gravitational perturbations. Next, let us solve the scalar part of the momentum constraint, Eqn.\ (\ref{eqn:momconstrscal}) 
in the same approximation. We find
\begin{align}\label{eqn:largelamham}
\delta N  \sim  \frac{1}{a^2\h}\frac{g^3\phi^5}{(-\partial^2 +2g^2\phi^2)}\frac{\lambda}{ f}\daxion.
\end{align}
Note that after one expresses the various background quantities in this expression in terms of $\mpsi$ and $\bflam = \psi \lambda/f$, one finds the contribution of this terms is suppressed by an additional factor of $\psi$ compared to the dominant terms in the quadratic action. Finally, we can use these solutions to find the leading behavior of $N^i$,
\begin{align}\label{eqn:largelammom}
2\frac{\h}{a^2} \,\partial_i \tilde N_i \sim &-6\h^{2}\dN + a^2 V'(\axion)\daxion +\frac{\dot{\phi}}{a}  \partial_i\tilde\da_{i0}.
\end{align}
Note that $N^i$ does not appear to be suppressed when it is written this way, since we are allowing for arbitrary $V'$. However, in the slow-roll limit, we can insert Eqn.\ (\ref{eqn:largelamgauss}) and Eqn.\ (\ref{eqn:largelamham}) and make use of the background equations of motion for $\psi$, Eqn. \ref{slowrollpsi},  to replace $V'(\axion)$;
after some rewriting, this yields
\begin{align}
2\frac{\h}{a^2} \,\partial_i \tilde N_i \sim  & -2\frac{\dot{\phi}}{a}  \partial_i\tilde\da_{i0} -12\h^{2}\dN,
\end{align}
which is of order $\mathcal{\psi}$. Intuitively, the reason this works out is that it is $V'$ that controls the size of $\psi$ on the background, so that the internal consistency
of the calculation guarantees that the $\daxion$ contribution to the momentum constraint cannot be larger than the $\dN$ and $\tilde \da_{i0}$ contributions.

%%%%%%%%%%%%%%%%%%%%%%
\subsection{Vector constraints}
%%%%%%%%%%%%%%%%%%%%%%

It remains to show that the vector parts of the constraints are also suppressed. In helicity basis, the vector part of the Gauss's law constraint can be written (recall that $u^{\pm} = (\chi^1 \pm i\chi^2)/\sqrt{2}$ )
\begin{align}\label{eqn:gaussvec}
\(k^2 +2g^2\phi^2\)\da^{\pm}_{0} +4g \partial_{\tau}\phi u^{\pm}  +g\frac{\phi^2}{a^2}(\pm k -2 g\phi)  N^{\pm} = 0.
\end{align}
The vector part of the linear order momentum constraint is
\begin{align}\label{eqn:momconstrvec}
\(k^2+2g^2\phi^2\)\frac{N^{\pm}}{a^2}+  g \psi^2 (  2\partial_\tau u^{\pm}& \pm k  \da^{\pm}_{0}  -2 g\phi \da^{\pm}_{0}) - \frac{\dot{\phi}}{a} ( i k (v^{\pm}\mp i u^{\pm}) +2 g \phi u^{\pm})  = 0.
\end{align}
While these two equations are not particularly difficult to solve, expicit solutions are unneccessary for our purposes. By inspection of Eqn.\ (\ref{eqn:gaussvec}) one can easily see the contribution of the vector part of the shift to the Gauss law constraint is suppressed relative to that of the gauge field perturbation by a factor of the gauge field VEV $\psi$. Further, examination of Eqn.\ (\ref{eqn:momconstrvec}) reveals that after substituting in the solution of Eqn.\ (\ref{eqn:gaussvec}), $N^{\pm}$ itself suppressed by factors of $\psi$ relative to the vector excitations of the gauge field $u^{\pm}$.

These results then justify our treatment in the main body of the paper, where we have worked to leading order and dropped all contributions to the dynamics generated by these constraints.

%%%%%%%%%%%%%%%%%%%%%%%%%%%%%%%%%%%%
%%%%%%%%%%%%%%%%%%%%%%%%%%%%%%%%%%%%
%%
\section{Higher dimensional embeddings}\label{sec:embeddings}
%%
%%%%%%%%%%%%%%%%%%%%%%%%%%%%%%%%%%%%
%%%%%%%%%%%%%%%%%%%%%%%%%%%%%%%%%%%%

In this appendix we study embeddings of the SU(2) gauge field configuration into higher dimensional Lie groups, SU(N).  We begin by noting that many terms in the Yang-Mills Lagrangian simplify if we choose $\Omega_k$ to satisfy the eigenvalue equation
\begin{align}
\Omega _{k} = \omega \Psi^{(\omega)}_{k}.% \quad \tr\[\da^{(\omega)}_i \da^{(\omega')}_i \] = \delta_{\omega \omega'}
\end{align}
The eigenvectors $\da_k^{(\omega)}$ are then orthogonal according to
\begin{align}
\tr\[\da^{(\omega)}_i \da^{(\omega')}_i \] = \delta_{\omega \omega'}\tr\[\da^{(\omega)}_i \da^{(\omega)}_i \].
\end{align}
Note that this choice of basis functions diagonalizes many terms in the action. The problem we then need to solve is for the eigenvalue and eigenvectors of the equation
\begin{align}\label{eqn:evaleqn}
\[J_{i}, \da^{(\omega)}_{j}\] = - \frac{i}{2}\omega \,\epsilon_{ijk}\da^{(\omega)}_{k}.
\end{align}
The analysis presented here mirrors that found in \cite{Dasgupta:2002hx};  we reproduce it here for completeness.

%%%%%%%%%%%%%%%%%%%%%%%%%%%%%%%%%%%%
%%
\subsection{Eigenvalues}
%%
%%%%%%%%%%%%%%%%%%%%%%%%%%%%%%%%%%%%

Suppressing the $\omega$ label on the eigenmodes, we begin by writing the eigenvalue equation, Eqn.\ (\ref{eqn:evaleqn}), in the form
\begin{align}
i \epsilon_{ijk}\[J_{i}, \da_{j}\] = \omega\da_{k}.
\end{align}
%Expanding, this then gives us three equations,
%\begin{align}
%i \[J_{2}, \da_{3}\]- i\[J_3, \da_{2}\]  = & \omega\da_{1}\\
%-i \[J_{1}, \da_{3}\]+ i\[J_3, \da_{1}\]  = & \omega\da_{2}\\
%i \[J_{1}, \da_{2}\]- i\[J_2, \da_{1}\]  =  &\omega\da_{3}.
%\end{align}
To proceed, we note that the $J^a$ are the angular momentum operators, and thus it will prove more useful to work with total angular momentum states. Introducing the linear combinations
\begin{align}
J^{\pm} = & J_1 +\pm i J_2,\\
\da^{\pm} = & \da_1 \pm i\da_2,
\end{align}
$J^{\pm}$ can be recognized as the familiar raising and lowering operators. In in this basis the eigenvalue equations can be written
\begin{align}\label{eqn:evalpm}
\[J^+, \da_{3}\] - \[J_{3}, \da^+\] = &\omega\da^+,\\
\[J^-, \da_{3}\] - \[J_{3},\da^-\] = &-\omega\da^-,\\
\frac{1}{2}\[J^{-}, \da^{+}\] - \frac{1}{2}\[J^{+}, \da^-\] = & \omega \da_{3}.
\end{align}
We can now make use of some familiar results from quantum mechanics in order to solve for these eigenvalues and eigenfunctions. In particular, we expand the modes $\da^{i}$ into (tensor) spherical harmonics
\begin{align}
\da^i = a_{j, m}^i Y_{j, m}.
\end{align}
These spherical harmonics are normalized according to
\begin{align}\label{eqn:sphharmorthog}
\tr\[Y^{\dagger}_{\ell, m}Y_{\ell', m'}\] = N \delta_{\ell\ell'}\delta_{mm'}.
\end{align}
Noting that the action of the raising and lowering operators on these spherical harmonics is given by
\begin{align}
\[J_{3}, Y_{j, m}\] = & m Y_{j, m}\\
\[J^{\pm}, Y_{j, m}\] = & \sqrt{j(j+1)-m(m\pm1)}Y_{j, m\pm 1}
\end{align}
the eigenvalue equations at Eqn.\ (\ref{eqn:evalpm}) can be written,
\begin{align}\nn\label{eqn:sphdecomp}
\sqrt{\ell(\ell+1) - m(m+1)}a^{3}_{\ell m} = &(\omega+m+1) a^+_{\ell, m+1} \\
\sqrt{\ell(\ell+1) - m(m-1)}a^{3}_{\ell m} = &(-\omega+m-1) a^-_{\ell, m-1}
\end{align}
and
\begin{align}\nn \label{eqn:sphdecomp2}
\frac{1}{2}\sqrt{\ell(\ell+1) - m(m+1)}a^+_{\ell, m+1}\quad\qquad& \\  - \frac{1}{2}\sqrt{\ell(\ell+1) - m(m-1)} a^-_{\ell, m-1}= & \omega a^{3}_{\ell, m}.
\end{align}
Eqns. (\ref{eqn:sphdecomp}) and (\ref{eqn:sphdecomp2}) can be combined to find a single equation for the eigenvalue $\omega$,
\begin{align}
 (\ell+1+\omega)(\omega - 1)(\omega - \ell ) = 0
\end{align}
and so the spectrum of eigenmodes has eigenvalues given by
\begin{align}
\omega = & -1, \\
\omega = & -\ell-1, \\
\omega = & \ell,
\end{align}
where $\ell \in \{1, 2, ...,  N-1\}$. 

%%%%%%%%%%%%%%%%%%%%%%%%%%%%%%%%%%%%
%%
\subsection{Infrared stability of higher dimensional embeddings}
%%
%%%%%%%%%%%%%%%%%%%%%%%%%%%%%%%%%%%%

With this decomposition in hand, we can now study the fluctuations of the SU(N) gauge fields about the specific SU(2) background configuration in Eqn.\ (\ref{eqn:gaugevev}). We will work in the limit $k \to 0$ to analyse the infrared stability of the gauge field configuration, and will henceforth drop spatial gradients. It will also prove useful to work in a gauge where the modes with eigenvalues $\omega = -1$ are chosen to vanish. Note that for the 2-dimensional representation considered in the text, this choice amounts to choosing $\chi^i = 0$, or $\da^a{}_{i} = \da_{i}{}^a$, i.e. symmetric gauge. In the limit $k \to 0$, this gauge choice implies that $\da_0  = 0$ and thus we can ignore the Coulomb conditions.

The generators of SU(2) can be written in terms of the tensor spherical harmonics in the following way.
\begin{align}\label{eqn:su2sphe}
J^+ = - \sqrt{\frac{N^2 - 1}{6}}Y_{1,1},& \quad
J^- =  \sqrt{\frac{N^2 - 1}{6}}Y_{1,-1},\\
J^3  = & \sqrt{\frac{N^2 - 1}{12}}Y_{1,0}.
\end{align}
The structure of the Chern-Simons interaction term, Eqn.\ (\ref{eqn:actCS}), and the orthogonality of the spherical harmonics Eqn.\ (\ref{eqn:sphharmorthog}), Eqn.\ (\ref{eqn:su2sphe}) taken together imply that only modes with $\ell = 1$ can couple to the axion fluctuations at linear order. Since the analyses presented in the main text have  already established the conditions under which these modes are stable, we will now turn to the higher spin modes, $\ell \geq 2$.

In the $k \to 0$ limit,  the quadratic action for the fluctuations with $\ell \geq 2$ is diagonal,
\begin{align}\label{actionellgeq2}
\delta^2\mathcal{L} = &\tr\[\da^{\omega}_i\(\partial^2_{\tau}-\(g^2\phi^2(\omega^2 - \omega )+\omega g\phi\frac{\lambda}{f}\partial_{\tau}\axion\)\)\da^{\omega}_i\].
%\\ &  +\(g\phi(\omega+\omega')+\frac{\lambda}{f}\partial_{\tau}\axion\)\tr\[\epsilon^{ijk}\da^{\omega}_i\partial_{j}\da^{\omega'}_{k}\].
\end{align}
The masses of these fluctuations are then given by
\begin{align}
\frac{m^{2}_{\omega}}{H^2 \tau^2} = & \(g^2\phi^2(\omega^2 - \omega )+\omega g\phi\frac{\lambda}{f}\partial_{\tau}\axion\).
\end{align}
Now, working in quasi-de Sitter space with $a \approx -1/(H\tau)$ and making use of the equations of motion for the background Eqns.\ (\ref{slowrollpsi}) and (\ref{slowrollX}), we can rewrite this as
\begin{align}
\frac{m^2_{\omega}}{H^2} = &  m_{\psi}^2(\omega^2 - \omega)+2\omega m_{\psi}\( m_{\psi}+  \frac{1}{m_{\psi}}\),
\end{align}
where recall $m_{\psi} = g\psi/H$ is a dimensionless ${\cal O}$(few) parameter. In terms of this mass, the equations of motion for these modes can be written
\begin{align}
\frac{d^2 \da^{\omega}_i }{dx^2}+ \frac{m^2_\omega}{x^2} \da^{\omega}_i = 0 ,
\end{align}
which in this limit admits a power law solution,
\begin{align}
\da^{\omega}_{i} = A_i^{\omega}x^{\alpha},\quad
\alpha = \frac{1}{2}\(1\pm\sqrt{1-4m^2_{\omega}}\).
\end{align}
Since the background is growing as $x^{-1}$, $m^2_{\omega} < -2$ would signal pathological behavior -- i.e., a perturbative mode growing faster than the background.  The absences of such behavior implies that there are no late time instabilities in these higher spin modes. However, the presence of  parity violating terms in the quadratic action means that some of these modes may become temporarily unstable during inflation, much as we previously found for the spin-2 excitations of the two-dimensional representation. The higher spin of these fluctuations may further exacerbate this instability, potentially leading to a breakdown of perturbation theory in such circumstances.

\end{document}